\documentclass[%
 aps,
 twocolumn,amsmath,amssymb,
 reprint,%
]{revtex4-1}

\usepackage{graphicx}
\usepackage{dcolumn}
\usepackage{bm}
\usepackage{xcolor}
\usepackage[utf8]{inputenc}
\usepackage[T1]{fontenc}
\usepackage{mathptmx}
\usepackage{empheq,etoolbox}
\usepackage{soul}

\patchcmd{\subequations}
  {\theparentequation\alph{equation}}
  {\theparentequation.\alph{equation}}
  {}{}

\makeatletter
\def\@email#1#2{%
 \endgroup
 \patchcmd{\titleblock@produce}
  {\frontmatter@RRAPformat}
  {\frontmatter@RRAPformat{\produce@RRAP{*#1\href{mailto:#2}{#2}}}\frontmatter@RRAPformat}
  {}{}
}%
\makeatother
\begin{document}

\preprint{AIP/123-QED}

\title[]{Emergent scales and spatial correlations at the yielding transition of glassy materials}
\author{Stefano Aime}
\email{stefano.aime@espci.fr}
 \affiliation{Laboratoire Matière Molle et Chimie, (UMR 7167, ESPCI-CNRS) ESPCI ParisTech, 10 rue Vauquelin, 75005 Paris, France}
\author{Domenico Truzzolillo}
\email{domenico.truzzolillo@umontpellier.fr}
\affiliation{Laboratoire Charles Coulomb (L2C), UMR 5221, CNRS-Universit\'{e} de Montpellier, France}

\date{\today}

\begin{abstract}
Glassy materials yield under large external mechanical solicitations. 
Under oscillatory shear, yielding shows a well-known rheological fingerprint, common to samples with widely different microstructures.
At the microscale, this corresponds to a transition between slow, solid-like dynamics and faster liquid-like dynamics, which can coexist at yielding in a finite range of strain amplitudes.
Here, we capture this phenomenology in a lattice model with two main parameters: glassiness and disorder, describing the average coupling between adjacent lattice sites, and their variance, respectively.
In absence of disorder, our model yields a law of correspondent states equivalent to trajectories on a cusp catastrophe manifold, a well-known class of problems including equilibrium liquid-vapour phase transitions.
Introducing a finite disorder in our model entails a qualitative change, to a continuous and rounded transition, whose extent is controlled by the magnitude of the disorder. We show that a spatial correlation length $\xi$ emerges spontaneously from the coupling between disorder and bifurcating dynamics. 
With vanishing disorder, $\xi$ diverges and yielding becomes discontinuous, suggesting that the abruptness of yielding can be rationalized in terms of a lengthscale of dynamic heterogeneities.
\end{abstract}

\maketitle

\section{Introduction}

Understanding the flow of glassy materials is a scientific challenge of great importance for many practical purposes that has been animating the research activity of a wide scientific community. 
Particular effort has been devoted in the last decades to understanding the onset of flow, from the solid-like response characterizing small deformations to progressive fluidization, with a response dominated by plastic dissipation as the deformation is augmented beyond the linear regime. This transition is called yielding, and exhibits similar rheological features in samples of interest for several research fields and applications, ranging from food processing and cosmetics to oil extraction, concrete formation and pharmaceutics \cite{nicolasDeformationFlowAmorphous2018,bonnYieldStressMaterials2017}.
To capture this common phenomenology, many attempts have been made to find a unifying description of yielding, and possibly predict the conditions under which industrially-relevant samples start to flow. 
The research of the basic ingredients needed to reproduce the observed yielding behavior has been helped by elastoplastic models \cite{nicolasDeformationFlowAmorphous2018, parleyMeanFieldTheoryYielding2022}, mode-coupling theories \cite{braderNonlinearResponseDense2010,voigtmannNonlinearGlassyRheology2014}, as well as other theoretical frameworks, such as the soft glassy rheology \cite{sollichRheologySoftGlassy1997} and fluidity models \cite{picardSimpleModelHeterogeneous2002, benziUnifiedTheoreticalExperimental2019, liuMeanFieldScenarioAthermal2018}. 
Starting from a minimal description of the dynamics common to all arrested systems, these models capture the evolution of the viscoelastic moduli as a function of the strain amplitude $\gamma_0$. 
This evolution is often gradual, and it takes place over an extended range of $\gamma_0$, apparently with no univoquely well-defined yield strain. 
Yet, the extent of this range may vary, depending not only on the sample studied, but also on many experimental parameters such as sample age, previous mechanical history, deformation geometry, boundary conditions \cite{divouxDuctiletobrittleTransitionYielding2024}.
Controlling the transition between gradual and abrupt yielding, sometimes referred to as ductile and brittle yielding in analogy to failure of hard materials, is of paramount importance in many industrial contexts. However, the physics underlying these different behaviors is still poorly understood.
Significant breakthrough came recently thanks to an advanced analysis of the rheological signal, showing that even gradual yielding corresponds to the well-defined onset of non-recoverable deformation, which increases with $\gamma_0$ and becomes predominant in the regime of large deformations \cite{donleyElucidatingOvershootSoft2020,kamaniUnificationRheologicalPhysics2021}.

The notion of yielding as a transition to irreversible, or unrecoverable, deformation has been particularly explored by probing the microscopic dynamics induced by shear in rheo-optical experiments and particle-based simulations. 
The large susceptibility typical of soft materials makes microscopic dynamics a very sensitive probe of nonlinear deformations: their acceleration with increasing $\gamma_0$ revealed that, at the microscale, yielding corresponds to a well-defined transition to irreversible particle motion \cite{pineChaosThresholdIrreversibility2005,himanagamanasaExperimentalSignaturesNonequilibrium2014,knowltonMicroscopicViewYielding2014,gopalNonlinearBubbleDynamics1995}. Numerical simulations confirmed this scenario~\cite{fioccoOscillatoryAthermalQuasistatic2013,kawasakiMacroscopicYieldingJammed2016,regevOnsetIrreversibilityChaos2013,bhaumikRoleAnnealingDetermining2021,nessAbsorbingStateTransitionsGranular2020,mariAbsorbingPhaseTransitions2022,munganCyclicAnnealingIterated2019}, suggesting that it can be described in the broader framework of absorbing phase transitions \cite{hinrichsenNonequilibriumCriticalPhenomena2000,lubeckUniversalScalingBehavior2004,dickmanSelforganizedCriticalityAbsorbingstate1998,maireInterplayAbsorbingPhase2024,ziffKineticPhaseTransitions1986,ziffKineticPhaseTransitions1986,corteRandomOrganizationPeriodically2008,jeanneretGeometricallyProtectedReversibility2014,keimMechanicalMicroscopicProperties2014,tjhungHyperuniformDensityFluctuations2015,weijsEmergentHyperuniformityPeriodically2015,takeuchiDirectedPercolationCriticality2007,mobiusIrreversibilityDenseGranular2014,neelDynamicsFirstorderTransition2014,chantryUniversalContinuousTransition2017}.
Yet, the nature of the yielding transition has been the object of a longstanding debate due to apparently contrasting results: on the one hand, some experiments reveal  features typical of a discontinuous (first-order) transition, such as the abrupt jump of a microscopic order parameter \cite{kawasakiMacroscopicYieldingJammed2016, jeanneretGeometricallyProtectedReversibility2014, knowltonMicroscopicViewYielding2014, rogersMicroscopicSignaturesYielding2018, jaiswalMechanicalYieldAmorphous2016,karmakarStatisticalPhysicsYielding2010,leishangthemYieldingTransitionAmorphous2017}, the coexistence of states with dissimilar dynamics \cite{jeanneretGeometricallyProtectedReversibility2014}, and hysteretic behavior \cite{divouxRheologicalHysteresisSoft2013,munganNetworksHierarchiesHow2019}, while on the other hand, other experiments probe features typical of a continuous (second-order) transition, such as the gradual evolution of the order parameter~\cite{pineChaosThresholdIrreversibility2005,himanagamanasaExperimentalSignaturesNonequilibrium2014,fioccoOscillatoryAthermalQuasistatic2013}, sluggish relaxations towards steady viscoelastic moduli or microscopic dynamics after preshear \cite{keimYieldingMicrostructure2D2013, himanagamanasaExperimentalSignaturesNonequilibrium2014}, and emerging large fluctuations \cite{himanagamanasaExperimentalSignaturesNonequilibrium2014, knowltonMicroscopicViewYielding2014,nordstromDynamicalHeterogeneitySoftparticle2011}.

This apparent contradiction has been recently rationalized in terms of an equation of state relating the microscopic dynamics to the applied deformation~\cite{aimeUnifiedStateDiagram2023}: in this framework, yielding is regarded as a dynamic transition that is discontinuous for samples deeply quenched into the glassy state, and that becomes continuous at the glass transition.
Implemented on a lattice, this model predicts the coexistence of solid-like and fluid-like relaxation modes at yielding, giving rise to dynamic heterogeneities whose magnitude and characteristic length scale depend on the glassiness of the sample.
This scenario, confirmed experimentally by dynamic light scattering and microscopy on a variety of soft glassy samples \cite{aimeUnifiedStateDiagram2023}, suggests that gradual yielding observed at the macro scale may result from locally discontinuous processes: the relationship between microscopic dynamics and macroscopic rheology emerges therefore as a major knowledge gap towards a comprehensive understanding of yielding in soft materials.

A route towards filling this gap is to inspect and to rationalize the appearance of spatial correlation of the dynamics at the rheological yielding.
In the recent few years, experiments and simulations have unveiled the emergence of a dynamic length scale in yielding soft glasses \cite{goyonSpatialCooperativitySoft2008,bonnoitMesoscopicLengthScale2010,benziUnifiedTheoreticalExperimental2019,himanagamanasaExperimentalSignaturesNonequilibrium2014, sentjabrskajaCreepFlowGlasses2015, mukherjiStrengthMechanicalMemories2019,jopMicroscaleRheologySoft2012,vinuthaMemoryShearFlow2024,martensSpontaneousFormationPermanent2012,keimGlobalMemoryLocal2020,bhaumikAvalanchesClustersStructural2022,bhaumikYieldingTransitionTwo2022,delgadoNonaffineDeformationInherent2008,fioccoOscillatoryAthermalQuasistatic2013,manningVibrationalModesIdentify2011,schallStructuralRearrangementsThat2007} that has been related to the extent of localized plastic events. 
This length scale, also observed upon melting \cite{jackMeltingStableGlasses2016,kearnsOneMicrometerLength2010, swallenStableGlassTransformation2009,herreroTwostepDevitrificationUltrastable2023}, may be significantly larger than that of dynamic heterogeneities at rest \cite{keysExcitationsAreLocalized2011,edigerSpatiallyHeterogeneousDynamics2000,donthCharacteristicLengthGlass2001}, and may even exceed the size of the whole system as it flows in tight geometries, giving rise to size-dependent rheological behavior \cite{goyonSpatialCooperativitySoft2008}.
Similarly, abrupt yielding in soft amorphous materials has been observed as individual plastic events organize in system-spanning shear bands \cite{ederaYieldingMicroscopeMultiscale2024,ozawaRandomCriticalPoint2018,pollardYieldingShearBanding2022,pommellaRoleNormalStress2020,lebouilEmergenceCooperativityPlasticity2014, houdouxPlasticFlowLocalization2018, argonPlasticFlowDisordered1979, maloneyUniversalBreakdownElasticity2004, dasguptaMicroscopicMechanismShear2012,antonaglia_bulk_2014}.
This behavior exhibits strong analogies with quasi-brittle failure in disordered hard materials, such as wood, concrete, composites, rocks and metallic alloys \cite{alegreExtensionMonkmanGrantModel2014,koivistoPredictingSampleLifetimes2016, garcimartinStatisticalPropertiesFracture1997, kandulaDynamicsMicroscalePrecursors2019, renardMicroscaleCharacterizationRupture2017, locknerQuasistaticFaultGrowth1991, locknerRoleAcousticEmission1993, schuhAtomisticBasisPlastic2003}, which has been described as a critical-like phenomenon with a characteristic length scale that grows and eventually diverges as macroscopic failure is approached 
\cite{zollerObservationGrowingCorrelation2001,potirakisNaturalTimeAnalysis2013,girardDamageClusterDistributionsSize2012,sornettePredictabilityCatastrophicEvents2002,sornetteScalingRespectDisorder1998a, zapperiPlasticityAvalancheBehaviour1997}.
Similar ideas guide the research of precursory signals anticipating granular or frictional slip \cite{johnsonAcousticEmissionMicroslip2013, houdouxMicroslipsExperimentalGranular2021, ostapchukMechanismLaboratoryEarthquake2020, gvirtzmanNucleationFrontsIgnite2021} and even rockfalls and earthquakes \cite{kromerIdentifyingRockSlope2015, scuderiPrecursoryChangesSeismic2016, dixonEarthquakeTsunamiForecasts2014}
Conversely, when failure occurs through sparse, uncorrelated plastic events, ductile behavior is observed \cite{vasseurHeterogeneityKeyFailure2015, kadarRecordStatisticsBursts2020, ozawaRandomCriticalPoint2018, yuAtomicscaleHomogeneousPlastic2021,sunPlasticityDuctileMetallic2010}: this further confirms that the length scale characterizing the spatial correlation of plastic events is key to determine the yielding behavior of amorphous solids.

Yet, despite its relevance, little is known about the origin of this length scale, its connection to the properties of the sample, and its emergence under a given applied deformation. Structural heterogeneity has been shown to suppress spatial correlations and to enhance ductility \cite{vasseurHeterogeneityKeyFailure2015, idrissiAtomicscaleViscoplasticityMechanisms2019, richardPredictingPlasticityDisordered2020,rossiEmergenceRandomField2022}, but even without significant structural differences, the same effect can be obtained through mechanical history and sample equilibration \cite{rodneyModelingMechanicsAmorphous2011,didioTransientDynamicsSoft2022,ederaTuningResidualStress2024,vasoyaNotchFractureToughness2016,fanEffectsCoolingRate2017,ozawaRandomCriticalPoint2018, keimGlobalMemoryLocal2020,bhaumikRoleAnnealingDetermining2021,bhaumikYieldingTransitionTwo2022}.
Recent experimental works on soft and metallic glasses \cite{kumarCriticalFictiveTemperature2013,sopuRejuvenationEngineeringMetallic2023, bianSignatureLocalStress2020, ederaTuningResidualStress2024}, highlighed the role of structural and mechanical disorder for the sharpness of mechanical yielding. Theoretical and numerical works \cite{ozawaRandomCriticalPoint2018,richardPredictingPlasticityDisordered2020,patinetConnectingLocalYield2016,rossiFarfromequilibriumCriticalityRandomfield2023,sethnaCracklingNoise2001,rossiFiniteDisorderCriticalPoint2022a} have rationalized this result, showing that more disordered samples yield through a larger number of smaller plastic events, resulting in the smooth evolution of the stress. By contrast, better-annealed glassy samples, characterized by a reduced disorder in local mechanical properties \cite{patinetConnectingLocalYield2016}, yield more abruptly, and tend to develop shear bands or fractures \cite{parmarStrainLocalizationYielding2019, ozawaRandomCriticalPoint2018,rossiFarfromequilibriumCriticalityRandomfield2023,ozawaRareEventsDisorder2022}.

Here, we study the effects of disorder on the sharpness of the dynamic transition observed at the microscale upon yielding. Building upon the on-lattice model developed to reproduce the stroboscopic dynamics measured by rheo-DLS \cite{aimeUnifiedStateDiagram2023}, we study how disorder affects dynamic correlations at yielding.
In absence of disorder, our model yields a law of correspondent states equivalent to trajectories on a cusp catastrophe manifold, a well-known class of problems including equilibrium liquid-vapour phase transitions.
Introducing a finite disorder in our model entails a qualitative change, to a continuous and rounded transition, whose extent is radically affected by the magnitude of the disorder. We show that a spatial correlation length $\xi$ emerges spontaneously from the coupling between disorder and bifurcating dynamics. 
With vanishing disorder, $\xi$ diverges and yielding becomes discontinuous, suggesting that the abruptness of yielding can be rationalized in terms of a diverging lengthscale of dynamic heterogeneities.

The reminder of this paper is structured as follows. In section \ref{exps} we highlight the experimental grounds on which our model is built. In section \ref{latticegen} we formulate the model in its general form and we detail its physical bases. In section \ref{sec:mf} we discuss the results obtained in mean field approximation, and we formulate a law of corresponding states allowing to map distinct fluids on the same equation of state.
In section \ref{Simul} we discuss the role of disorder in our model by implementing it on a 2D lattice. Simulations indicate that with increasing disorder, yielding undergoes a sharp-to-gradual transition that exhibits strong system size effects. 
This transition is rationalized in section \ref{catastrophe} by recasting our model as an elementary catastrophe, and showing that sharp yielding is a manifestation of metastability, quantified in terms of an effective potential.
In section \ref{brittleductile} we show that yielding is associated to a dynamic correlation length that grows as coupling disorder is reduced, and eventually exceeds the system size, entailing the transition to abrupt failure, which therefore emerges as a finite-size effect.
Finally, in section \ref{conclusions} we make some concluding remarks, we summarize the key results and we highlight the potential perspectives of our work.

\begin{figure*}[htbp]
  \includegraphics[width=0.8\linewidth]{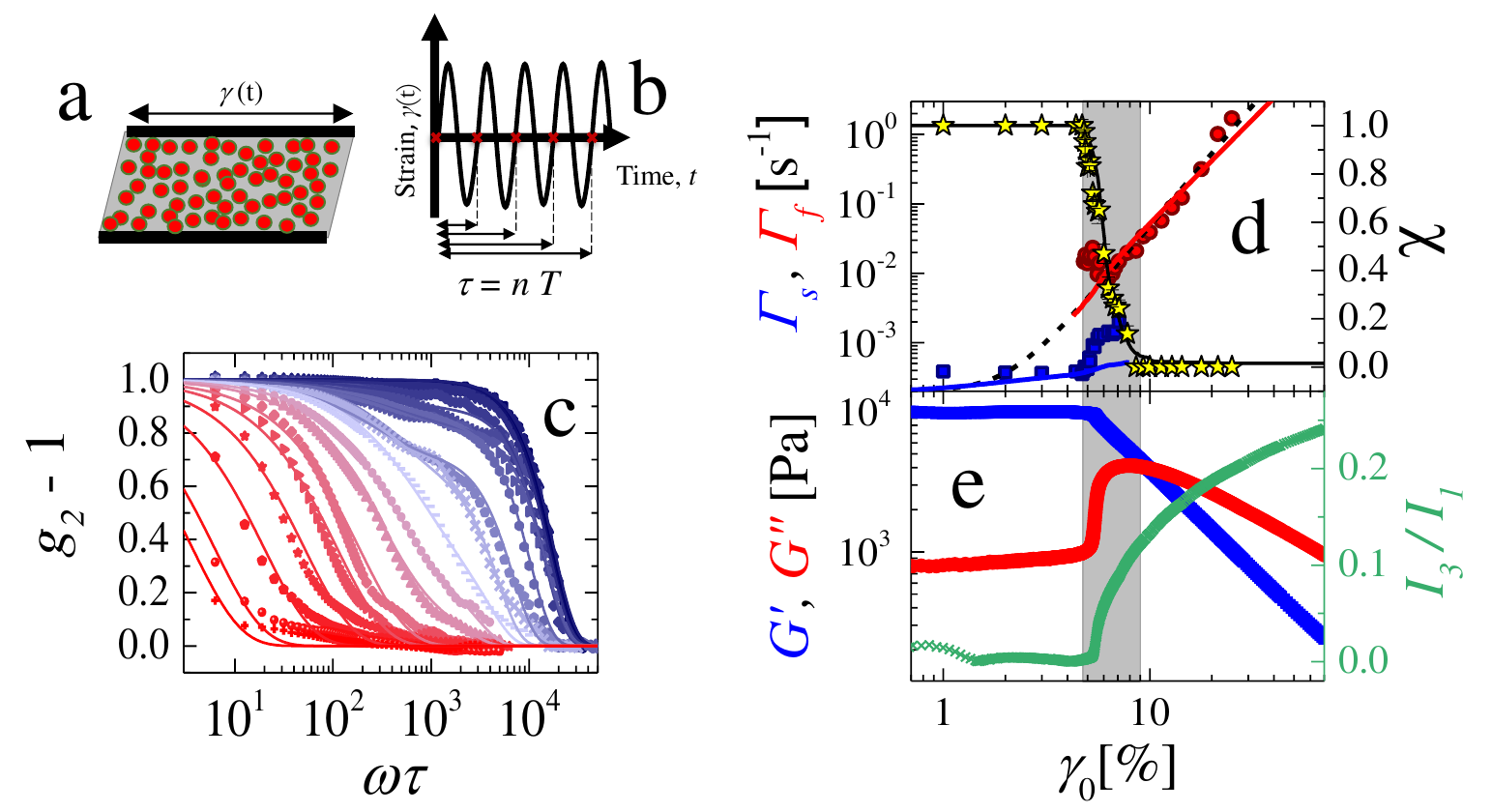}\\
  \caption{\textbf{Yielding as a dynamic transition}. a) Sketch of the parallel plate geometry employed in rheo-DLS experiments. b) Schematic of the stroboscopic collection of speckle images during one oscillatory experiment: One image is recorderd every oscillation period $T$. The lag time is then a multiple of a period, $\tau=nT$. c): Intensity autocorrelation functions under oscillatory shear, for concentrated Ludox nanoparticles ($\varphi=0.45$), plotted \textit{vs} the normalized time delay $\omega \tau$, with $2\pi/\omega$ the period of the oscillations. The strain amplitude $\gamma_0$ increases from blue to red shades, spanning the rheological yielding transition. Symbols: experimental data. Lines: fits using Eq.~\ref{eqn:g1_dblexp}. The compressed exponent $\beta_s$ of the slow mode is 1.9, fixed by the spontaneous dynamics measured on the samples at rest. The stretching exponent $\beta_f$ is 0.9. The fitting parameters for the same samples are shown in panel d). Left axis: $\gamma_0$ dependence of the rates $\Gamma_s$ (slow mode, blue squares) and $\Gamma_f$ (fast mode, red circles). Right axes: relative amplitude $\chi$ of the slow mode  (stars). Lines: numerical results of the general model for the yielding transition discussed in the text. The best fit parameters are reported in \cite{aimeUnifiedStateDiagram2023}. 
  In panels d) and e), the gray shaded rectangles indicate the range of $\gamma_0$ from the onset of the third harmonic in stress response to the strain crossover where $G'=G''$.\label{fig:exps}}
\end{figure*}
\twocolumngrid

\section{Experimental findings}\label{exps}
Dynamic Light Scattering (DLS) quantifies the microscopic dynamics via the intensity correlation function $g_2(\tau)-1$, which decays from 1 to 0 as microscopic displacements grow beyond a length scale $\pi/q \approx (0.1-1)\mu m$, set by the scattering vector $q$ \cite{berneDynamicLightScattering1990}.
To investigate the microscopic dynamics under deformation through rheo-DLS, we apply an oscillatory shear with angular frequency $\omega$ and strain amplitude $\gamma_0$, and stroboscopically measure $g_2(\tau)-1$, for lag times $\tau$ that are multiples of the oscillation period.
We find that for small enough $\gamma_0$ the dynamics are independent of the applied shear, and that dynamics accelerate with growing strain at larger $\gamma_0$. Concomitantly, the shape of $g_2-1$ evolves from a compressed exponential at low $\gamma_0$ to a stretched exponential at large $\gamma_0$, with an intermediate regime where correlation functions exhibit a two-step decay well described by a linear combination of the two modes, as shown in Figure \ref{fig:exps}-c.
To model this evolution, correlation functions measured at all strain amplitudes are fitted by the following expression:

\begin{multline}
	g_2(\tau)-1 =\left\{ \chi\exp\left[-\left(\Gamma_s \tau\right)^{\beta_s}\right]\right.\\
	\left.+\left(1-\chi\right)\exp\left[-\left(\Gamma_f \tau\right)^{\beta_f}\right] \right\}^2 \,,
	\label{eqn:g1_dblexp}
\end{multline}

where $\Gamma_s$ and $\Gamma_f$ are the relaxation rates of the slow and fast decay modes, respectively; $\chi$ is the relative weight of the slow decay mode, and $\beta_f$ and $\beta_s$ are the stretching and compressing exponents of the two modes, shared by all correlation functions.
The fit results suggest that, as $\gamma_0$ is increased, three distinct regimes can be identified:
\begin{enumerate}
    \item[i] For $\gamma_0<5\%$, the correlation function decays as a single compressed exponential ($\chi=1$), with an exponent $\beta_s \approx 1.9$ and a slow relaxation rate $\Gamma_s$, hardly dependent on $\gamma_0$. We find that $\Gamma_s$ is approximately equal to $\Gamma_0=10^{-4}~\mathrm{s}^{-1}$, the relaxation rate measured in absence of shear. Therefore, we refer to this slow, compressed-exponential relaxation as the \emph{solid-like} dynamics.
    \item[ii] For $5\% < \gamma_0 < 9\%$, a second, faster mode, characterized by a stretched exponential relaxation, with $\beta_f \approx 0.9$, adds to the spontaneous relaxation mode, and becomes increasingly dominant with increasing $\gamma_0$, as $\chi$ decays from 1 to 0.
    \item[iii] Finally, for $\gamma_0 > 9\%$, the correlation functions decay as single stretched exponentials ($\chi=0$), with a relaxation rate $\Gamma_f$ growing as a power law of $\gamma_0$, with a sample-dependent exponent $n\approx 3$. Because such stretched-exponential dynamics are similar to those found in dense suspensions just below the glass transition~\cite{philippeGlassTransitionSoft2018}, we refer to this faster relaxation mode as \emph{fluid-like} dynamics.
\end{enumerate}
Comparison with the rheological response indicates that these three regimes correspond to: (i) the linear regime, characterized by a purely harmonic stress response and strain-independent viscoelastic moduli, with $G'>G"$; (ii) the yielding regime, with an increasingly inharmonic stress response and an overshoot in the first-harmonic $G"(\gamma_0)$; (iii) the fluidized regime, with $G">G'$ and both nonlinear moduli decaying as power laws of $\gamma_0$, as shown in Figure~\ref{fig:exps}-e.

Comparing the experimental results reported in Figure~\ref{fig:exps} to the result of analogous experiments on different soft glassy materials~\cite{aimeUnifiedStateDiagram2023}, we find that that this scenario does not depend on sample details such as the nature of the microscopic constituents, their soft repulsive interactions or their concentration, as long as the sample exhibits glassy dynamics at rest. 
The most remarkable feature changing from sample to sample is the abruptness of the transition, quantified by the extent of regime (ii). For instance, we find that, for a given class of samples, say, charged nanoparticles as in Fig.~\ref{fig:exps}, more concentrated samples exhibit more abrupt yielding. A similar trend is found as one given sample is probed at increasingly large angular frequency $\omega$, or if it is aged for longer time prior to being tested. 
These findings suggests that a deeper understanding of the dynamic transition observed at the microscale can reveal the very nature of yielding: this motivates the search for a general model able to describe this behavior.


\section{Lattice model with facilitated advection}\label{latticegen}

To model the microscopic relaxation rate $\Gamma$, we assume that it can be written as the sum of two independent processes: the spontaneous dynamics and the shear-induced dynamics, characterized by relaxation rates $\Gamma_0$ and $\Gamma_{sh}(\gamma_0)$, respectively. 
Previous works \cite{wyssStrainRateFrequencySuperposition2007, hessYieldingStructuralRelaxation2011a} have postulated a generic power-law dependence of $\Gamma_{sh}$ on the macroscopic shear rate. Here, we adapt this ansatz to oscillatory rheology, taking $\dot{\gamma}_0=\omega\gamma_0$ as a representative shear rate, and using the experimental observation that $\Gamma_{sh}\propto \omega$ to properly capture the dependence on $\omega$~\cite{aimeUnifiedStateDiagram2023}. We obtain:

\begin{equation}\label{eq:eos_nocoupling}
    \Gamma=\Gamma_0 + K\omega\gamma_0^n \,.
\end{equation}

This result captures well the limits of small and large $\gamma_0$, but fails describing the well-defined transition found in experiments, rather predicting a continuous, smooth acceleration of $\Gamma$ with increasing $\gamma_0$, as shown by the black dashed line in Fig. \ref{fig:exps}-d.
To improve the model predictions in the yielding region, we consider the role of dynamic and structural heterogeneities in the sample~\cite{hallettLocalStructureDeeply2018, gokhaleGrowingDynamicalFacilitation2014, mishraDynamicalFacilitationGoverns2014} arising from dynamic correlations between nearby sample regions.
Indeed, glassy dynamics at rest are known to be governed by dynamic facilitation~\cite{mishraDynamicalFacilitationGoverns2014, gokhaleGrowingDynamicalFacilitation2014}: portions of the system with faster dynamics promote structural relaxation of nearby regions. By analogy, we introduce a facilitated advection (FA) model that couples the shear-induced dynamics of nearby sample regions in an Ising-like fashion. 
We implement this model on a lattice and we assume that the shear-induced relaxation of a generic site $i$ does not depend only on the applied macroscopic shear field but also on the local environment, that screens the imposed global deformation and delays local rearrangements:  

\begin{equation}\label{eq:concept}
    \Gamma_i=\Gamma_0 + K\left[\tau_{sh}(\gamma_0)+\tau_{FA}(i,j)\right]^{-1}.
\end{equation}

Here $\tau_{sh}(\gamma_0)=(\omega\gamma_0^n)^{-1}$, whereas $\tau_{FA}(i,j)$ is a characteristic time that depends on the dynamics of two adjacent lattice sites $i$ and $j$, and represents the leading-order expansion of the perturbation of the shear-imposed time scale due to local dynamics.  
$\Gamma_i$ is therefore the local relaxation of the $i$-th lattice cell representing a mesoscopic region of the sample, that reduces to the rate given by Eq.~\ref{eq:eos_nocoupling} only when the dynamics are uncorrelated. This is the case of ergodic fluids with no dynamical heterogeneities and very poor spatial correlation of local dynamics, characterized by $\tau_{FA}(i,j)=0$.
More generally, we model the effect of FA by introducing a term, $\tau_{FA}(i,j)$, that penalizes strong differences in the dynamics of nearby cells, while preserving the small and large amplitude limits of Eq.~\ref{eq:eos_nocoupling}:

\begin{equation}\label{eq:ising}
    \Gamma_i=\Gamma_0 + \omega K\left(\frac{1}{\gamma_0^n}+\frac{\omega^2}{N_z}\sum\limits_{j=1}^{N_z}\frac{\alpha_{ij}}{\Gamma_i\Gamma_j}\right)^{-1}
\end{equation}

Here $\alpha_{ij}$ are positive constants coupling the dynamics of the site $i$ to the ones of its $N_z$ nearest neighbors. 
In the limit of $\alpha_{ij}=0$, Eq.~\ref{eq:ising} reduces to the original ansatz of Eq.~\ref{eq:eos_nocoupling} with $\Gamma_i=\Gamma$ for all lattice sites. 
With $\alpha_{ij}>0$, the local rates vary between lattice sites: lower-than-average $\Gamma_j$ result in larger FA term for its neighbors, decreasing their rate relative to sites in faster-relaxing neighborhoods.
Therefore, the FA term introduced in Eq.~\ref{eq:ising} suppresses differences between neighbor lattice sites, like in dynamic facilitation \cite{chandlerDynamicsWayForming2010,speckDynamicFacilitationTheory2019}.

\section{Mean field approximation\label{sec:mf}}

To illustrate the new features introduced by FA, we first set all coupling constants to the same value $\alpha$, and we study the resulting mean-field equation of state (EOS) for the average relaxation rate $\Gamma(\gamma_0)$:

\begin{equation}\label{eq:ising_MF_2}
    \Gamma= \Gamma_0 + \omega K \gamma_0^n\left(1+\frac{\alpha\omega^2\gamma_0^n}{\Gamma^2}\right)^{-1}
\end{equation}

For any given set of model parameters $\{\Gamma_0, \omega, K, n, \alpha\}$, this expression reduces to Eq.~\ref{eq:eos_nocoupling} both in the limit of small strains, $\gamma_0^n \ll \Gamma_0^2/\alpha\omega^2$, and of large strains, $\gamma_0^n \gg \alpha/K^2$, where $\Gamma^2\sim \omega^2 K^2 \gamma_0^{2n} \gg \alpha\omega^2\gamma_0^n$. However, for $K\lesssim \alpha\omega/\Gamma_0$, there exists an intermediate range of strain amplitudes where the model with FA exhibits a qualitatively new behavior. 
To show this, we recast Eq.~\ref{eq:ising_MF_2} in a form that s formally identical to the Van der Waals (VdW) equation of state for real gases, with $\Gamma$, $1/\gamma_0^n$ and $\omega K$ playing the role of volume, pressure and temperature, respectively~\cite{landauStatisticalPhysicsLifshitz2011}:

\begin{equation}\label{eq:ising_MF}
    \left(\Gamma-\Gamma_0\right)\left(\frac{1}{\gamma_0^n}+\frac{\alpha\omega^2}{\Gamma^2}\right)=\omega K
\end{equation}

\begin{figure}[htbp]
  \includegraphics[width=9.8cm]{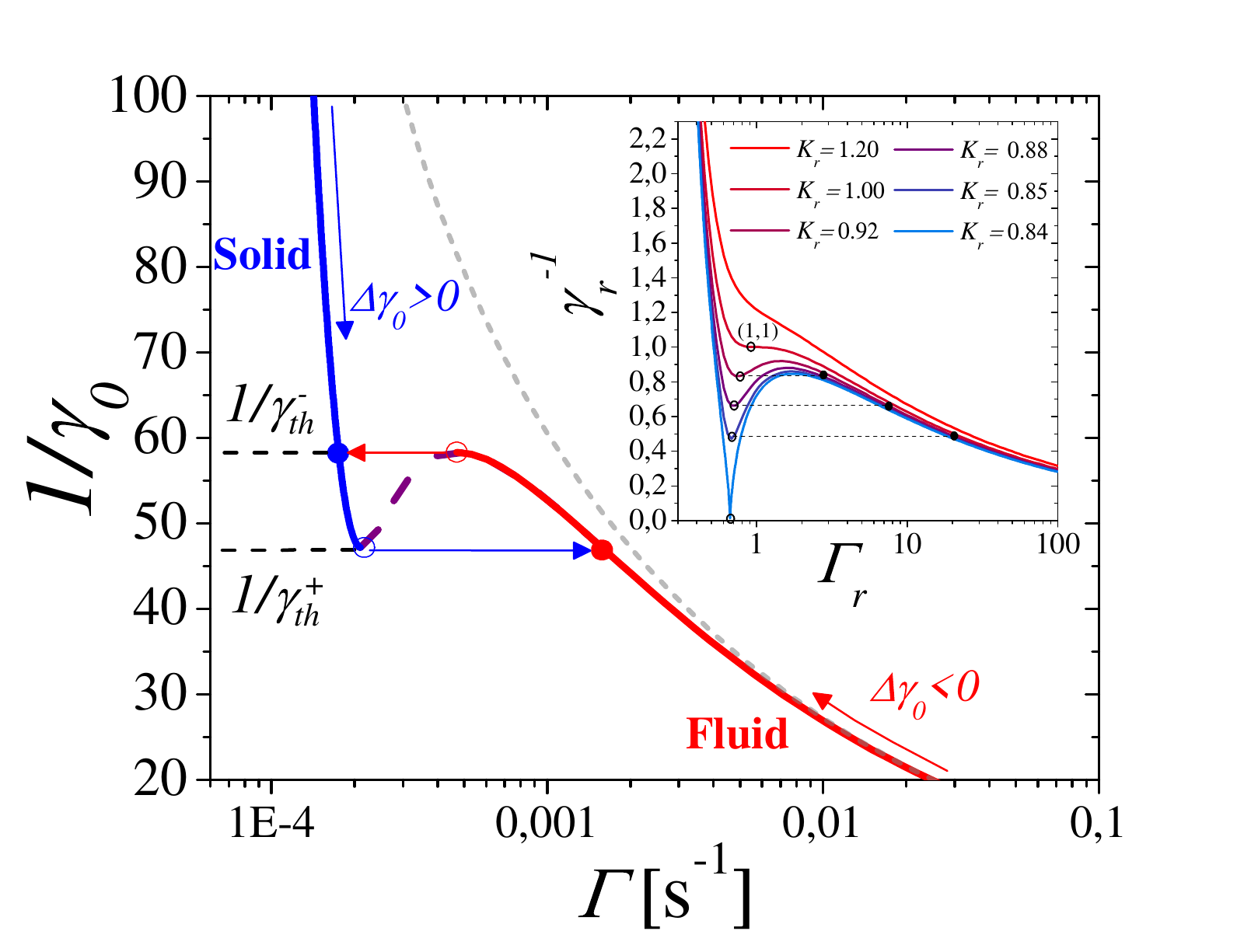}\\
  \caption{Inverse strain amplitude $\gamma_0^{-1}(\Gamma)$ obtained according to equation \ref{eq:eos_withcoupling}. Transitions following an ascending ramp ($\Delta\gamma_0>0$, increasing $\gamma_0$ from the solid branch) and a descending ramp ($\Delta\gamma_0<0$, decreasing $\gamma_0$ from the fluid branch) occur at the two extrema of the curve for $\gamma_0=\gamma_{th}^+$ and $\gamma_0=\gamma_{th}^-$, as indicated by the horizontal blue and red arrows respectively. Here $K=300$, $n=3$, $\beta=2$, $\alpha=0.17$, $\omega=0.66$ rad/s and $\Gamma_0=10^{-4}$ s$^{-1}$. The gray dashed lines shows as reference the EOS for $\alpha=0$ with the other parameters left unaltered.  
  Inset: Reduced equation of state $\gamma_r^{-1}(\Gamma_r)$ obtained for $n=3$, $\beta=2$ and different values of $K_r$ as indicated in the figure. The horizontal dotted lines indicate the predicted yielding for ascending strain amplitude ramps, $\Delta\gamma_0>0$ (decreasing $\gamma_r^{-1}$), starting from a sample at rest.}\label{fig:eos}
\end{figure}

Guided by this analogy, we represent the mean-field equation of state (EOS) as curves in the $1/\gamma_0$ vs $\Gamma$ plane, analogous to isotherms in $p-V$ diagrams for real gases.
These are curves defined by the equation:

\begin{equation}\label{eq:eos_withcoupling}
    \frac{1}{\gamma_0}=\left[\frac{\omega K}{\Gamma-\Gamma_0}
    -\alpha\left(\frac{\omega}{\Gamma}\right)^{\beta}\right]^{1/n}\,,
\end{equation}

\noindent where the exponent $\beta > 1$ is introduced to generalize Eq.\ref{eq:ising} (in which $\beta=2$) to longer-range dynamic coupling, extending to $\beta-1$ lattice sites (see \cite{SeeSupplementalMaterial} for the expression on lattice). 
Like in real gases, the EOS of Eq.~\ref{eq:eos_withcoupling} describes $\gamma_0^{-1}(\Gamma)$ curves that decrease monotonically for sufficiently large $K$, and that become nonmonotonic as $K$ is reduced below a critical value, as shown in Fig.~\ref{fig:eos}. 

In a typical oscillatory strain sweep, the system is prepared in an aged and stationary glassy state at rest, and then it is subject to oscillatory deformation at increasingly large strain amplitude. We model the evolution of the microscopic dynamics by fixing $K$ and $\alpha$, and following the increase in $\Gamma$ as $\gamma_0$ is increased in Eq.~\ref{eq:eos_withcoupling}. This corresponds to following a path going from the top-left to the bottom-right corner of the $\gamma_0^{-1}-\Gamma$ diagram in Fig.~\ref{fig:eos}. For small $\gamma_0$, corresponding to the linear regime probed by rheology, $\Gamma(\gamma_0)$ is nearly constant as the EOS exhibits a very steep, solid-like branch analogous to that of nearly-incompressible VdW liquids.
The presence of a local minimum in Eq.~\ref{eq:eos_withcoupling} sets an upper bound, $\gamma_{th}^+$, for the strain amplitude that can be attained along this section of the EOS.
Beyond $\gamma_{th}^+$, $\Gamma$ can be no longer increased continuously with increasing $\gamma_0$. Instead, a small increase of $\gamma_0$ causes $\Gamma$ to jump discontinuously to the \textit{fluid} branch of the EOS, $\Gamma_f(\gamma_0)$, shown as a red line in Fig.~\ref{fig:eos}. This discontinuous jump represents an abrupt yielding event.
Conversely, if $\gamma_0$ is decreased starting from the fluidized state in a reverse strain sweep, the system will follow the fluid branch until the local maximum at a second threshold strain amplitude, $\gamma_{th}^-<\gamma_{th}^+$, where it discontinuously jumps back on the solid branch. Therefore, in mean field, our model predicts an abrupt, discontinuous yielding with hysteresis for a range of model parameters such that Eq.~\ref{eq:eos_withcoupling} gives nonmonotonic $\gamma_0^{-1}(\Gamma)$.

While still unable to fully capture the coexistence of solid-like and liquid-like dynamics for a finite range of $\gamma_0$, this mean field model substantially improves Eq.~\ref{eq:eos_nocoupling} by introducing a well-defined dynamic transition at yielding. Remarkably, this is achieved with the introduction of one single additional parameter, the coupling constant $\alpha$, which can be adjusted to quantitatively reproduce the magnitude of the dynamic acceleration measured upon yielding for each sample. In an attempt to rationalize the impact of model parameters on the predicted dynamic acceleration, we refer once more to the formal analogy with VdW gases: there, the liquid-vapor phase transition of different gases at different temperatures can be described in a universal manner by the so-called law of corresponding states, stating that the behavior of a real gas in a given thermodynamic state $(p,V,T)$ only depends on the distance from the critical point.
In an analogous way, here we seek for a normalized description in terms of rescaled variables $\{\gamma_r^{-1}=\gamma_c/\gamma_0$, $\Gamma_r=\Gamma/\Gamma_c$, $K_r=K/K_c\}$, where $(\gamma_c^{-1}, \Gamma_c, K_c)$ defines the critical point, i.e. the horizontal inflection point of the EOS, satisfying the condition: $\partial\gamma_0^{-1}/\partial\Gamma = \partial^2\gamma_0^{-1}/\partial\Gamma^2=0$. Imposing this condition in Eq.~\ref{eq:eos_withcoupling} we obtain: 

\begin{subequations}
  \label{eq:criticalparams}
  \begin{empheq}{align}
    \Gamma_c&=\Gamma_0\frac{\beta+1}{\beta-1} \label{eq:criticalrate}\\
    \gamma_c^{-1}&=\left[\alpha\left(\frac{\omega}{\Gamma_0}\right)^\beta\left(\frac{\beta-1}{\beta+1}\right)^{\beta+1}\right]^{1/n} \label{eq:criticalgamma}\\
    K_c&=4\alpha\left(\frac{\omega}{\Gamma_0}\right)^{\beta-1}\frac{\beta(\beta-1)^{\beta-1}}{(\beta+1)^{\beta+1}} \label{eq:criticalK}
  \end{empheq}
\end{subequations}

Rewriting Eq.~\ref{eq:ising_MF} in terms of these rescaled variables, we obtain a reduced equation of state:

\begin{equation}\label{gen-EOS-beta1}
    \left(\Gamma_r-\frac{1}{\mu}\right)\left(\gamma_r^{-n}+\frac{\mu}{\Gamma_r^{\beta}}\right)=\frac{4\beta}{\beta^2-1}K_r
\end{equation}

\noindent where $\mu=(\beta+1)/(\beta-1)$, and $K_r$ plays the role of a reduced temperature: $K_r<1$ defines the range of model parameters for which "iso-$K_r$" $\gamma^{-1}(\Gamma)$ curves are nonmonotonic and therefore describe yielding soft solids.
Quantitative differences in the yielding behavior of different materials are incorporated in the value of $K_r$: for $K_r \lesssim 1$, $\Gamma(\gamma_0)$ undergoes a sizeable increase along the solid branch prior to yielding, upon which it experiences a relatively small jump to reach the fluid branch. This behavior mimics the gradual yielding observed for soft materials close to the glass transition \cite{knowltonMicroscopicViewYielding2014}. 
Conversely, for smaller $K_r$, $\Gamma(\gamma_0)$ is nearly constant in the solid branch, and the discontinuity at $\gamma_{th}^+$ grows, mimicking the larger dynamic acceleration upon yielding that is observed in experiments~\cite{knowltonMicroscopicViewYielding2014} and numerical simulations~\cite{ozawaRandomCriticalPoint2018} on samples that have been well equilibrated deep in the glassy regime.
Therefore, we refer to $1-K_r$ as the sample \textit{glassiness}.
From Eq.~\ref{eq:criticalK}, we find that $K_r$ decreases with both increasing $\alpha$, indicating that stronger FA coupling enhances dynamic acceleration upon yielding, and with increasing $\omega$, indicating that yielding is time-dependent, like the glassy state itself, which is characterized by a finite relaxation time $\Gamma_0$.
As longer timescales are probed by reducing the Deborah number $\mathrm{De}=\omega/\Gamma_0$, the sample will be effectively less glassy, with a progressively smaller jump of the relaxation rate $\Gamma$ at yielding, until $\omega/\Gamma_0=K/4\alpha$, a threshold beyond which the model predicts fluid-like behavior at all strain amplitudes, as $K_r>1$.

The amount of dynamic acceleration prior to yielding and the jump of relaxation rate at the transition also depend on $\beta$, which represents the spatial range of FA coupling. In particular, for increasing $\beta$ we find that the discontinuity of $\Gamma$ at yielding increases, and that the value of $\Gamma$ just before yielding decreases, approaching $\Gamma_0$, as detailed in~\cite{SeeSupplementalMaterial}. 
The result of an augmented range of FA coupling is in line with enhanced fluidization and stress drops that emerge in numerical simulations of deeply quenched glasses and percolating networks, where long-ranged stress correlations determine rigidity at rest and possibly brittleness under shear \cite{ tongEmergentSolidityAmorphous2020,colomboMicroscopicPictureCooperative2013,ozawaRandomCriticalPoint2018}.

On a final note, we observe that as $K_r$ is decreased, the normalized threshold strain $\gamma_{r,th}^+$ increases. Interestingly, we find that $\gamma_{r,th}^+$ diverges for a critical, finite value of $K_r$: $K_{rm}=(1+\beta^{-1})^{\beta+1}/4$. This singularity cannot be properly discussed in the framework of the mean field model, but hints at a second transition, to a different, more catastrophic failure mechanism.\\

\section{Effect of disorder: numerical simulations\label{Simul}}

The mean-field model described by Eqs.~\ref{eq:ising_MF_2}-\ref{gen-EOS-beta1} accounts for both the linear and the fully-fluidized regimes, predicting an abrupt transition between the two states. This is in contrast with the experimental observation that yielding may manifest as a gradual process, with fast and slow relaxation modes coexisting in a finite range of strain amplitudes \cite{aimeUnifiedStateDiagram2023}. 
To account for this coexistence, we discuss model predictions beyond mean field, by considering the effect of structural disorder on the local heterogeneity of shear-induced relaxations in glassy materials \cite{ortliebProbingExcitationsCooperatively2023,cubukStructurepropertyRelationshipsUniversal2017, tongStructuralOrderGenuine2019}. Hereafter, without loss of generality, we will focus on the particular case $\beta=2$, describing short-range FA coupling.

\begin{figure}[htbp]
  \includegraphics[width=\columnwidth]{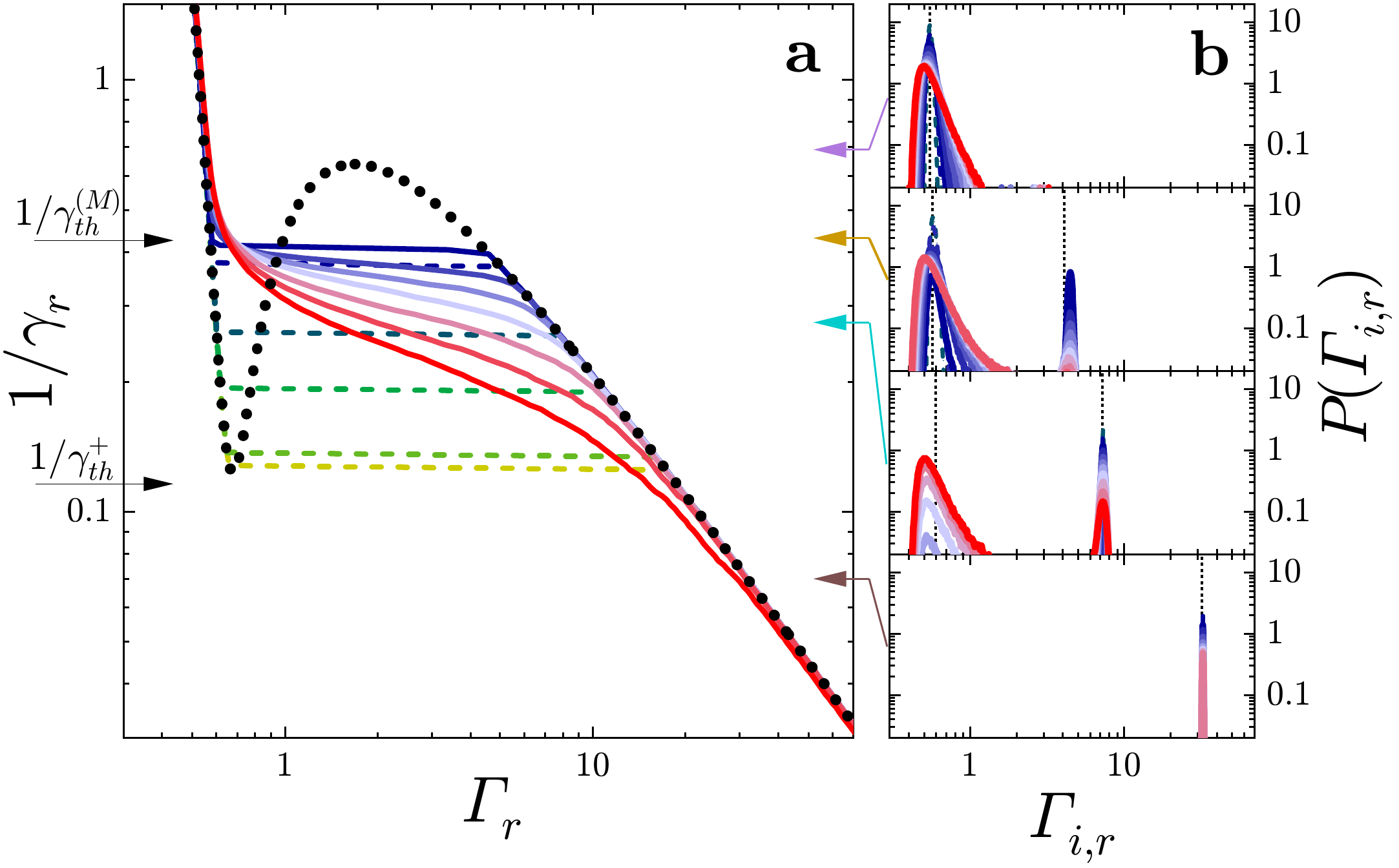}\\
  \includegraphics[width=\columnwidth]{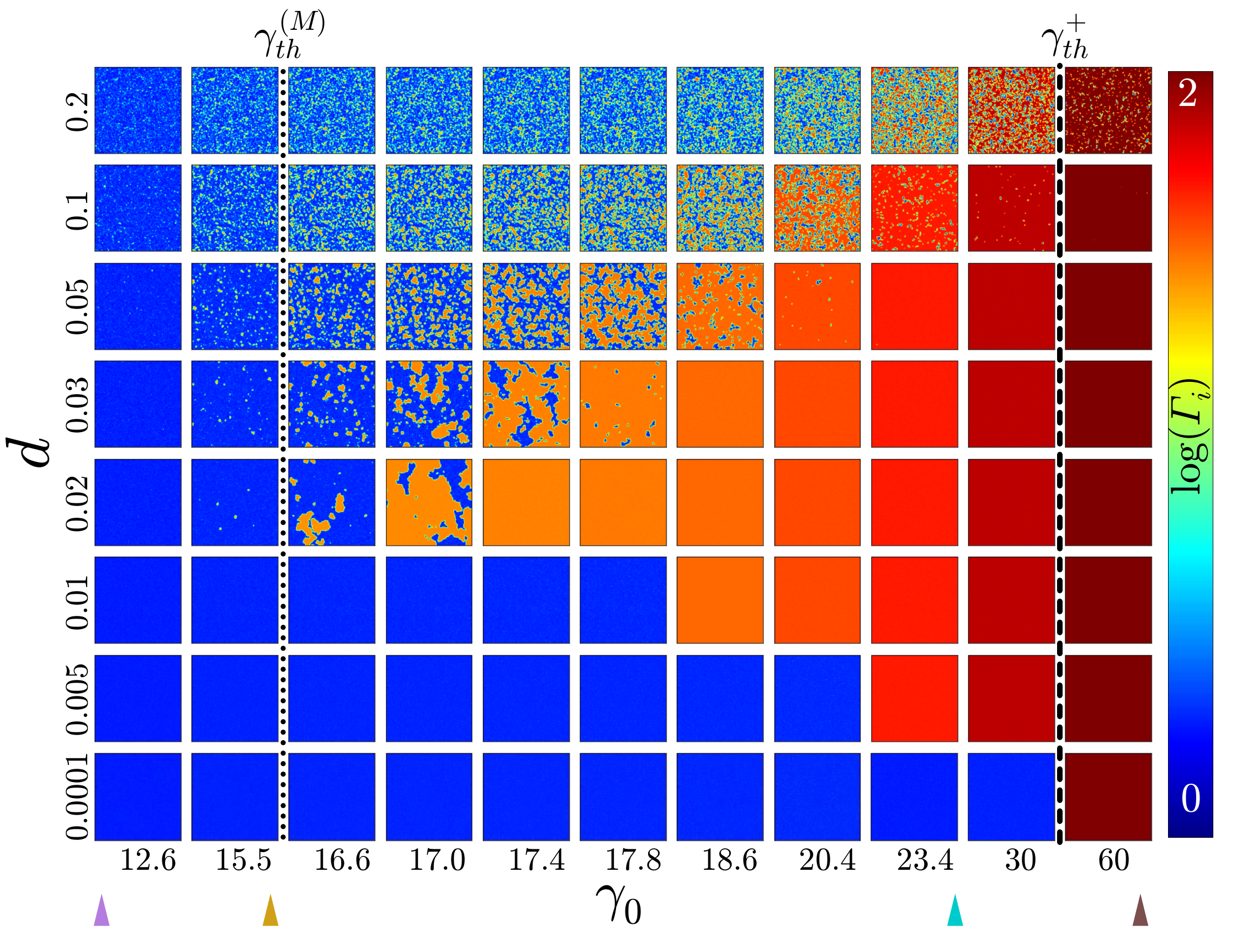}\\
  \caption{\textbf{Effect of disorder in numerical simulations}. (a) $\gamma_r^{-1}(\Gamma_r)$ diagrams obtained by averaging local rates $\Gamma_i$ in simulations with $K_r=0.86$ and disorder $d$ increasing from $10^{-5}$ to 0.175 (green to blue to red shades). Dashed lines represent simulations where yielding was abrupt due to finite size effects. Dotted line: mean-field equation of state, eq.~\ref{gen-EOS-beta1}, with $\beta=2$ and $K_r=0.86$. 
  (b) Histogram of local rates $\Gamma_i$, with $K_r=0.86$ and $\gamma_r$ increasing from top to bottom, as denoted by the colored arrows: $1/\gamma_r=0.69$ (purple), 0.43 (gold), 0.275 (cyan), 0.07 (brown). Vertical dashed black lines represent mean field values $\Gamma^s$, $\Gamma^f$. Bottom panel: snapshots of 2D configuration of local rates. Rate increases from blue ($\Gamma=1$) to red shades ($\Gamma=100$) in a logarithmic fashion, as shown by the color bar. Colored arrows mark strain amplitudes shown in panel b, dashed and dotted black vertical lines mark the reference threshold strain amplitudes predicted by mean-field, $\gamma_{th}^+$, and by the Maxwell rule, $\gamma_{th}^{(M)}$, respectively. 
  }\label{sim_overview}
\end{figure}

In presence of disorder, the local rates $\Gamma_i$ vary from site to site, and to find their values, one needs to solve $N$ coupled nonlinear equations, each of which may have multiple solutions. Tackling this problem analytically goes beyond the scope of this paper, though a simple and preliminary analysis based on the central limit theorem \cite{fleermannCENTRALLIMITTHEOREM} helps in setting the scenario, as discussed in \cite{SeeSupplementalMaterial}. Here, to get insights into the new behavior arising from disorder, we study it numerically, on a $N\times N$ square lattice with periodic boundary conditions. Each site, $i$, is assigned a local relaxation rate, $\Gamma_i$, and each couple of neighbor sites is assigned a coupling constant, $\alpha_{ij}$, randomly drawn from a probability distribution, $P(\alpha_{ij})$.
For a given strain amplitude $\gamma_0$, solving our model numerically consists in finding a configuration of local rates $\{\Gamma_i\}$ that satisfies Eq.~\ref{eq:ising} for all sites. 
In our implementation, $\{\Gamma_i\}$ is approached iteratively: starting from an initial configuration of local rates $\{\tilde\Gamma_i^{(0)}\}$, we compute a set of target rates $\{\tilde\Gamma_i^{(1)}\}$ through Eq.~\ref{eq:ising} using $\{\tilde\Gamma_i^{(0)}\}$ to evaluate the FA coupling term. If all $\tilde\Gamma_i^{(0)}$ satisfied Eq.~\ref{eq:ising}, we would obtain $\{\tilde\Gamma_i^{(1)}\}=\{\tilde\Gamma_i^{(0)}\}=\{\Gamma_i\}$. If this is not the case, we use $\{\tilde\Gamma_i^{(1)}\}$ to compute a new set of target rates $\{\tilde\Gamma_i^{(2)}\}$, and keep iterating until the algorithm converges on a stationary set of rates $\{\tilde\Gamma_i^{(s)}\}$, for which the loss function $\mathcal{L}(s)=\sum_i\left[\left(\log\tilde\Gamma_i^{(s+1)}-\log\tilde\Gamma_i^{(s)}\right)/\log\tilde\Gamma_i^{(s)}\right]^2$ vanishes: we then take $\{\Gamma_i\}=\{\tilde\Gamma_i^{(s)}\}$. Although close to yielding many iterations may be needed to find this solution, we find that this algorithm was always converging, as shown in \cite{SeeSupplementalMaterial}.

Numerical simulations with disorder can quantitatively reproduce the gradual yielding observed in experiments, as shown by the solid lines in Figure~\ref{fig:exps}d. 
To highlight the role of glassiness and disorder on the yielding transition, in the following we fix $n=\omega=\Gamma_0=K=1$, and we draw the coupling parameters $\alpha_{ij}$ from a Gaussian distribution with average $\bar\alpha$ and variance $\sigma_\alpha^2$, which we use to define the disorder parameter $d=\sigma_\alpha^2/\bar{\alpha}^2$. To avoid unphysical negative coupling constants, we truncate $P(\alpha_{ij})$ to ensure that $P(\alpha_{ij}\leq 0)=0$. 
To mimic the effect of a strain sweep applied on a sample initially at rest, we start from a \textit{solid-like} condition, with $\tilde{\Gamma}_i^{(0)}=\Gamma_0$, and solve the numerical model with a set of strain amplitudes $\gamma_0$ spanning the physically interesting range across the yielding transition.
For each $\gamma_0$, we then compute the average relaxation rate $\bar\Gamma=\langle \Gamma_i\rangle$, and we compare it to the mean-field EOS.
In the limit of $d \rightarrow 0$, we find that our simulations reproduce the mean field results: $\bar\Gamma$ closely follows Eq.~\ref{eq:ising_MF}, with $\alpha=\bar\alpha$ and $\beta=2$, and it jumps abruptly as $\gamma_0$ reaches a threshold value $\gamma_y$ very close to $\gamma_{th}^+$, the local minimum of Eq.~\ref{eq:eos_withcoupling}, as shown by the light green dashed lines in Figure \ref{sim_overview}a.

Increasing $d$, we find that $\gamma_y$ decreases, seemingly approaching a second characteristic strain, marked as $\gamma_{th}^{(M)}$ in Fig.~\ref{sim_overview}a.
This suggests that it exists a regime of strain amplitudes smaller than $\gamma_{th}^+$ for which the solid branch of the mean-field EOS is actually metastable, similarly to what occurs in first order thermodynamic phase transitions \cite{landauStatisticalPhysicsLifshitz2011} between binodal and spinodal lines.

As $d$ is further increased, simulations show a qualitative change, to a continuous, gradual increase of $\bar\Gamma(\gamma_0)$, as shown by solid lines in Figure~\ref{sim_overview}a. 
To understand the origin of this qualitative change, we analyze the distribution of local relaxation rates, $P(\Gamma_i)$. For small $d$, we find that $P(\Gamma_i)$ is sharply peaked around $\bar\Gamma$, with a variance proportional to $d$, reflecting the heterogeneity of local couplings. As $d$ increases and the discontinuity in $\Gamma(\gamma_0)$ disappears, we find that, within a finite interval of strain amplitudes around $\gamma_y$, $P(\Gamma_i)$ becomes bimodal: a second peak appears at larger $\Gamma_i$, roughly located at $\Gamma^f(\gamma_0)$, the fluidized branch of the mean-field EOS, as shown in Fig.~\ref{sim_overview}b. 
In this regime, two distinct populations of local rates can be identified: slower ones, following the solid branch of the EOS, and faster ones, which jump to the fluidized branch without entailing the transition of the entire lattice.

As $\gamma_0$ is increased, we find that the two peaks in $P(\Gamma_i)$ change their relative amplitude: 
the \textit{fluid} peak increases in amplitude and slightly shifts towards larger rates, while the \textit{solid} peak gradually disappears without significant shift in its position. 
This indicates that the continuous evolution of $\gamma^{-1}(\bar\Gamma)$ reported in Fig.~\ref{sim_overview}a actually reflects a spatially heterogeneous process, which is locally discontinuous. 
This behavior recalls the dynamic coexistence of fast and slow relaxation modes observed experimentally in Fig.~\ref{fig:exps}, therefore we refer to the range of model parameters for which $P(\Gamma_i)$ is bimodal as the \textit{coexistence} region.
As $\gamma_0$ is increased within this region, an increasing fraction of local rates $\Gamma_i$ jumps from the solid to the fluidized branch of the EOS, reproducing the continuous decay of $\chi$ in Figure~\ref{fig:exps}d.

For the most disordered configurations, this coexistence extends over a wide range of strain amplitudes, as liquid-like local rates appear at small $\gamma_0$, while solid-like ones persist until larger $\gamma_0$: this entails a more gradual increase of $\bar\Gamma(\gamma_0)$ as shown by the red solid lines in Fig.~\ref{sim_overview}. In this sense, we observe that increasing disorder in our model produces more gradual, "ductile-like" yielding. Its ability to control the abruptness of yielding through one single model parameter makes our model a convenient playground to study the impact of disorder on the yielding transition, which has been the focus of recent theoretical work \cite{ozawaRandomCriticalPoint2018,richardPredictingPlasticityDisordered2020,patinetConnectingLocalYield2016}. Comparing dashed and solid lines in Fig.~\ref{sim_overview}a, it is tempting to conclude that our model predicts a transition from continuous to discontinuous yielding at a well-defined critical disorder $d^*$.
To better understand this transition, we look at how the local rates $\Gamma_i$ are distributed in space, with particular focus on the coexistence region. We find that slow and fast-relaxing sites are not distributed randomly, following the spatially-uncorrelated coupling constants, but they are organized in finite-sized domains separated by relatively sharp interfaces, as shown in the color maps of Fig.~\ref{sim_overview}.
In particular, we find that the size of these domains grows with decreasing $d$, and becomes comparable with the size of the whole lattice for $d\approx d^*$, at the transition to discontinuous yielding.

We interpret this result in analogy with nucleation and growth in first-order phase transitions: the FA term couples nearby lattice sites, penalizing gradients in the local rate, therefore it favors the formation of domains with homogeneous local dynamics.
In absence of disorder, this penalty produces the hysteresis shown in Figure~\ref{fig:eos}, analogous to the behavior observed in metastable systems close to a first-order transition such as supercooled liquids \cite{petrovStudyFreezingMelting2011} or ferromagnetic materials under an external magnetic field \cite{chikazumiPhysicsFerromagnetism1997}, 
and also reported for colloidal glasses under oscillatory shear \cite{dangReversibilityHysteresisSharp2016}. With disorder, we observe that solid-to-liquid transition is first nucleated at sites where the local coupling is weak, which then act as nucleation seeds. If the disorder is small, this nucleation occurs when $\gamma_0$ is in the metastable region, and triggers the abrupt transition of the entire lattice. Conversely, for large disorders, liquid domains form at $\gamma_0<\gamma_y$, and coexist with solid-like domains pinned to sites with strong enough local coupling. 
As disorder increases, $P(\alpha_{ij})$ gets broader, and the number density of sites susceptible of pinning solid or liquid domains increases. This yields a larger number of smaller-sized domains, coexisting for a wider range of strain amplitudes. Conversely, for smaller disorders, the reduced variance of local couplings produces larger-sized domains, only observable in a narrower range of $\gamma_0$.
In the extreme case of very small disorder, we interpret the transition to discontinuous yielding as a result of the scarcity of sites with weak enough coupling to seed the nucleation of liquid domains at $\gamma_0<\gamma_y$. 
This interpretation is analogous to the one emerging in numerical simulations of dense 2D soft disks, where brittleness is related to the density of ”softer” elements \cite{richardPredictingPlasticityDisordered2020}. 

\begin{figure}[htbp]
  \includegraphics[width=0.8\columnwidth]{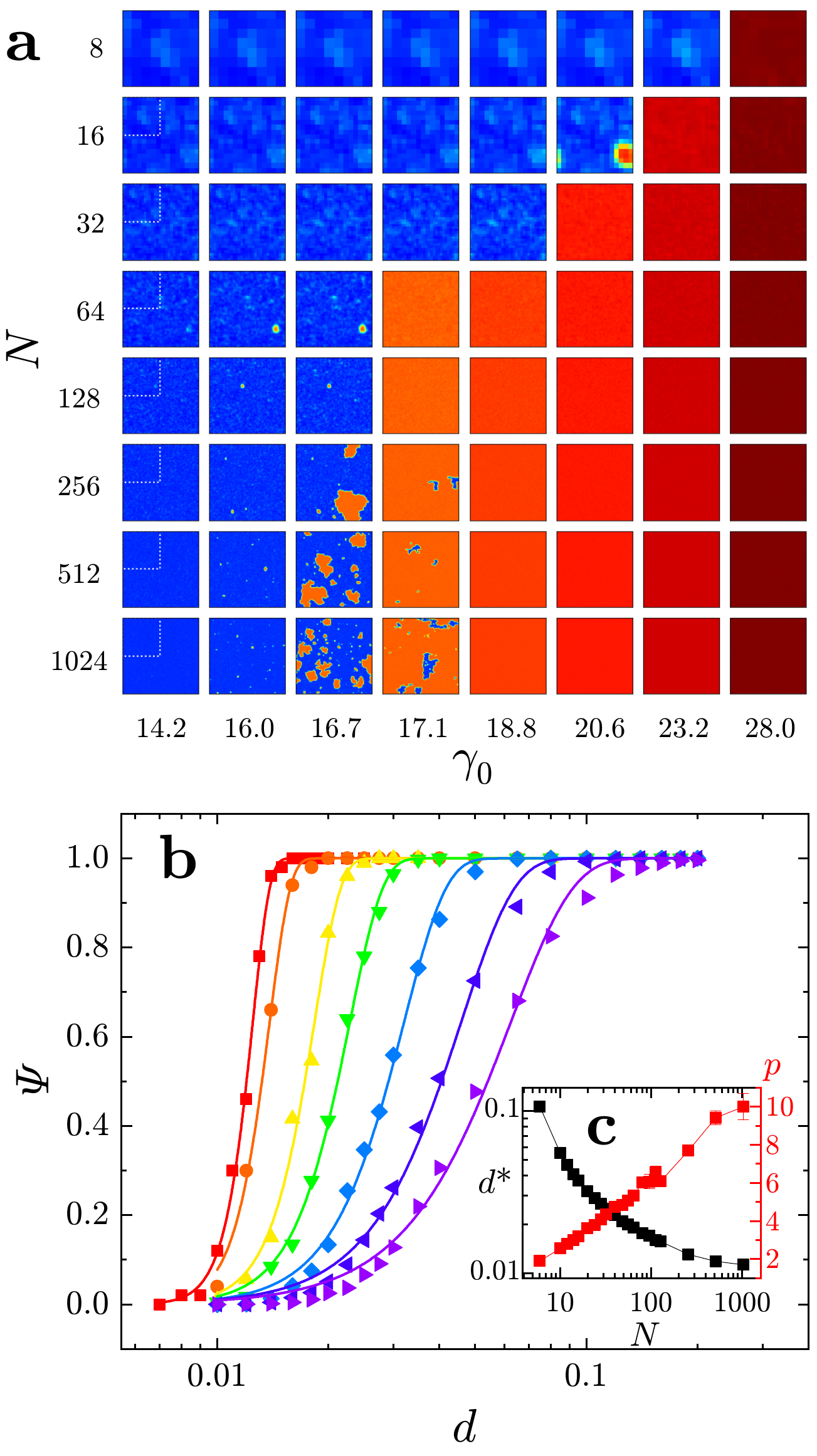}
  \caption{\textbf{Effect of lattice size}. (a) snapshots of 2D simulations on increasingly small subset of an original 1024x1024 lattice (bottom row). Local rate increases from blue to red shades. Simulation parameters are: $K_r=0.86$ and $d=0.02$. The white dashed square in each left column snapshot represent the subset of the lattice that has been used for the smaller scale simulations reported in the row above. (b) Symbols: fraction of simulations exibiting gradual yielding, $\Psi$, as a function of quenched disorder and for different lattice sites, increasing from $N=10$ (purple) to $N=512$ (red). Lines: Weibull cumulative distributions fitted to the data. (c) Weibull fit parameters: shape function, $p$ (red, right axis) and threshold disorder, $d^*$ (black, left axis), as a function of simulation lattice size.
  }\label{sim_size}
\end{figure}

To test this hypothesis in our model, we select a representative disorder, $d=0.02$, and we run a series of simulations on increasingly smaller subsets of a $1024\times 1024$ lattice with fixed $\alpha_{ij}$. 
We find that in full-scale simulations $\bar\Gamma(\gamma_0)$ increases gradually, and that slow and fast domains coexist in a finite range of strains around $\gamma_0\approx 17$. We then extract a subset of the lattice with $N=512$ and run the simulation again. The resulting $\bar\Gamma(\gamma_0)$ is fully consistent with the one obtained on the larger lattice, and slow and fast domains are reproduced with the same spatial patterns, as shown in the bottom rows of Figure~\ref{sim_size}a. The same result holds for $N=256$, proving that in this range of $N$ finite size effects are negligible. However, we find that, as $N$ if reduced below a threshold value $N^*\sim 128$, roughly corresponding to the size of the largest domains, $\bar\Gamma(\gamma_0)$ becomes discontinuous, yielding is delayed to larger $\gamma_0$, and coexistence is no longer observed.
This corroborates the hypothesis that the coexistence of solid-like and fluidized domains is intrinsic to our model with disorder, and that the discontinuous jump observed for smaller $N$ or $d$ is an effect of the finite size of the simulation lattice.

To better assess these finite size effects, we repeat this series of simulations for various disorders, and for each disorder we run many independent simulations, as detailed in \cite{SeeSupplementalMaterial}. Close to the transition between abrupt and gradual yielding, we observe that, depending on the randomly generated configuration of $\alpha_{ij}$, some simulations exhibit gradual yielding with coexistence, and some do not. 
For each $N$ and $d$, we then compute the fraction, $\Psi$, of simulations exhibiting gradual yielding. We find that $\Psi$ increases with both $N$ and $d$, reflecting the higher probability to find weakly-coupled sites able to nucleate fast domains before the solid state becomes metastable.
For any given $N$, we find that $\Psi_N(d)$ has a sigmoidal shape well-described by a cumulative Weibull function, $1-\exp[-(d/d^*)^p]$, further confirming that the onset of gradudal yielding with coexistence depends on the low-$\alpha$ tail of the distribution of coupling constants, which makes it an extreme value problem, analogous to rupture nucleated by defects \cite{weibullStatisticalTheoryStrength1939, colemanStatisticsTimeDependence1958, aimeModelFailureThermoplastic2017,ozawaRareEventsDisorder2022}. 
Fitting $\Psi_N(d)$, we extract the characteristic disorder $d^*$ and the exponent $p$. We find that the larger variability of smaller-scale simulations reflects in broader Weibull distributions, characterized by a lower exponent $p$. Moreover, we find that $d^*$ decreases with increasing $N$, confirming that for smaller disorders, larger-scale simulations are needed to capture the larger domains coexisting at the yield point, as shown in Fig.~\ref{sim_size}c.
We conclude that in the thermodynamic limit, $N\rightarrow \infty$, our model predicts gradual yielding with coexistence for all finite disorders, and that simulations showing discontinuous yielding, marked by dashed lines in Figure~\ref{sim_overview}a, are biased by finite size effects.

\section{Elementary catastrophe and Maxwell rule\label{catastrophe}}

Restricting our analysis to gradually-yielding simulations, we observe that decreasing disorder (red to blue solid lines in Figure~\ref{sim_overview}a) highlights the existence of a well-defined yield strain, $\gamma_{th}^{(M)}$, distinct from the local extrema of the mean field EOS, $\gamma_{th}^\pm$. 
Although our simulation data can only track gradual yielding down to $d=0.02$, our results strongly suggest that in the limit of infinitely large simulation lattices, $N\rightarrow \infty$, a discontinuity of $\bar\Gamma(\gamma_0)$ at $\gamma_0=\gamma_{th}^{(M)}$ arises in the limit of $d \rightarrow 0^+$. 
The presence of this second yield strain 
was not predicted by the mean-field model discussed above, as it can only be observed with a finite disorder.
However, its persistence for $d \rightarrow 0^+$ suggests that $\gamma_{th}^{(M)}$ is somehow related to some hidden property of the mean-field model itself.
Once more exploiting the formal analogy with VdW gases, we speculate that this yield strain, laying in between $\gamma_{th}^-$ and $\gamma_{th}^+$, may arise from an optimum criterion analogous to the Maxwell rule, which selects the exact phase transition pressure within the interval where multiple solutions are analytically admitted, based on the minimization of the free energy.
Indeed, we find that the value of $\gamma_{th}^{(M)}$ estimated using numerical simulations closely matches the one computed applying the Maxwell rule to Eq.~\ref{eq:eos_withcoupling}: for the model parameters used in Fig.~\ref{sim_size} we find $\gamma_{th}^{(M)}=15.6$, closely matching simulation results with small disorders as shown in Fig.~\ref{sim_overview} and as further documented in \cite{SeeSupplementalMaterial}. This correspondence corroborates our hypothesis, suggesting the existence of an underlying potential function playing the role of a free energy.
This is particularly surprising because, in contrast with VdW, here Equation~\ref{gen-EOS-beta1} describes a dynamic transition in systems out of equilibrium, and is not derived from energy minimization.

To illustrate how something similar to a free energy can be derived in our model, we take $\beta=2$, and we recast Eq.~\ref{gen-EOS-beta1} in the form: $\gamma_r^{-n}\Gamma_r^3-\left(\gamma_r^{-n}+8K_r\right)\Gamma_r^2/3+ 3\Gamma_r-1 = 0$.
This expression takes an even simpler 
form in terms of a new variable, $\rho$, defined such that the critical point is in $\rho=0$:

\begin{equation}\label{manifold2}
    \rho^{3}+a\rho+b=0
\end{equation}

with:

\begin{subequations}
  \label{eq:abrho-params}
  \begin{empheq}{align}
    \rho&=\frac{1}{\Gamma_r}-1 \label{eq:rho-param}\\
    a&=\frac{8K_r}{3}+
\frac{1}{3}\gamma_r^{-n}-3 \label{eq:a-param}\\
    b&=\frac{8K_r}{3}-\frac{2}{3}\gamma_r^{-n} -2 \label{eq:b-param}
  \end{empheq}
\end{subequations}

Eq.~\ref{manifold2} defines a manifold unifying all trajectories followed by samples at various $K_r$ subject to varying strain amplitudes. Its functional form is well-known in statistical physics and bifurcation theory, as it represents the manifold of a so-called \textit{cusp catastrophe}, the same describing the liquid-vapour transition of gases~\cite{okninskiCatastropheTheory1992}.

\begin{figure}[htbp]
  \includegraphics[width=8.7cm]{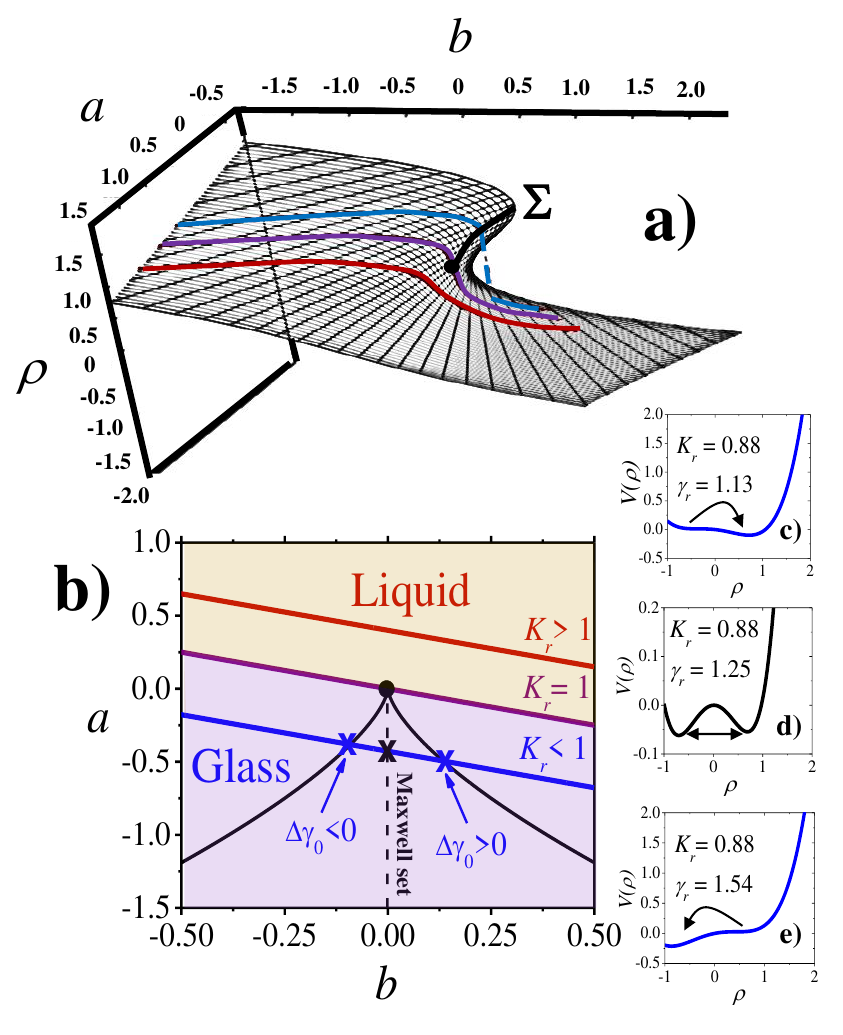}
\caption{a) Cusp catastrophe manifold and trajectories. Three trajectories ($K_r=0.915, 1.000, 1.428$) are shown and represent ascending strain sweeps $\Delta\gamma_0>0$ for a glassy (blue), critical (purple) and liquid (red). 
The trajectory of the glassy system undergoes a discontinuous jump as it meets the critical set $\Sigma$. 
b): Bifurcation set $B$ (black solid line) and three representative trajectories $a(b)$ for $K_r>1$ (red, liquid), $K_r=1$ (purple, critical iso-K) and $K_r<1$ (blue, glass) in the control parameter space $(a,b)$ defined by eqs (\ref{eq:a-param}) and (\ref{eq:b-param}). The cusp point coincides to the critical point ($\gamma_r$,$\Gamma_r$,$K_r$)=$(1,1,1)$. The two intersections between the curve at $K_r<1$ and the bifurcation set correspond to the yield points in the absence of coupling disorder for ascending (right) and descending (left) strain ramps. c-e) Potential $V(\rho)$ for $K_r=0.88$, $n=3$ and the three critical strains corresponding to the critical set $\Sigma$ and zero-order transition (c and e), with a jump occurring from the vanishing minimum to the other (indicated by the arrows), and the Maxwell set (d) where the transition is reversible and first-order. 
\label{cusp}} 
\end{figure}

To highlight its features, we show it in Figure~\ref{cusp} as a function of the new set of variables $\{\rho, a, b\}$.
The non-monotonic EOS characterizing glassy samples defined by $K_r<1$ correspond to paths crossing the region where the manifold folds over itself, such that for a given strain, fixing the couple $(a, b)$, multiple values of $\rho$, thus of the microscopic rate $\Gamma$, are allowed.
In this region, iso-$K_r$ paths have local extrema, where the manifold is locally perpendicular to planes of constant $\rho$.
Differentiating Eq.~\ref{manifold2}, we find that this occurs for $3\rho^2+a=0$, which defines the so-called bifurcation set, $\Sigma$~\cite{okninskiCatastropheTheory1992}. 
The projection of $\Sigma$ on the $(a,b)$ plane is defined by $a=-3(b/2)^{3/2}$, which has a cusp in the origin, as shown in Fig.~\ref{cusp}b. 
Hence, in our model yielding falls into the broader class of critical phenomena described as a cusp catastrophe \cite{okninskiCatastropheTheory1992,demazureBifurcationsCatastrophes2000}. Projected on the $(a,b)$ plane, strain sweeps are straight lines defined by $a=4(K_r-1)-b/2$, with $\gamma$ increasing with $b$. For $K_r<1$, these lines intercept $\Sigma$ in two points corresponding to $\gamma_{th}^+$ and $\gamma_{th}^-$. 
Between these two points lays the critical strain resulting from simulations with disorder, $\gamma_{th}^{(M)}$. Unlike $\gamma_{th}^\pm$, this new critical strain does not correspond to a critical point of the manifold. To relate it to the mean field model, we integrate Eq.~\ref{manifold2} to obtain a potential function $V(\rho;a,b)$ such that $\partial V/\partial \rho=0$ yields the equation of state. For Van der Waals gases, $V$ would correspond to Helmholtz free energy. Here, we obtain:

\begin{equation}\label{potential2}
    V(\rho;a,b)=\frac{1}{4}\rho^{4}+\frac{a}{2}\rho^2+
b\rho
\end{equation}

The shape of this function is dictated by the parameters $a$ and $b$. In particular, we observe that $V$ is convex for $a>0$, whereas for $a<0$ its second derivative becomes negative for $a/3<\rho^2<-a/3$, and the function develops two minima, corresponding to the two branches of the critical set $\Sigma$. 
This evolution of $V$ recalls the behavior of the free energy functional in Landau theory for magnetic phase transitions in the presence of an external field \cite{landauStatisticalPhysicsLifshitz2011}, with $\rho$ as the order parameter. 
In that framework, the requirement that the free energy should be minimized provides a stability criterion that selects the state of the system when the equation of state has multiple solutions. Here, for $a<0$, this criterion corresponds to selecting the deepest of the two minima, which depends on the sign of $b$, the coefficient of the only odd term in $V$. For $b=0$, the potential is an even function of $\rho$ and the two minima are equivalent (Fig. \ref{cusp}d): this criterion defines the so-called Maxwell set, which is shown as a dashed line in Fig.~\ref{cusp}b. 
This second critical set is profoundly different from $\gamma_{th}^\pm$, as it can only be highlighted by tracking the potential functional $V(\rho; a,b)$ across yielding. When yielding occurs at $\gamma_{th}^{(M)}$, $V$ evolves continuously, as the two minima are equally deep, when $\rho$ jumps from one to the other. This is what characterizes first-order phase transitions, in which the free energy is continuous, although its derivative is not. On the contrary, when yielding occurs at $\gamma_{th}^+$, $V$ jumps discontinuously to the deeper minimum as its second local minimum disappears. This defines a different class of transitions, of the zeroth order. 

In the case of thermodynamic phase transitions, the Maxwell set corresponds to the phase boundary. Here, $V(\rho; a,b)$ was introduced by formal analogy to thermodynamic free energy: although by construction its \textit{local} minimization yields the equation of state, its \textit{global} minimization does not have a straightforward physical relevance. Yet, surprisingly, simulations indicate that the Maxwell set, which we can write from Eq.~\ref{eq:b-param} as $\gamma_{th}^{(M)}=\gamma_c(4K_r-3)^{-1/n}$, does correspond to the yield point in the limit of large-scale simulations and for $d\rightarrow 0$. 
This result suggests that $V$ has an unanticipated physical relevance, which is revealed by a finite, albeit small, disorder. 
The profound physical meaning of this functional is yet to be unveiled, and goes beyond the scope of this paper. 

Here, we emphasize that disorder, which was initially introduced to account for the gradual onset of yielding as observed in experiments, plays a key role in determining the nature of the yielding transition itself.
According to our FA model, yielding is thus a rounded and continuous first-order transition with an emergent criticality at the yield strain, sharing features with both pure first- and second-order transitions. The yield point is set by a Maxwell rule, describing the strain at which solid-like dynamics become metastable and fluid-like domains can nucleate at weakly-coupled lattice sites, playing the role of soft spots~\cite{widmer-cooperIrreversibleReorganizationSupercooled2008}. The density of such soft spots sets a characteristic length scale that diverges at $d=0$, where yielding becomes discontinuous.

\section{Emerging dynamic length scale}\label{brittleductile}

\begin{figure}[h]
  \includegraphics[width=8cm]{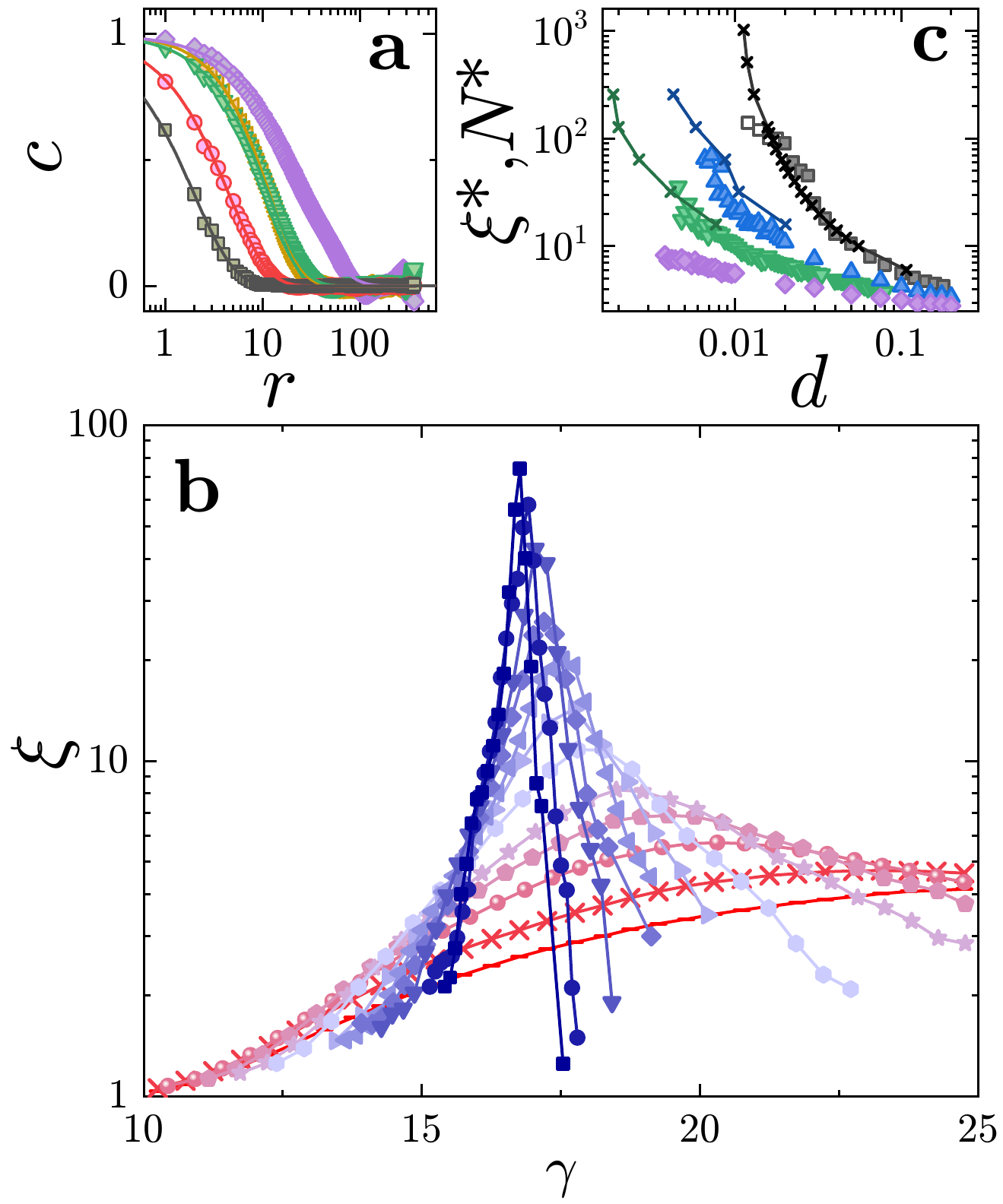}\\
  \caption{\textbf{Spatial heterogeneities} (a) symbols: spatial correlation of local rates for $K_r=0.86$ and $d=0.03$, for $\gamma_r$ across the yielding transition: $\gamma_r=2.15$ (black), 2.25 (red), 2.41 (green), 2.53 (purple), 2.59 (gold). Lines: stretched exponential fits. 
  (b) characteristic correlation length, $\xi$, as a function strain amplitudes in the coexistence region, for disorder $d$ increasing from 0.02 (blue) to 0.2 (red).
  (c) Symbols: maximum correlation length, $\xi^*$, as a function of disorder for $K_r$ decreasing from 0.98 (purple) to 0.86 (black), for $N=512$. Full datapoints are averaged on 10 independent simulations, all of which exhibited gradual yielding. Open black symbols denote disorders for which one or more simulations exhibited abrupt yielding: these simulations were excluded from the averaged dataset. Lines and crosses: threshold simulation lattice size, $N^*$, extracted for each disorder by inverting $d^*(N)$ data from Figure~\ref{sim_size}c.}
  \label{sim_xcorr}
\end{figure}

This characteristic length scale emerges spontaneously from our on-lattice simulations, and dictates whether yielding manifests as an abrupt or a continuous transition in finite systems.
To characterize this length scale, we compute the connected spatial correlation function of local rates: $c(r)\propto \langle\log\Gamma_i\log\Gamma_j\rangle-\langle\log\Gamma_i\rangle\langle\log\Gamma_j\rangle$, by averaging on all pairs of lattice sites at a given Euclidean distance $r$. 
We normalize $c(r)$, such that all correlations decay from 1 to 0, and we define the correlation length as its integral average $\xi=\int c(r) dr$. Operationally, we compute the integral by fitting $c(r)$ to Kohlrausch–Williams–Watts (KWW) functions $c(r)=\exp[-(\lambda r)^p]$, and expressing the integral as a function of the fit parameters $\xi=\lambda^{-1}\Gamma_{\varepsilon}(p^{-1} +1)$, where $\Gamma_{\varepsilon}(x)$ denotes Euler's Gamma function. Figure~\ref{sim_xcorr}a shows representative $c(r)$ for $K_r = 0.86$ and $d = 0.03$ for strains spanning the coexistence region.   
For simulations exhibiting abrupt yielding, we find that $c(r)$ decays very fast for all values of $\gamma_0$: $\xi$ is as small as a single lattice spacing, reflecting the lack of spatial correlations in the coupling constants. 
By contrast, for simulations exhibiting gradual yielding, $\xi$ depends strongly on $\gamma_0$, and is significantly larger than the lattice spacing in the coexistence region. It peaks to a maximum value, $\xi^*$, quantifying the growth and coalescence of fluid domains, 
as shown in Figure~\ref{sim_xcorr}b. 
We find that $\xi^*$ increases with decreasing $d$, reflecting the larger size of slow and fast domains coexisting at the transition, as shown previously in the snapshots of Fig.~\ref{sim_overview}. Moreover, we observe an increase of the same peak height with decreasing reduced temperature $K_r$ as shown in figure ~\ref{sim_xcorr}c, that is coherent with the experimental evidence that the more a system is quenched in the glassy phase, the more it will be characterized by long-ranged correlated dynamics and sharp yielding. 

This increase of $\xi^*$ with decreasing $d$ requires increasingly large simulation lattices to be properly captured. 
To find a more direct connection between $\xi^*$ and $N$, we invert $d^*(N)$ from Figure \ref{sim_size}, to obtain a threshold lattice size for a given disorder, $N^*(d)$. We observe that the growth of $N^*$ with decreasing $d$ follows closely that of the largest lengthscale, $\xi^*$ and that this holds for all the investigated temperatures, as shown by solid lines in Fig.~\ref{sim_xcorr}c. 
We thus conclude that the transition from abrupt yielding to gradual yielding is dictated by the competition of two lengthscales: the system size, $N$, and the characteristic lengthscale of shear-induced dynamic heterogeneities, $\xi$. For $N>\xi^*$, the system yields gradually through the nucleation and growth of fast-relaxing domains, whereas for smaller $N$ the scarcity of weakly-coupled lattice sites delays the transition to larger strain amplitudes, where the system is metastable, and eventually yields abruptly.

This change emerges in our simulations as a finite size effect, rather than an intrinsic feature of our model itself, which predicts gradual yielding, independent of the sample size, for any $d>0$. However, this effect highlights that a diverging correlation length may have tangible consequences on the manifestation of yielding at the macro scale.
Exploiting our model to investigate the conditions under which a soft glassy sample may nucleate system-spanning heterogeneities such as shear bands animates our current research effort.



\section{Conclusions}\label{conclusions}

In this work we have presented a new model reproducing the key features of microscopic yielding in glassy systems under oscillatory shear, as reported by recent rheo-optical experiments 
\cite{aimeUnifiedStateDiagram2023}. We model the microscopic relaxation rate as a function of the imposed strain amplitude, by accounting for both the ultraslow, \textit{solid-like} aging dynamics found in samples at rest or under small-amplitude deformations, and the faster, \textit{liquid-like} dynamics measured stroboscopically in samples subject to large-amplitude oscillatory shear. 
We show that these two behaviors are captured by established models postulating a generic power-law dependence of shear-induced dynamics on the macroscopic shear rate, which however fail predicting the well-defined transition observed upon yielding. To properly describe this transition, we borrow ideas from dynamic facilitation of quiescent glasses, and we introduce a new dynamic coupling, which we called facilitated advection (FA), between neighboring sample regions. 
The mean field solution of this model yields an equation of state predicting an abrupt transition between glassy and fluidized states as a function of strain amplitude. This equation of state shares the same cusp catastrophe manifold as Van der Waals gases, and predicts the existence of a critical point with a genuine second order transition for a set of model parameters corresponding to samples at the glass transition. 
We are currently working to test this prediction in experiments on colloidal suspensions at different concentrations, across the glass transition.

To describe gradual yielding as observed in experiments, we performed numerical simulations introducing disorder in the FA coupling. We showed that a finite disorder changes qualitatively the nature of the transition: in presence of disorder, we find that solid and liquid-like domains coexist in a disorder-dependent range of strain amplitudes. Within this range, we measure a dynamic correlation length, $\xi$, that peaks at yielding and increases with decreasing $d$, diverging for $d\rightarrow 0$. 
The system size therefore plays a key role in determining the nature of the transition: for large enough systems, yielding is a rounded and continuous transition sharing features with both first-order and second-order transitions, akin to what has been observed in spin systems with coupling disorder \cite{bellafardEffectQuenchedBond2015a}. 
Conversely, as $\xi$ exceeds the system size, we observe a qualitative change to abrupt yielding. This change emerges in our simulations as a finite size effect, rather than a feature of our model itself. However, it highlights that a diverging correlation length may have tangible consequences on the manifestation of yielding at the macro scale. 
To explore the relationship between local dynamics and rheological behavior, our ongoing effort is devoted to integrate our model with constitutive equations defining the time evolution of the stress within each mesoscopic portion of sheared materials. This will allow us to test the generality of the FA model to other rheological tests beyond oscillatory shear. 

Finally, although our model has been developed to describe yielding in soft glasses, its formulation in terms of a strain-dependent local relaxation rate allows its further generalization to systems with more complex structure and rheology, such as gels, attractive glasses or anisotropic systems, in a broader attempt to achieve a unified description of yielding in soft amorphous materials.

\begin{acknowledgments}
This work was financially supported by CNES. SA acknowledges the support of the French Agence Nationale de la Recherche (ANR), under grant ANR-22-CE06-0025. 
We thank L. Cipelletti and W. Kob for illuminating discussions.
\end{acknowledgments}


\begin{thebibliography}{145}%
\makeatletter
\providecommand \@ifxundefined [1]{%
 \@ifx{#1\undefined}
}%
\providecommand \@ifnum [1]{%
 \ifnum #1\expandafter \@firstoftwo
 \else \expandafter \@secondoftwo
 \fi
}%
\providecommand \@ifx [1]{%
 \ifx #1\expandafter \@firstoftwo
 \else \expandafter \@secondoftwo
 \fi
}%
\providecommand \natexlab [1]{#1}%
\providecommand \enquote  [1]{``#1''}%
\providecommand \bibnamefont  [1]{#1}%
\providecommand \bibfnamefont [1]{#1}%
\providecommand \citenamefont [1]{#1}%
\providecommand \href@noop [0]{\@secondoftwo}%
\providecommand \href [0]{\begingroup \@sanitize@url \@href}%
\providecommand \@href[1]{\@@startlink{#1}\@@href}%
\providecommand \@@href[1]{\endgroup#1\@@endlink}%
\providecommand \@sanitize@url [0]{\catcode `\\12\catcode `\$12\catcode
  `\&12\catcode `\#12\catcode `\^12\catcode `\_12\catcode `\%12\relax}%
\providecommand \@@startlink[1]{}%
\providecommand \@@endlink[0]{}%
\providecommand \url  [0]{\begingroup\@sanitize@url \@url }%
\providecommand \@url [1]{\endgroup\@href {#1}{\urlprefix }}%
\providecommand \urlprefix  [0]{URL }%
\providecommand \Eprint [0]{\href }%
\providecommand \doibase [0]{http://dx.doi.org/}%
\providecommand \selectlanguage [0]{\@gobble}%
\providecommand \bibinfo  [0]{\@secondoftwo}%
\providecommand \bibfield  [0]{\@secondoftwo}%
\providecommand \translation [1]{[#1]}%
\providecommand \BibitemOpen [0]{}%
\providecommand \bibitemStop [0]{}%
\providecommand \bibitemNoStop [0]{.\EOS\space}%
\providecommand \EOS [0]{\spacefactor3000\relax}%
\providecommand \BibitemShut  [1]{\csname bibitem#1\endcsname}%
\let\auto@bib@innerbib\@empty
\bibitem [{\citenamefont {Nicolas}\ \emph {et~al.}(2018)\citenamefont
  {Nicolas}, \citenamefont {Ferrero}, \citenamefont {Martens},\ and\
  \citenamefont {Barrat}}]{nicolasDeformationFlowAmorphous2018}%
  \BibitemOpen
  \bibfield  {author} {\bibinfo {author} {\bibfnamefont {A.}~\bibnamefont
  {Nicolas}}, \bibinfo {author} {\bibfnamefont {E.~E.}\ \bibnamefont
  {Ferrero}}, \bibinfo {author} {\bibfnamefont {K.}~\bibnamefont {Martens}}, \
  and\ \bibinfo {author} {\bibfnamefont {J.-L.}\ \bibnamefont {Barrat}},\
  }\href {\doibase 10.1103/RevModPhys.90.045006} {\bibfield  {journal}
  {\bibinfo  {journal} {Rev. Mod. Phys.}\ }\textbf {\bibinfo {volume} {90}},\
  \bibinfo {pages} {045006} (\bibinfo {year} {2018})}\BibitemShut {NoStop}%
\bibitem [{\citenamefont {Bonn}\ \emph {et~al.}(2017)\citenamefont {Bonn},
  \citenamefont {Denn}, \citenamefont {Berthier}, \citenamefont {Divoux},\ and\
  \citenamefont {Manneville}}]{bonnYieldStressMaterials2017}%
  \BibitemOpen
  \bibfield  {author} {\bibinfo {author} {\bibfnamefont {D.}~\bibnamefont
  {Bonn}}, \bibinfo {author} {\bibfnamefont {M.~M.}\ \bibnamefont {Denn}},
  \bibinfo {author} {\bibfnamefont {L.}~\bibnamefont {Berthier}}, \bibinfo
  {author} {\bibfnamefont {T.}~\bibnamefont {Divoux}}, \ and\ \bibinfo {author}
  {\bibfnamefont {S.}~\bibnamefont {Manneville}},\ }\href {\doibase
  10.1103/RevModPhys.89.035005} {\bibfield  {journal} {\bibinfo  {journal}
  {Rev. Mod. Phys.}\ }\textbf {\bibinfo {volume} {89}},\ \bibinfo {pages}
  {035005} (\bibinfo {year} {2017})},\ \Eprint
  {http://arxiv.org/abs/1502.05281} {arXiv:1502.05281} \BibitemShut {NoStop}%
\bibitem [{\citenamefont {Parley}\ \emph {et~al.}(2022)\citenamefont {Parley},
  \citenamefont {Sastry},\ and\ \citenamefont
  {Sollich}}]{parleyMeanFieldTheoryYielding2022}%
  \BibitemOpen
  \bibfield  {author} {\bibinfo {author} {\bibfnamefont {J.~T.}\ \bibnamefont
  {Parley}}, \bibinfo {author} {\bibfnamefont {S.}~\bibnamefont {Sastry}}, \
  and\ \bibinfo {author} {\bibfnamefont {P.}~\bibnamefont {Sollich}},\ }\href
  {\doibase 10.1103/PhysRevLett.128.198001} {\bibfield  {journal} {\bibinfo
  {journal} {Phys. Rev. Lett.}\ }\textbf {\bibinfo {volume} {128}},\ \bibinfo
  {pages} {198001} (\bibinfo {year} {2022})}\BibitemShut {NoStop}%
\bibitem [{\citenamefont {Brader}\ \emph {et~al.}(2010)\citenamefont {Brader},
  \citenamefont {Siebenb{\"u}rger}, \citenamefont {Ballauff}, \citenamefont
  {Reinheimer}, \citenamefont {Wilhelm}, \citenamefont {Frey}, \citenamefont
  {Weysser},\ and\ \citenamefont {Fuchs}}]{braderNonlinearResponseDense2010}%
  \BibitemOpen
  \bibfield  {author} {\bibinfo {author} {\bibfnamefont {J.~M.}\ \bibnamefont
  {Brader}}, \bibinfo {author} {\bibfnamefont {M.}~\bibnamefont
  {Siebenb{\"u}rger}}, \bibinfo {author} {\bibfnamefont {M.}~\bibnamefont
  {Ballauff}}, \bibinfo {author} {\bibfnamefont {K.}~\bibnamefont
  {Reinheimer}}, \bibinfo {author} {\bibfnamefont {M.}~\bibnamefont {Wilhelm}},
  \bibinfo {author} {\bibfnamefont {S.~J.}\ \bibnamefont {Frey}}, \bibinfo
  {author} {\bibfnamefont {F.}~\bibnamefont {Weysser}}, \ and\ \bibinfo
  {author} {\bibfnamefont {M.}~\bibnamefont {Fuchs}},\ }\href {\doibase
  10.1103/PhysRevE.82.061401} {\bibfield  {journal} {\bibinfo  {journal} {Phys.
  Rev. E}\ }\textbf {\bibinfo {volume} {82}},\ \bibinfo {pages} {061401}
  (\bibinfo {year} {2010})}\BibitemShut {NoStop}%
\bibitem [{\citenamefont
  {Voigtmann}(2014)}]{voigtmannNonlinearGlassyRheology2014}%
  \BibitemOpen
  \bibfield  {author} {\bibinfo {author} {\bibfnamefont {T.}~\bibnamefont
  {Voigtmann}},\ }\href {\doibase 10.1016/j.cocis.2014.11.001} {\bibfield
  {journal} {\bibinfo  {journal} {Current Opinion in Colloid \& Interface
  Science}\ }\textbf {\bibinfo {volume} {19}},\ \bibinfo {pages} {549}
  (\bibinfo {year} {2014})}\BibitemShut {NoStop}%
\bibitem [{\citenamefont {Sollich}\ \emph {et~al.}(1997)\citenamefont
  {Sollich}, \citenamefont {Lequeux}, \citenamefont {H{\'e}braud},\ and\
  \citenamefont {Cates}}]{sollichRheologySoftGlassy1997}%
  \BibitemOpen
  \bibfield  {author} {\bibinfo {author} {\bibfnamefont {P.}~\bibnamefont
  {Sollich}}, \bibinfo {author} {\bibfnamefont {F.}~\bibnamefont {Lequeux}},
  \bibinfo {author} {\bibfnamefont {P.}~\bibnamefont {H{\'e}braud}}, \ and\
  \bibinfo {author} {\bibfnamefont {M.~E.}\ \bibnamefont {Cates}},\ }\href@noop
  {} {\bibfield  {journal} {\bibinfo  {journal} {Physical review letters}\
  }\textbf {\bibinfo {volume} {78}},\ \bibinfo {pages} {2020} (\bibinfo {year}
  {1997})}\BibitemShut {NoStop}%
\bibitem [{\citenamefont {Picard}\ \emph {et~al.}(2002)\citenamefont {Picard},
  \citenamefont {Ajdari}, \citenamefont {Bocquet},\ and\ \citenamefont
  {Lequeux}}]{picardSimpleModelHeterogeneous2002}%
  \BibitemOpen
  \bibfield  {author} {\bibinfo {author} {\bibfnamefont {G.}~\bibnamefont
  {Picard}}, \bibinfo {author} {\bibfnamefont {A.}~\bibnamefont {Ajdari}},
  \bibinfo {author} {\bibfnamefont {L.}~\bibnamefont {Bocquet}}, \ and\
  \bibinfo {author} {\bibfnamefont {F.}~\bibnamefont {Lequeux}},\ }\href
  {\doibase 10.1103/PhysRevE.66.051501} {\bibfield  {journal} {\bibinfo
  {journal} {Physical Review E}\ }\textbf {\bibinfo {volume} {66}} (\bibinfo
  {year} {2002}),\ 10.1103/PhysRevE.66.051501}\BibitemShut {NoStop}%
\bibitem [{\citenamefont {Benzi}\ \emph {et~al.}(2019)\citenamefont {Benzi},
  \citenamefont {Divoux}, \citenamefont {Barentin}, \citenamefont {Manneville},
  \citenamefont {Sbragaglia},\ and\ \citenamefont
  {Toschi}}]{benziUnifiedTheoreticalExperimental2019}%
  \BibitemOpen
  \bibfield  {author} {\bibinfo {author} {\bibfnamefont {R.}~\bibnamefont
  {Benzi}}, \bibinfo {author} {\bibfnamefont {T.}~\bibnamefont {Divoux}},
  \bibinfo {author} {\bibfnamefont {C.}~\bibnamefont {Barentin}}, \bibinfo
  {author} {\bibfnamefont {S.}~\bibnamefont {Manneville}}, \bibinfo {author}
  {\bibfnamefont {M.}~\bibnamefont {Sbragaglia}}, \ and\ \bibinfo {author}
  {\bibfnamefont {F.}~\bibnamefont {Toschi}},\ }\href {\doibase
  10.1103/PhysRevLett.123.248001} {\bibfield  {journal} {\bibinfo  {journal}
  {Phys. Rev. Lett.}\ }\textbf {\bibinfo {volume} {123}},\ \bibinfo {pages}
  {248001} (\bibinfo {year} {2019})}\BibitemShut {NoStop}%
\bibitem [{\citenamefont {Liu}\ \emph {et~al.}(2018)\citenamefont {Liu},
  \citenamefont {Martens},\ and\ \citenamefont
  {Barrat}}]{liuMeanFieldScenarioAthermal2018}%
  \BibitemOpen
  \bibfield  {author} {\bibinfo {author} {\bibfnamefont {C.}~\bibnamefont
  {Liu}}, \bibinfo {author} {\bibfnamefont {K.}~\bibnamefont {Martens}}, \ and\
  \bibinfo {author} {\bibfnamefont {J.-L.}\ \bibnamefont {Barrat}},\ }\href
  {\doibase 10.1103/PhysRevLett.120.028004} {\bibfield  {journal} {\bibinfo
  {journal} {Phys. Rev. Lett.}\ }\textbf {\bibinfo {volume} {120}},\ \bibinfo
  {pages} {028004} (\bibinfo {year} {2018})}\BibitemShut {NoStop}%
\bibitem [{\citenamefont {Divoux}\ \emph {et~al.}(2024)\citenamefont {Divoux},
  \citenamefont {Agoritsas}, \citenamefont {Aime}, \citenamefont {Barentin},
  \citenamefont {Barrat}, \citenamefont {Benzi}, \citenamefont {Berthier},
  \citenamefont {Bi}, \citenamefont {Biroli}, \citenamefont {Bonn},
  \citenamefont {Bourrianne}, \citenamefont {Bouzid}, \citenamefont {Del~Gado},
  \citenamefont {{Delano{\"e}-Ayari}}, \citenamefont {Farain}, \citenamefont
  {Fielding}, \citenamefont {Fuchs}, \citenamefont {Van Der~Gucht},
  \citenamefont {Henkes}, \citenamefont {Jalaal}, \citenamefont {Joshi},
  \citenamefont {Lema{\^i}tre}, \citenamefont {Leheny}, \citenamefont
  {Manneville}, \citenamefont {Martens}, \citenamefont {Poon}, \citenamefont
  {Popovic}, \citenamefont {Procaccia}, \citenamefont {Ramos}, \citenamefont
  {Richards}, \citenamefont {Rogers}, \citenamefont {Rossi}, \citenamefont
  {Sbragaglia}, \citenamefont {Tarjus}, \citenamefont {Toschi}, \citenamefont
  {Trappe}, \citenamefont {Vermant}, \citenamefont {Wyart}, \citenamefont
  {Zamponi},\ and\ \citenamefont
  {Zare}}]{divouxDuctiletobrittleTransitionYielding2024}%
  \BibitemOpen
  \bibfield  {author} {\bibinfo {author} {\bibfnamefont {T.}~\bibnamefont
  {Divoux}}, \bibinfo {author} {\bibfnamefont {E.}~\bibnamefont {Agoritsas}},
  \bibinfo {author} {\bibfnamefont {S.}~\bibnamefont {Aime}}, \bibinfo {author}
  {\bibfnamefont {C.}~\bibnamefont {Barentin}}, \bibinfo {author}
  {\bibfnamefont {J.-L.}\ \bibnamefont {Barrat}}, \bibinfo {author}
  {\bibfnamefont {R.}~\bibnamefont {Benzi}}, \bibinfo {author} {\bibfnamefont
  {L.}~\bibnamefont {Berthier}}, \bibinfo {author} {\bibfnamefont
  {D.}~\bibnamefont {Bi}}, \bibinfo {author} {\bibfnamefont {G.}~\bibnamefont
  {Biroli}}, \bibinfo {author} {\bibfnamefont {D.}~\bibnamefont {Bonn}},
  \bibinfo {author} {\bibfnamefont {P.}~\bibnamefont {Bourrianne}}, \bibinfo
  {author} {\bibfnamefont {M.}~\bibnamefont {Bouzid}}, \bibinfo {author}
  {\bibfnamefont {E.}~\bibnamefont {Del~Gado}}, \bibinfo {author}
  {\bibfnamefont {H.}~\bibnamefont {{Delano{\"e}-Ayari}}}, \bibinfo {author}
  {\bibfnamefont {K.}~\bibnamefont {Farain}}, \bibinfo {author} {\bibfnamefont
  {S.~M.}\ \bibnamefont {Fielding}}, \bibinfo {author} {\bibfnamefont
  {M.}~\bibnamefont {Fuchs}}, \bibinfo {author} {\bibfnamefont
  {J.}~\bibnamefont {Van Der~Gucht}}, \bibinfo {author} {\bibfnamefont
  {S.}~\bibnamefont {Henkes}}, \bibinfo {author} {\bibfnamefont
  {M.}~\bibnamefont {Jalaal}}, \bibinfo {author} {\bibfnamefont {Y.~M.}\
  \bibnamefont {Joshi}}, \bibinfo {author} {\bibfnamefont {A.}~\bibnamefont
  {Lema{\^i}tre}}, \bibinfo {author} {\bibfnamefont {R.~L.}\ \bibnamefont
  {Leheny}}, \bibinfo {author} {\bibfnamefont {S.}~\bibnamefont {Manneville}},
  \bibinfo {author} {\bibfnamefont {K.}~\bibnamefont {Martens}}, \bibinfo
  {author} {\bibfnamefont {W.}~\bibnamefont {Poon}}, \bibinfo {author}
  {\bibfnamefont {M.}~\bibnamefont {Popovic}}, \bibinfo {author} {\bibfnamefont
  {I.}~\bibnamefont {Procaccia}}, \bibinfo {author} {\bibfnamefont
  {L.}~\bibnamefont {Ramos}}, \bibinfo {author} {\bibfnamefont {J.~A.}\
  \bibnamefont {Richards}}, \bibinfo {author} {\bibfnamefont {S.~A.}\
  \bibnamefont {Rogers}}, \bibinfo {author} {\bibfnamefont {S.}~\bibnamefont
  {Rossi}}, \bibinfo {author} {\bibfnamefont {M.}~\bibnamefont {Sbragaglia}},
  \bibinfo {author} {\bibfnamefont {G.}~\bibnamefont {Tarjus}}, \bibinfo
  {author} {\bibfnamefont {F.}~\bibnamefont {Toschi}}, \bibinfo {author}
  {\bibfnamefont {V.}~\bibnamefont {Trappe}}, \bibinfo {author} {\bibfnamefont
  {J.}~\bibnamefont {Vermant}}, \bibinfo {author} {\bibfnamefont
  {M.}~\bibnamefont {Wyart}}, \bibinfo {author} {\bibfnamefont
  {F.}~\bibnamefont {Zamponi}}, \ and\ \bibinfo {author} {\bibfnamefont
  {D.}~\bibnamefont {Zare}},\ }\href {\doibase 10.1039/D3SM01740K} {\bibfield
  {journal} {\bibinfo  {journal} {Soft Matter}\ ,\ \bibinfo {pages}
  {10.1039.D3SM01740K}} (\bibinfo {year} {2024})}\BibitemShut {NoStop}%
\bibitem [{\citenamefont {Donley}\ \emph {et~al.}(2020)\citenamefont {Donley},
  \citenamefont {Singh}, \citenamefont {Shetty},\ and\ \citenamefont
  {Rogers}}]{donleyElucidatingOvershootSoft2020}%
  \BibitemOpen
  \bibfield  {author} {\bibinfo {author} {\bibfnamefont {G.~J.}\ \bibnamefont
  {Donley}}, \bibinfo {author} {\bibfnamefont {P.~K.}\ \bibnamefont {Singh}},
  \bibinfo {author} {\bibfnamefont {A.}~\bibnamefont {Shetty}}, \ and\ \bibinfo
  {author} {\bibfnamefont {S.~A.}\ \bibnamefont {Rogers}},\ }\href {\doibase
  10.1073/pnas.2003869117} {\bibfield  {journal} {\bibinfo  {journal} {Proc
  Natl Acad Sci U.S.A.}\ }\textbf {\bibinfo {volume} {117}},\ \bibinfo {pages}
  {21945} (\bibinfo {year} {2020})}\BibitemShut {NoStop}%
\bibitem [{\citenamefont {Kamani}\ \emph {et~al.}(2021)\citenamefont {Kamani},
  \citenamefont {Donley},\ and\ \citenamefont
  {Rogers}}]{kamaniUnificationRheologicalPhysics2021}%
  \BibitemOpen
  \bibfield  {author} {\bibinfo {author} {\bibfnamefont {K.}~\bibnamefont
  {Kamani}}, \bibinfo {author} {\bibfnamefont {G.~J.}\ \bibnamefont {Donley}},
  \ and\ \bibinfo {author} {\bibfnamefont {S.~A.}\ \bibnamefont {Rogers}},\
  }\href {\doibase 10.1103/PhysRevLett.126.218002} {\bibfield  {journal}
  {\bibinfo  {journal} {Phys. Rev. Lett.}\ }\textbf {\bibinfo {volume} {126}},\
  \bibinfo {pages} {218002} (\bibinfo {year} {2021})}\BibitemShut {NoStop}%
\bibitem [{\citenamefont {Pine}\ \emph {et~al.}(2005)\citenamefont {Pine},
  \citenamefont {Gollub}, \citenamefont {Brady},\ and\ \citenamefont
  {Leshansky}}]{pineChaosThresholdIrreversibility2005}%
  \BibitemOpen
  \bibfield  {author} {\bibinfo {author} {\bibfnamefont {D.~J.}\ \bibnamefont
  {Pine}}, \bibinfo {author} {\bibfnamefont {J.~P.}\ \bibnamefont {Gollub}},
  \bibinfo {author} {\bibfnamefont {J.~F.}\ \bibnamefont {Brady}}, \ and\
  \bibinfo {author} {\bibfnamefont {A.~M.}\ \bibnamefont {Leshansky}},\ }\href
  {\doibase 10.1038/nature04380} {\bibfield  {journal} {\bibinfo  {journal}
  {Nature}\ }\textbf {\bibinfo {volume} {438}},\ \bibinfo {pages} {997}
  (\bibinfo {year} {2005})}\BibitemShut {NoStop}%
\bibitem [{\citenamefont {Hima~Nagamanasa}\ \emph {et~al.}(2014)\citenamefont
  {Hima~Nagamanasa}, \citenamefont {Gokhale}, \citenamefont {Sood},\ and\
  \citenamefont
  {Ganapathy}}]{himanagamanasaExperimentalSignaturesNonequilibrium2014}%
  \BibitemOpen
  \bibfield  {author} {\bibinfo {author} {\bibfnamefont {K.}~\bibnamefont
  {Hima~Nagamanasa}}, \bibinfo {author} {\bibfnamefont {S.}~\bibnamefont
  {Gokhale}}, \bibinfo {author} {\bibfnamefont {A.~K.}\ \bibnamefont {Sood}}, \
  and\ \bibinfo {author} {\bibfnamefont {R.}~\bibnamefont {Ganapathy}},\ }\href
  {\doibase 10.1103/PhysRevE.89.062308} {\bibfield  {journal} {\bibinfo
  {journal} {Phys. Rev. E}\ }\textbf {\bibinfo {volume} {89}},\ \bibinfo
  {pages} {062308} (\bibinfo {year} {2014})}\BibitemShut {NoStop}%
\bibitem [{\citenamefont {Knowlton}\ \emph {et~al.}(2014)\citenamefont
  {Knowlton}, \citenamefont {Pine},\ and\ \citenamefont
  {Cipelletti}}]{knowltonMicroscopicViewYielding2014}%
  \BibitemOpen
  \bibfield  {author} {\bibinfo {author} {\bibfnamefont {E.~D.}\ \bibnamefont
  {Knowlton}}, \bibinfo {author} {\bibfnamefont {D.~J.}\ \bibnamefont {Pine}},
  \ and\ \bibinfo {author} {\bibfnamefont {L.}~\bibnamefont {Cipelletti}},\
  }\href {\doibase 10.1039/c4sm00531g} {\bibfield  {journal} {\bibinfo
  {journal} {Soft Matter}\ }\textbf {\bibinfo {volume} {10}},\ \bibinfo {pages}
  {6931} (\bibinfo {year} {2014})},\ \Eprint {http://arxiv.org/abs/1403.4433}
  {arXiv:1403.4433} \BibitemShut {NoStop}%
\bibitem [{\citenamefont {Gopal}\ and\ \citenamefont
  {Durian}(1995)}]{gopalNonlinearBubbleDynamics1995}%
  \BibitemOpen
  \bibfield  {author} {\bibinfo {author} {\bibfnamefont {A.~D.}\ \bibnamefont
  {Gopal}}\ and\ \bibinfo {author} {\bibfnamefont {D.~J.}\ \bibnamefont
  {Durian}},\ }\href {\doibase 10.1103/PhysRevLett.75.2610} {\bibfield
  {journal} {\bibinfo  {journal} {Phys. Rev. Lett.}\ }\textbf {\bibinfo
  {volume} {75}},\ \bibinfo {pages} {2610} (\bibinfo {year}
  {1995})}\BibitemShut {NoStop}%
\bibitem [{\citenamefont {Fiocco}\ \emph {et~al.}(2013)\citenamefont {Fiocco},
  \citenamefont {Foffi},\ and\ \citenamefont
  {Sastry}}]{fioccoOscillatoryAthermalQuasistatic2013}%
  \BibitemOpen
  \bibfield  {author} {\bibinfo {author} {\bibfnamefont {D.}~\bibnamefont
  {Fiocco}}, \bibinfo {author} {\bibfnamefont {G.}~\bibnamefont {Foffi}}, \
  and\ \bibinfo {author} {\bibfnamefont {S.}~\bibnamefont {Sastry}},\ }\href
  {\doibase 10.1103/PhysRevE.88.020301} {\bibfield  {journal} {\bibinfo
  {journal} {Phys. Rev. E}\ }\textbf {\bibinfo {volume} {88}},\ \bibinfo
  {pages} {020301} (\bibinfo {year} {2013})}\BibitemShut {NoStop}%
\bibitem [{\citenamefont {Kawasaki}\ and\ \citenamefont
  {Berthier}(2016)}]{kawasakiMacroscopicYieldingJammed2016}%
  \BibitemOpen
  \bibfield  {author} {\bibinfo {author} {\bibfnamefont {T.}~\bibnamefont
  {Kawasaki}}\ and\ \bibinfo {author} {\bibfnamefont {L.}~\bibnamefont
  {Berthier}},\ }\href {\doibase 10.1103/PhysRevE.94.022615} {\bibfield
  {journal} {\bibinfo  {journal} {Phys. Rev. E}\ }\textbf {\bibinfo {volume}
  {94}},\ \bibinfo {pages} {022615} (\bibinfo {year} {2016})}\BibitemShut
  {NoStop}%
\bibitem [{\citenamefont {Regev}\ \emph {et~al.}(2013)\citenamefont {Regev},
  \citenamefont {Lookman},\ and\ \citenamefont
  {Reichhardt}}]{regevOnsetIrreversibilityChaos2013}%
  \BibitemOpen
  \bibfield  {author} {\bibinfo {author} {\bibfnamefont {I.}~\bibnamefont
  {Regev}}, \bibinfo {author} {\bibfnamefont {T.}~\bibnamefont {Lookman}}, \
  and\ \bibinfo {author} {\bibfnamefont {C.}~\bibnamefont {Reichhardt}},\
  }\href {\doibase 10.1103/PhysRevE.88.062401} {\bibfield  {journal} {\bibinfo
  {journal} {Phys. Rev. E}\ }\textbf {\bibinfo {volume} {88}},\ \bibinfo
  {pages} {062401} (\bibinfo {year} {2013})}\BibitemShut {NoStop}%
\bibitem [{\citenamefont {Bhaumik}\ \emph {et~al.}(2021)\citenamefont
  {Bhaumik}, \citenamefont {Foffi},\ and\ \citenamefont
  {Sastry}}]{bhaumikRoleAnnealingDetermining2021}%
  \BibitemOpen
  \bibfield  {author} {\bibinfo {author} {\bibfnamefont {H.}~\bibnamefont
  {Bhaumik}}, \bibinfo {author} {\bibfnamefont {G.}~\bibnamefont {Foffi}}, \
  and\ \bibinfo {author} {\bibfnamefont {S.}~\bibnamefont {Sastry}},\ }\href
  {\doibase 10.1073/pnas.2100227118} {\bibfield  {journal} {\bibinfo  {journal}
  {Proc Natl Acad Sci USA}\ }\textbf {\bibinfo {volume} {118}},\ \bibinfo
  {pages} {e2100227118} (\bibinfo {year} {2021})}\BibitemShut {NoStop}%
\bibitem [{\citenamefont {Ness}\ and\ \citenamefont
  {Cates}(2020)}]{nessAbsorbingStateTransitionsGranular2020}%
  \BibitemOpen
  \bibfield  {author} {\bibinfo {author} {\bibfnamefont {C.}~\bibnamefont
  {Ness}}\ and\ \bibinfo {author} {\bibfnamefont {M.~E.}\ \bibnamefont
  {Cates}},\ }\href {\doibase 10.1103/PhysRevLett.124.088004} {\bibfield
  {journal} {\bibinfo  {journal} {Phys. Rev. Lett.}\ }\textbf {\bibinfo
  {volume} {124}},\ \bibinfo {pages} {088004} (\bibinfo {year}
  {2020})}\BibitemShut {NoStop}%
\bibitem [{\citenamefont {Mari}\ \emph {et~al.}(2022)\citenamefont {Mari},
  \citenamefont {Bertin},\ and\ \citenamefont
  {Nardini}}]{mariAbsorbingPhaseTransitions2022}%
  \BibitemOpen
  \bibfield  {author} {\bibinfo {author} {\bibfnamefont {R.}~\bibnamefont
  {Mari}}, \bibinfo {author} {\bibfnamefont {E.}~\bibnamefont {Bertin}}, \ and\
  \bibinfo {author} {\bibfnamefont {C.}~\bibnamefont {Nardini}},\ }\href
  {\doibase 10.1103/PhysRevE.105.L032602} {\bibfield  {journal} {\bibinfo
  {journal} {Phys. Rev. E}\ }\textbf {\bibinfo {volume} {105}},\ \bibinfo
  {pages} {L032602} (\bibinfo {year} {2022})}\BibitemShut {NoStop}%
\bibitem [{\citenamefont {Mungan}\ and\ \citenamefont
  {Witten}(2019)}]{munganCyclicAnnealingIterated2019}%
  \BibitemOpen
  \bibfield  {author} {\bibinfo {author} {\bibfnamefont {M.}~\bibnamefont
  {Mungan}}\ and\ \bibinfo {author} {\bibfnamefont {T.~A.}\ \bibnamefont
  {Witten}},\ }\href {\doibase 10.1103/PhysRevE.99.052132} {\bibfield
  {journal} {\bibinfo  {journal} {Phys. Rev. E}\ }\textbf {\bibinfo {volume}
  {99}},\ \bibinfo {pages} {052132} (\bibinfo {year} {2019})}\BibitemShut
  {NoStop}%
\bibitem [{\citenamefont
  {Hinrichsen}(2000)}]{hinrichsenNonequilibriumCriticalPhenomena2000}%
  \BibitemOpen
  \bibfield  {author} {\bibinfo {author} {\bibfnamefont {H.}~\bibnamefont
  {Hinrichsen}},\ }\href {\doibase 10.1080/00018730050198152} {\bibfield
  {journal} {\bibinfo  {journal} {Advances in Physics}\ }\textbf {\bibinfo
  {volume} {49}},\ \bibinfo {pages} {815} (\bibinfo {year} {2000})}\BibitemShut
  {NoStop}%
\bibitem [{\citenamefont {Lübeck}(2004)}]{lubeckUniversalScalingBehavior2004}%
  \BibitemOpen
  \bibfield  {author} {\bibinfo {author} {\bibfnamefont {S.}~\bibnamefont
  {Lübeck}},\ }\href {\doibase 10.1142/S0217979204027748} {\bibfield
  {journal} {\bibinfo  {journal} {Int. J. Mod. Phys. B}\ }\textbf {\bibinfo
  {volume} {18}},\ \bibinfo {pages} {3977} (\bibinfo {year}
  {2004})}\BibitemShut {NoStop}%
\bibitem [{\citenamefont {Dickman}\ \emph {et~al.}(1998)\citenamefont
  {Dickman}, \citenamefont {Vespignani},\ and\ \citenamefont
  {Zapperi}}]{dickmanSelforganizedCriticalityAbsorbingstate1998}%
  \BibitemOpen
  \bibfield  {author} {\bibinfo {author} {\bibfnamefont {R.}~\bibnamefont
  {Dickman}}, \bibinfo {author} {\bibfnamefont {A.}~\bibnamefont {Vespignani}},
  \ and\ \bibinfo {author} {\bibfnamefont {S.}~\bibnamefont {Zapperi}},\ }\href
  {\doibase 10.1103/PhysRevE.57.5095} {\bibfield  {journal} {\bibinfo
  {journal} {Phys. Rev. E}\ }\textbf {\bibinfo {volume} {57}},\ \bibinfo
  {pages} {5095} (\bibinfo {year} {1998})}\BibitemShut {NoStop}%
\bibitem [{\citenamefont {Maire}\ \emph {et~al.}(2024)\citenamefont {Maire},
  \citenamefont {Plati}, \citenamefont {Stockinger}, \citenamefont {Trizac},
  \citenamefont {Smallenburg},\ and\ \citenamefont
  {Foffi}}]{maireInterplayAbsorbingPhase2024}%
  \BibitemOpen
  \bibfield  {author} {\bibinfo {author} {\bibfnamefont {R.}~\bibnamefont
  {Maire}}, \bibinfo {author} {\bibfnamefont {A.}~\bibnamefont {Plati}},
  \bibinfo {author} {\bibfnamefont {M.}~\bibnamefont {Stockinger}}, \bibinfo
  {author} {\bibfnamefont {E.}~\bibnamefont {Trizac}}, \bibinfo {author}
  {\bibfnamefont {F.}~\bibnamefont {Smallenburg}}, \ and\ \bibinfo {author}
  {\bibfnamefont {G.}~\bibnamefont {Foffi}},\ }\href {\doibase
  10.1103/PhysRevLett.132.238202} {\bibfield  {journal} {\bibinfo  {journal}
  {Phys. Rev. Lett.}\ }\textbf {\bibinfo {volume} {132}},\ \bibinfo {pages}
  {238202} (\bibinfo {year} {2024})}\BibitemShut {NoStop}%
\bibitem [{\citenamefont {Ziff}\ \emph {et~al.}(1986)\citenamefont {Ziff},
  \citenamefont {Gulari},\ and\ \citenamefont
  {Barshad}}]{ziffKineticPhaseTransitions1986}%
  \BibitemOpen
  \bibfield  {author} {\bibinfo {author} {\bibfnamefont {R.~M.}\ \bibnamefont
  {Ziff}}, \bibinfo {author} {\bibfnamefont {E.}~\bibnamefont {Gulari}}, \ and\
  \bibinfo {author} {\bibfnamefont {Y.}~\bibnamefont {Barshad}},\ }\href
  {\doibase 10.1103/PhysRevLett.56.2553} {\bibfield  {journal} {\bibinfo
  {journal} {Phys. Rev. Lett.}\ }\textbf {\bibinfo {volume} {56}},\ \bibinfo
  {pages} {2553} (\bibinfo {year} {1986})}\BibitemShut {NoStop}%
\bibitem [{\citenamefont {Cort{\'e}}\ \emph {et~al.}(2008)\citenamefont
  {Cort{\'e}}, \citenamefont {Chaikin}, \citenamefont {Gollub},\ and\
  \citenamefont {Pine}}]{corteRandomOrganizationPeriodically2008}%
  \BibitemOpen
  \bibfield  {author} {\bibinfo {author} {\bibfnamefont {L.}~\bibnamefont
  {Cort{\'e}}}, \bibinfo {author} {\bibfnamefont {P.~M.}\ \bibnamefont
  {Chaikin}}, \bibinfo {author} {\bibfnamefont {J.~P.}\ \bibnamefont {Gollub}},
  \ and\ \bibinfo {author} {\bibfnamefont {D.~J.}\ \bibnamefont {Pine}},\
  }\href {\doibase 10.1038/nphys891} {\bibfield  {journal} {\bibinfo  {journal}
  {Nature Phys}\ }\textbf {\bibinfo {volume} {4}},\ \bibinfo {pages} {420}
  (\bibinfo {year} {2008})}\BibitemShut {NoStop}%
\bibitem [{\citenamefont {Jeanneret}\ and\ \citenamefont
  {Bartolo}(2014)}]{jeanneretGeometricallyProtectedReversibility2014}%
  \BibitemOpen
  \bibfield  {author} {\bibinfo {author} {\bibfnamefont {R.}~\bibnamefont
  {Jeanneret}}\ and\ \bibinfo {author} {\bibfnamefont {D.}~\bibnamefont
  {Bartolo}},\ }\href {\doibase 10.1038/ncomms4474} {\bibfield  {journal}
  {\bibinfo  {journal} {Nat. Commun.}\ }\textbf {\bibinfo {volume} {5}},\
  \bibinfo {pages} {3474} (\bibinfo {year} {2014})}\BibitemShut {NoStop}%
\bibitem [{\citenamefont {Keim}\ and\ \citenamefont
  {Arratia}(2014)}]{keimMechanicalMicroscopicProperties2014}%
  \BibitemOpen
  \bibfield  {author} {\bibinfo {author} {\bibfnamefont {N.~C.}\ \bibnamefont
  {Keim}}\ and\ \bibinfo {author} {\bibfnamefont {P.~E.}\ \bibnamefont
  {Arratia}},\ }\href {\doibase 10.1103/PhysRevLett.112.028302} {\bibfield
  {journal} {\bibinfo  {journal} {Phys. Rev. Lett.}\ }\textbf {\bibinfo
  {volume} {112}},\ \bibinfo {pages} {028302} (\bibinfo {year}
  {2014})}\BibitemShut {NoStop}%
\bibitem [{\citenamefont {Tjhung}\ and\ \citenamefont
  {Berthier}(2015)}]{tjhungHyperuniformDensityFluctuations2015}%
  \BibitemOpen
  \bibfield  {author} {\bibinfo {author} {\bibfnamefont {E.}~\bibnamefont
  {Tjhung}}\ and\ \bibinfo {author} {\bibfnamefont {L.}~\bibnamefont
  {Berthier}},\ }\href {\doibase 10.1103/PhysRevLett.114.148301} {\bibfield
  {journal} {\bibinfo  {journal} {Phys. Rev. Lett.}\ }\textbf {\bibinfo
  {volume} {114}},\ \bibinfo {pages} {148301} (\bibinfo {year}
  {2015})}\BibitemShut {NoStop}%
\bibitem [{\citenamefont {Weijs}\ \emph {et~al.}(2015)\citenamefont {Weijs},
  \citenamefont {Jeanneret}, \citenamefont {Dreyfus},\ and\ \citenamefont
  {Bartolo}}]{weijsEmergentHyperuniformityPeriodically2015}%
  \BibitemOpen
  \bibfield  {author} {\bibinfo {author} {\bibfnamefont {J.~H.}\ \bibnamefont
  {Weijs}}, \bibinfo {author} {\bibfnamefont {R.}~\bibnamefont {Jeanneret}},
  \bibinfo {author} {\bibfnamefont {R.}~\bibnamefont {Dreyfus}}, \ and\
  \bibinfo {author} {\bibfnamefont {D.}~\bibnamefont {Bartolo}},\ }\href
  {\doibase 10.1103/PhysRevLett.115.108301} {\bibfield  {journal} {\bibinfo
  {journal} {Phys. Rev. Lett.}\ }\textbf {\bibinfo {volume} {115}},\ \bibinfo
  {pages} {108301} (\bibinfo {year} {2015})}\BibitemShut {NoStop}%
\bibitem [{\citenamefont {Takeuchi}\ \emph {et~al.}(2007)\citenamefont
  {Takeuchi}, \citenamefont {Kuroda}, \citenamefont {Chaté},\ and\
  \citenamefont {Sano}}]{takeuchiDirectedPercolationCriticality2007}%
  \BibitemOpen
  \bibfield  {author} {\bibinfo {author} {\bibfnamefont {K.~A.}\ \bibnamefont
  {Takeuchi}}, \bibinfo {author} {\bibfnamefont {M.}~\bibnamefont {Kuroda}},
  \bibinfo {author} {\bibfnamefont {H.}~\bibnamefont {Chaté}}, \ and\ \bibinfo
  {author} {\bibfnamefont {M.}~\bibnamefont {Sano}},\ }\href {\doibase
  10.1103/PhysRevLett.99.234503} {\bibfield  {journal} {\bibinfo  {journal}
  {Phys. Rev. Lett.}\ }\textbf {\bibinfo {volume} {99}},\ \bibinfo {pages}
  {234503} (\bibinfo {year} {2007})}\BibitemShut {NoStop}%
\bibitem [{\citenamefont {Möbius}\ and\ \citenamefont
  {Heussinger}()}]{mobiusIrreversibilityDenseGranular2014}%
  \BibitemOpen
  \bibfield  {author} {\bibinfo {author} {\bibfnamefont {R.}~\bibnamefont
  {Möbius}}\ and\ \bibinfo {author} {\bibfnamefont {C.}~\bibnamefont
  {Heussinger}},\ }\href {\doibase 10.1039/C4SM00178H} {\bibfield  {journal}
  {\bibinfo  {journal} {Soft Matter}\ }\textbf {\bibinfo {volume} {10}},\
  \bibinfo {pages} {4806}}\BibitemShut {NoStop}%
\bibitem [{\citenamefont {Néel}\ \emph {et~al.}(2014)\citenamefont {Néel},
  \citenamefont {Rondini}, \citenamefont {Turzillo}, \citenamefont {Mujica},\
  and\ \citenamefont {Soto}}]{neelDynamicsFirstorderTransition2014}%
  \BibitemOpen
  \bibfield  {author} {\bibinfo {author} {\bibfnamefont {B.}~\bibnamefont
  {Néel}}, \bibinfo {author} {\bibfnamefont {I.}~\bibnamefont {Rondini}},
  \bibinfo {author} {\bibfnamefont {A.}~\bibnamefont {Turzillo}}, \bibinfo
  {author} {\bibfnamefont {N.}~\bibnamefont {Mujica}}, \ and\ \bibinfo {author}
  {\bibfnamefont {R.}~\bibnamefont {Soto}},\ }\href {\doibase
  10.1103/PhysRevE.89.042206} {\bibfield  {journal} {\bibinfo  {journal} {Phys.
  Rev. E}\ }\textbf {\bibinfo {volume} {89}},\ \bibinfo {pages} {042206}
  (\bibinfo {year} {2014})}\BibitemShut {NoStop}%
\bibitem [{\citenamefont {Chantry}\ \emph {et~al.}(2017)\citenamefont
  {Chantry}, \citenamefont {Tuckerman},\ and\ \citenamefont
  {Barkley}}]{chantryUniversalContinuousTransition2017}%
  \BibitemOpen
  \bibfield  {author} {\bibinfo {author} {\bibfnamefont {M.}~\bibnamefont
  {Chantry}}, \bibinfo {author} {\bibfnamefont {L.~S.}\ \bibnamefont
  {Tuckerman}}, \ and\ \bibinfo {author} {\bibfnamefont {D.}~\bibnamefont
  {Barkley}},\ }\href {\doibase 10.1017/jfm.2017.405} {\bibfield  {journal}
  {\bibinfo  {journal} {J. Fluid Mech.}\ }\textbf {\bibinfo {volume} {824}},\
  \bibinfo {pages} {R1} (\bibinfo {year} {2017})}\BibitemShut {NoStop}%
\bibitem [{\citenamefont {Rogers}\ \emph {et~al.}(2018)\citenamefont {Rogers},
  \citenamefont {Chen}, \citenamefont {Pagenkopp}, \citenamefont {Mason},
  \citenamefont {Narayanan}, \citenamefont {Harden},\ and\ \citenamefont
  {Leheny}}]{rogersMicroscopicSignaturesYielding2018}%
  \BibitemOpen
  \bibfield  {author} {\bibinfo {author} {\bibfnamefont {M.~C.}\ \bibnamefont
  {Rogers}}, \bibinfo {author} {\bibfnamefont {K.}~\bibnamefont {Chen}},
  \bibinfo {author} {\bibfnamefont {M.~J.}\ \bibnamefont {Pagenkopp}}, \bibinfo
  {author} {\bibfnamefont {T.~G.}\ \bibnamefont {Mason}}, \bibinfo {author}
  {\bibfnamefont {S.}~\bibnamefont {Narayanan}}, \bibinfo {author}
  {\bibfnamefont {J.~L.}\ \bibnamefont {Harden}}, \ and\ \bibinfo {author}
  {\bibfnamefont {R.~L.}\ \bibnamefont {Leheny}},\ }\href {\doibase
  10.1103/PhysRevMaterials.2.095601} {\bibfield  {journal} {\bibinfo  {journal}
  {Physical Review Materials}\ }\textbf {\bibinfo {volume} {2}},\ \bibinfo
  {pages} {095601} (\bibinfo {year} {2018})}\BibitemShut {NoStop}%
\bibitem [{\citenamefont {Jaiswal}\ \emph {et~al.}(2016)\citenamefont
  {Jaiswal}, \citenamefont {Procaccia}, \citenamefont {Rainone},\ and\
  \citenamefont {Singh}}]{jaiswalMechanicalYieldAmorphous2016}%
  \BibitemOpen
  \bibfield  {author} {\bibinfo {author} {\bibfnamefont {P.~K.}\ \bibnamefont
  {Jaiswal}}, \bibinfo {author} {\bibfnamefont {I.}~\bibnamefont {Procaccia}},
  \bibinfo {author} {\bibfnamefont {C.}~\bibnamefont {Rainone}}, \ and\
  \bibinfo {author} {\bibfnamefont {M.}~\bibnamefont {Singh}},\ }\href
  {\doibase 10.1103/PhysRevLett.116.085501} {\bibfield  {journal} {\bibinfo
  {journal} {Phys. Rev. Lett.}\ }\textbf {\bibinfo {volume} {116}},\ \bibinfo
  {pages} {085501} (\bibinfo {year} {2016})}\BibitemShut {NoStop}%
\bibitem [{\citenamefont {Karmakar}\ \emph {et~al.}(2010)\citenamefont
  {Karmakar}, \citenamefont {Lerner},\ and\ \citenamefont
  {Procaccia}}]{karmakarStatisticalPhysicsYielding2010}%
  \BibitemOpen
  \bibfield  {author} {\bibinfo {author} {\bibfnamefont {S.}~\bibnamefont
  {Karmakar}}, \bibinfo {author} {\bibfnamefont {E.}~\bibnamefont {Lerner}}, \
  and\ \bibinfo {author} {\bibfnamefont {I.}~\bibnamefont {Procaccia}},\ }\href
  {\doibase 10.1103/PhysRevE.82.055103} {\bibfield  {journal} {\bibinfo
  {journal} {Phys. Rev. E}\ }\textbf {\bibinfo {volume} {82}},\ \bibinfo
  {pages} {055103} (\bibinfo {year} {2010})}\BibitemShut {NoStop}%
\bibitem [{\citenamefont {Leishangthem}\ \emph {et~al.}(2017)\citenamefont
  {Leishangthem}, \citenamefont {Parmar},\ and\ \citenamefont
  {Sastry}}]{leishangthemYieldingTransitionAmorphous2017}%
  \BibitemOpen
  \bibfield  {author} {\bibinfo {author} {\bibfnamefont {P.}~\bibnamefont
  {Leishangthem}}, \bibinfo {author} {\bibfnamefont {A.~D.~S.}\ \bibnamefont
  {Parmar}}, \ and\ \bibinfo {author} {\bibfnamefont {S.}~\bibnamefont
  {Sastry}},\ }\href {\doibase 10.1038/ncomms14653} {\bibfield  {journal}
  {\bibinfo  {journal} {Nat Commun}\ }\textbf {\bibinfo {volume} {8}},\
  \bibinfo {pages} {14653} (\bibinfo {year} {2017})}\BibitemShut {NoStop}%
\bibitem [{\citenamefont {Divoux}\ \emph {et~al.}(2013)\citenamefont {Divoux},
  \citenamefont {Grenard},\ and\ \citenamefont
  {Manneville}}]{divouxRheologicalHysteresisSoft2013}%
  \BibitemOpen
  \bibfield  {author} {\bibinfo {author} {\bibfnamefont {T.}~\bibnamefont
  {Divoux}}, \bibinfo {author} {\bibfnamefont {V.}~\bibnamefont {Grenard}}, \
  and\ \bibinfo {author} {\bibfnamefont {S.}~\bibnamefont {Manneville}},\
  }\href {\doibase 10.1103/PhysRevLett.110.018304} {\bibfield  {journal}
  {\bibinfo  {journal} {Phys. Rev. Lett.}\ }\textbf {\bibinfo {volume} {110}},\
  \bibinfo {pages} {018304} (\bibinfo {year} {2013})}\BibitemShut {NoStop}%
\bibitem [{\citenamefont {Mungan}\ \emph {et~al.}()\citenamefont {Mungan},
  \citenamefont {Sastry}, \citenamefont {Dahmen},\ and\ \citenamefont
  {Regev}}]{munganNetworksHierarchiesHow2019}%
  \BibitemOpen
  \bibfield  {author} {\bibinfo {author} {\bibfnamefont {M.}~\bibnamefont
  {Mungan}}, \bibinfo {author} {\bibfnamefont {S.}~\bibnamefont {Sastry}},
  \bibinfo {author} {\bibfnamefont {K.}~\bibnamefont {Dahmen}}, \ and\ \bibinfo
  {author} {\bibfnamefont {I.}~\bibnamefont {Regev}},\ }\href {\doibase
  10.1103/PhysRevLett.123.178002} {\bibfield  {journal} {\bibinfo  {journal}
  {Phys. Rev. Lett.}\ }\textbf {\bibinfo {volume} {123}},\ \bibinfo {pages}
  {178002}}\BibitemShut {NoStop}%
\bibitem [{\citenamefont {Keim}\ and\ \citenamefont
  {Arratia}(2013)}]{keimYieldingMicrostructure2D2013}%
  \BibitemOpen
  \bibfield  {author} {\bibinfo {author} {\bibfnamefont {N.~C.}\ \bibnamefont
  {Keim}}\ and\ \bibinfo {author} {\bibfnamefont {P.~E.}\ \bibnamefont
  {Arratia}},\ }\href {\doibase 10.1039/c3sm51014j} {\bibfield  {journal}
  {\bibinfo  {journal} {Soft Matter}\ }\textbf {\bibinfo {volume} {9}},\
  \bibinfo {pages} {6222} (\bibinfo {year} {2013})}\BibitemShut {NoStop}%
\bibitem [{\citenamefont {Nordstrom}\ \emph {et~al.}(2011)\citenamefont
  {Nordstrom}, \citenamefont {Gollub},\ and\ \citenamefont
  {Durian}}]{nordstromDynamicalHeterogeneitySoftparticle2011}%
  \BibitemOpen
  \bibfield  {author} {\bibinfo {author} {\bibfnamefont {K.~N.}\ \bibnamefont
  {Nordstrom}}, \bibinfo {author} {\bibfnamefont {J.~P.}\ \bibnamefont
  {Gollub}}, \ and\ \bibinfo {author} {\bibfnamefont {D.~J.}\ \bibnamefont
  {Durian}},\ }\href {\doibase 10.1103/PhysRevE.84.021403} {\bibfield
  {journal} {\bibinfo  {journal} {Phys. Rev. E}\ }\textbf {\bibinfo {volume}
  {84}},\ \bibinfo {pages} {021403} (\bibinfo {year} {2011})}\BibitemShut
  {NoStop}%
\bibitem [{\citenamefont {Aime}\ \emph {et~al.}(2023)\citenamefont {Aime},
  \citenamefont {Truzzolillo}, \citenamefont {Pine}, \citenamefont {Ramos},\
  and\ \citenamefont {Cipelletti}}]{aimeUnifiedStateDiagram2023}%
  \BibitemOpen
  \bibfield  {author} {\bibinfo {author} {\bibfnamefont {S.}~\bibnamefont
  {Aime}}, \bibinfo {author} {\bibfnamefont {D.}~\bibnamefont {Truzzolillo}},
  \bibinfo {author} {\bibfnamefont {D.~J.}\ \bibnamefont {Pine}}, \bibinfo
  {author} {\bibfnamefont {L.}~\bibnamefont {Ramos}}, \ and\ \bibinfo {author}
  {\bibfnamefont {L.}~\bibnamefont {Cipelletti}},\ }\href {\doibase
  10.1038/s41567-023-02153-w} {\bibfield  {journal} {\bibinfo  {journal} {Nat.
  Phys.}\ } (\bibinfo {year} {2023}),\ 10.1038/s41567-023-02153-w}\BibitemShut
  {NoStop}%
\bibitem [{\citenamefont {Goyon}\ \emph {et~al.}(2008)\citenamefont {Goyon},
  \citenamefont {Colin}, \citenamefont {Ovarlez}, \citenamefont {Ajdari},\ and\
  \citenamefont {Bocquet}}]{goyonSpatialCooperativitySoft2008}%
  \BibitemOpen
  \bibfield  {author} {\bibinfo {author} {\bibfnamefont {J.}~\bibnamefont
  {Goyon}}, \bibinfo {author} {\bibfnamefont {A.}~\bibnamefont {Colin}},
  \bibinfo {author} {\bibfnamefont {G.}~\bibnamefont {Ovarlez}}, \bibinfo
  {author} {\bibfnamefont {A.}~\bibnamefont {Ajdari}}, \ and\ \bibinfo {author}
  {\bibfnamefont {L.}~\bibnamefont {Bocquet}},\ }\href {\doibase
  10.1038/nature07026} {\bibfield  {journal} {\bibinfo  {journal} {Nature}\
  }\textbf {\bibinfo {volume} {454}},\ \bibinfo {pages} {84} (\bibinfo {year}
  {2008})}\BibitemShut {NoStop}%
\bibitem [{\citenamefont {Bonnoit}\ \emph {et~al.}(2010)\citenamefont
  {Bonnoit}, \citenamefont {Lanuza}, \citenamefont {Lindner},\ and\
  \citenamefont {Clement}}]{bonnoitMesoscopicLengthScale2010}%
  \BibitemOpen
  \bibfield  {author} {\bibinfo {author} {\bibfnamefont {C.}~\bibnamefont
  {Bonnoit}}, \bibinfo {author} {\bibfnamefont {J.}~\bibnamefont {Lanuza}},
  \bibinfo {author} {\bibfnamefont {A.}~\bibnamefont {Lindner}}, \ and\
  \bibinfo {author} {\bibfnamefont {E.}~\bibnamefont {Clement}},\ }\href
  {\doibase 10.1103/PhysRevLett.105.108302} {\bibfield  {journal} {\bibinfo
  {journal} {Phys. Rev. Lett.}\ }\textbf {\bibinfo {volume} {105}},\ \bibinfo
  {pages} {108302} (\bibinfo {year} {2010})}\BibitemShut {NoStop}%
\bibitem [{\citenamefont {Sentjabrskaja}\ \emph {et~al.}(2015)\citenamefont
  {Sentjabrskaja}, \citenamefont {Chaudhuri}, \citenamefont {Hermes},
  \citenamefont {Poon}, \citenamefont {Horbach}, \citenamefont {Egelhaaf},\
  and\ \citenamefont {Laurati}}]{sentjabrskajaCreepFlowGlasses2015}%
  \BibitemOpen
  \bibfield  {author} {\bibinfo {author} {\bibfnamefont {T.}~\bibnamefont
  {Sentjabrskaja}}, \bibinfo {author} {\bibfnamefont {P.}~\bibnamefont
  {Chaudhuri}}, \bibinfo {author} {\bibfnamefont {M.}~\bibnamefont {Hermes}},
  \bibinfo {author} {\bibfnamefont {W.~C.~K.}\ \bibnamefont {Poon}}, \bibinfo
  {author} {\bibfnamefont {J.}~\bibnamefont {Horbach}}, \bibinfo {author}
  {\bibfnamefont {S.~U.}\ \bibnamefont {Egelhaaf}}, \ and\ \bibinfo {author}
  {\bibfnamefont {M.}~\bibnamefont {Laurati}},\ }\href {\doibase
  10.1038/srep11884} {\bibfield  {journal} {\bibinfo  {journal} {Sci Rep}\
  }\textbf {\bibinfo {volume} {5}},\ \bibinfo {pages} {11884} (\bibinfo {year}
  {2015})}\BibitemShut {NoStop}%
\bibitem [{\citenamefont {Mukherji}\ \emph {et~al.}(2019)\citenamefont
  {Mukherji}, \citenamefont {Kandula}, \citenamefont {Sood},\ and\
  \citenamefont {Ganapathy}}]{mukherjiStrengthMechanicalMemories2019}%
  \BibitemOpen
  \bibfield  {author} {\bibinfo {author} {\bibfnamefont {S.}~\bibnamefont
  {Mukherji}}, \bibinfo {author} {\bibfnamefont {N.}~\bibnamefont {Kandula}},
  \bibinfo {author} {\bibfnamefont {A.~K.}\ \bibnamefont {Sood}}, \ and\
  \bibinfo {author} {\bibfnamefont {R.}~\bibnamefont {Ganapathy}},\ }\href
  {\doibase 10.1103/PhysRevLett.122.158001} {\bibfield  {journal} {\bibinfo
  {journal} {Phys. Rev. Lett.}\ }\textbf {\bibinfo {volume} {122}},\ \bibinfo
  {pages} {158001} (\bibinfo {year} {2019})}\BibitemShut {NoStop}%
\bibitem [{\citenamefont {Jop}\ \emph {et~al.}(2012)\citenamefont {Jop},
  \citenamefont {Mansard}, \citenamefont {Chaudhuri}, \citenamefont {Bocquet},\
  and\ \citenamefont {Colin}}]{jopMicroscaleRheologySoft2012}%
  \BibitemOpen
  \bibfield  {author} {\bibinfo {author} {\bibfnamefont {P.}~\bibnamefont
  {Jop}}, \bibinfo {author} {\bibfnamefont {V.}~\bibnamefont {Mansard}},
  \bibinfo {author} {\bibfnamefont {P.}~\bibnamefont {Chaudhuri}}, \bibinfo
  {author} {\bibfnamefont {L.}~\bibnamefont {Bocquet}}, \ and\ \bibinfo
  {author} {\bibfnamefont {A.}~\bibnamefont {Colin}},\ }\href {\doibase
  10.1103/PhysRevLett.108.148301} {\bibfield  {journal} {\bibinfo  {journal}
  {Phys. Rev. Lett.}\ }\textbf {\bibinfo {volume} {108}},\ \bibinfo {pages}
  {148301} (\bibinfo {year} {2012})}\BibitemShut {NoStop}%
\bibitem [{\citenamefont {Vinutha}\ \emph {et~al.}(2024)\citenamefont
  {Vinutha}, \citenamefont {Marchand}, \citenamefont {Caggioni}, \citenamefont
  {Vasisht}, \citenamefont {Del~Gado},\ and\ \citenamefont
  {Trappe}}]{vinuthaMemoryShearFlow2024}%
  \BibitemOpen
  \bibfield  {author} {\bibinfo {author} {\bibfnamefont {H.~A.}\ \bibnamefont
  {Vinutha}}, \bibinfo {author} {\bibfnamefont {M.}~\bibnamefont {Marchand}},
  \bibinfo {author} {\bibfnamefont {M.}~\bibnamefont {Caggioni}}, \bibinfo
  {author} {\bibfnamefont {V.~V.}\ \bibnamefont {Vasisht}}, \bibinfo {author}
  {\bibfnamefont {E.}~\bibnamefont {Del~Gado}}, \ and\ \bibinfo {author}
  {\bibfnamefont {V.}~\bibnamefont {Trappe}},\ }\href {\doibase
  10.1093/pnasnexus/pgae441} {\bibfield  {journal} {\bibinfo  {journal} {PNAS
  Nexus}\ }\textbf {\bibinfo {volume} {3}},\ \bibinfo {pages} {pgae441}
  (\bibinfo {year} {2024})}\BibitemShut {NoStop}%
\bibitem [{\citenamefont {Martens}\ \emph {et~al.}(2012)\citenamefont
  {Martens}, \citenamefont {Bocquet},\ and\ \citenamefont
  {Barrat}}]{martensSpontaneousFormationPermanent2012}%
  \BibitemOpen
  \bibfield  {author} {\bibinfo {author} {\bibfnamefont {K.}~\bibnamefont
  {Martens}}, \bibinfo {author} {\bibfnamefont {L.}~\bibnamefont {Bocquet}}, \
  and\ \bibinfo {author} {\bibfnamefont {J.-L.}\ \bibnamefont {Barrat}},\
  }\href {\doibase 10.1039/C2SM07090A} {\bibfield  {journal} {\bibinfo
  {journal} {Soft Matter}\ }\textbf {\bibinfo {volume} {8}},\ \bibinfo {pages}
  {4197} (\bibinfo {year} {2012})}\BibitemShut {NoStop}%
\bibitem [{\citenamefont {Keim}\ \emph {et~al.}(2020)\citenamefont {Keim},
  \citenamefont {Hass}, \citenamefont {Kroger},\ and\ \citenamefont
  {Wieker}}]{keimGlobalMemoryLocal2020}%
  \BibitemOpen
  \bibfield  {author} {\bibinfo {author} {\bibfnamefont {N.~C.}\ \bibnamefont
  {Keim}}, \bibinfo {author} {\bibfnamefont {J.}~\bibnamefont {Hass}}, \bibinfo
  {author} {\bibfnamefont {B.}~\bibnamefont {Kroger}}, \ and\ \bibinfo {author}
  {\bibfnamefont {D.}~\bibnamefont {Wieker}},\ }\href {\doibase
  10.1103/PhysRevResearch.2.012004} {\bibfield  {journal} {\bibinfo  {journal}
  {Phys. Rev. Res.}\ }\textbf {\bibinfo {volume} {2}},\ \bibinfo {pages}
  {012004} (\bibinfo {year} {2020})}\BibitemShut {NoStop}%
\bibitem [{\citenamefont {Bhaumik}\ \emph
  {et~al.}(2022{\natexlab{a}})\citenamefont {Bhaumik}, \citenamefont {Foffi},\
  and\ \citenamefont {Sastry}}]{bhaumikAvalanchesClustersStructural2022}%
  \BibitemOpen
  \bibfield  {author} {\bibinfo {author} {\bibfnamefont {H.}~\bibnamefont
  {Bhaumik}}, \bibinfo {author} {\bibfnamefont {G.}~\bibnamefont {Foffi}}, \
  and\ \bibinfo {author} {\bibfnamefont {S.}~\bibnamefont {Sastry}},\ }\href
  {\doibase 10.1103/PhysRevLett.128.098001} {\bibfield  {journal} {\bibinfo
  {journal} {Phys. Rev. Lett.}\ }\textbf {\bibinfo {volume} {128}},\ \bibinfo
  {pages} {098001} (\bibinfo {year} {2022}{\natexlab{a}})}\BibitemShut
  {NoStop}%
\bibitem [{\citenamefont {Bhaumik}\ \emph
  {et~al.}(2022{\natexlab{b}})\citenamefont {Bhaumik}, \citenamefont {Foffi},\
  and\ \citenamefont {Sastry}}]{bhaumikYieldingTransitionTwo2022}%
  \BibitemOpen
  \bibfield  {author} {\bibinfo {author} {\bibfnamefont {H.}~\bibnamefont
  {Bhaumik}}, \bibinfo {author} {\bibfnamefont {G.}~\bibnamefont {Foffi}}, \
  and\ \bibinfo {author} {\bibfnamefont {S.}~\bibnamefont {Sastry}},\ }\href
  {\doibase 10.1063/5.0085064} {\bibfield  {journal} {\bibinfo  {journal} {J.
  Chem. Phys.}\ }\textbf {\bibinfo {volume} {156}},\ \bibinfo {pages} {064502}
  (\bibinfo {year} {2022}{\natexlab{b}})}\BibitemShut {NoStop}%
\bibitem [{\citenamefont {Del~Gado}\ \emph {et~al.}(2008)\citenamefont
  {Del~Gado}, \citenamefont {Ilg}, \citenamefont {Kröger},\ and\ \citenamefont
  {Öttinger}}]{delgadoNonaffineDeformationInherent2008}%
  \BibitemOpen
  \bibfield  {author} {\bibinfo {author} {\bibfnamefont {E.}~\bibnamefont
  {Del~Gado}}, \bibinfo {author} {\bibfnamefont {P.}~\bibnamefont {Ilg}},
  \bibinfo {author} {\bibfnamefont {M.}~\bibnamefont {Kröger}}, \ and\
  \bibinfo {author} {\bibfnamefont {H.~C.}\ \bibnamefont {Öttinger}},\ }\href
  {\doibase 10.1103/PhysRevLett.101.095501} {\bibfield  {journal} {\bibinfo
  {journal} {Phys. Rev. Lett.}\ }\textbf {\bibinfo {volume} {101}},\ \bibinfo
  {pages} {095501} (\bibinfo {year} {2008})}\BibitemShut {NoStop}%
\bibitem [{\citenamefont {Manning}\ and\ \citenamefont
  {Liu}(2011)}]{manningVibrationalModesIdentify2011}%
  \BibitemOpen
  \bibfield  {author} {\bibinfo {author} {\bibfnamefont {M.~L.}\ \bibnamefont
  {Manning}}\ and\ \bibinfo {author} {\bibfnamefont {A.~J.}\ \bibnamefont
  {Liu}},\ }\href {\doibase 10.1103/PhysRevLett.107.108302} {\bibfield
  {journal} {\bibinfo  {journal} {Phys. Rev. Lett.}\ }\textbf {\bibinfo
  {volume} {107}},\ \bibinfo {pages} {108302} (\bibinfo {year}
  {2011})}\BibitemShut {NoStop}%
\bibitem [{\citenamefont {Schall}\ \emph {et~al.}(2007)\citenamefont {Schall},
  \citenamefont {Weitz},\ and\ \citenamefont
  {Spaepen}}]{schallStructuralRearrangementsThat2007}%
  \BibitemOpen
  \bibfield  {author} {\bibinfo {author} {\bibfnamefont {P.}~\bibnamefont
  {Schall}}, \bibinfo {author} {\bibfnamefont {D.~A.}\ \bibnamefont {Weitz}}, \
  and\ \bibinfo {author} {\bibfnamefont {F.}~\bibnamefont {Spaepen}},\ }\href
  {\doibase 10.1126/science.1149308} {\bibfield  {journal} {\bibinfo  {journal}
  {Science}\ }\textbf {\bibinfo {volume} {318}},\ \bibinfo {pages} {1895}
  (\bibinfo {year} {2007})}\BibitemShut {NoStop}%
\bibitem [{\citenamefont {Jack}\ and\ \citenamefont
  {Berthier}(2016)}]{jackMeltingStableGlasses2016}%
  \BibitemOpen
  \bibfield  {author} {\bibinfo {author} {\bibfnamefont {R.~L.}\ \bibnamefont
  {Jack}}\ and\ \bibinfo {author} {\bibfnamefont {L.}~\bibnamefont
  {Berthier}},\ }\href {\doibase 10.1063/1.4954327} {\bibfield  {journal}
  {\bibinfo  {journal} {The Journal of Chemical Physics}\ }\textbf {\bibinfo
  {volume} {144}},\ \bibinfo {pages} {244506} (\bibinfo {year}
  {2016})}\BibitemShut {NoStop}%
\bibitem [{\citenamefont {Kearns}\ \emph {et~al.}(2010)\citenamefont {Kearns},
  \citenamefont {Ediger}, \citenamefont {Huth},\ and\ \citenamefont
  {Schick}}]{kearnsOneMicrometerLength2010}%
  \BibitemOpen
  \bibfield  {author} {\bibinfo {author} {\bibfnamefont {K.~L.}\ \bibnamefont
  {Kearns}}, \bibinfo {author} {\bibfnamefont {M.~D.}\ \bibnamefont {Ediger}},
  \bibinfo {author} {\bibfnamefont {H.}~\bibnamefont {Huth}}, \ and\ \bibinfo
  {author} {\bibfnamefont {C.}~\bibnamefont {Schick}},\ }\href {\doibase
  10.1021/jz9002179} {\bibfield  {journal} {\bibinfo  {journal} {J. Phys. Chem.
  Lett.}\ }\textbf {\bibinfo {volume} {1}},\ \bibinfo {pages} {388} (\bibinfo
  {year} {2010})}\BibitemShut {NoStop}%
\bibitem [{\citenamefont {Swallen}\ \emph {et~al.}(2009)\citenamefont
  {Swallen}, \citenamefont {Traynor}, \citenamefont {McMahon}, \citenamefont
  {Ediger},\ and\ \citenamefont
  {Mates}}]{swallenStableGlassTransformation2009}%
  \BibitemOpen
  \bibfield  {author} {\bibinfo {author} {\bibfnamefont {S.~F.}\ \bibnamefont
  {Swallen}}, \bibinfo {author} {\bibfnamefont {K.}~\bibnamefont {Traynor}},
  \bibinfo {author} {\bibfnamefont {R.~J.}\ \bibnamefont {McMahon}}, \bibinfo
  {author} {\bibfnamefont {M.~D.}\ \bibnamefont {Ediger}}, \ and\ \bibinfo
  {author} {\bibfnamefont {T.~E.}\ \bibnamefont {Mates}},\ }\href {\doibase
  10.1103/PhysRevLett.102.065503} {\bibfield  {journal} {\bibinfo  {journal}
  {Phys. Rev. Lett.}\ }\textbf {\bibinfo {volume} {102}},\ \bibinfo {pages}
  {065503} (\bibinfo {year} {2009})}\BibitemShut {NoStop}%
\bibitem [{\citenamefont {Herrero}\ \emph {et~al.}(2023)\citenamefont
  {Herrero}, \citenamefont {Scalliet}, \citenamefont {Ediger},\ and\
  \citenamefont {Berthier}}]{herreroTwostepDevitrificationUltrastable2023}%
  \BibitemOpen
  \bibfield  {author} {\bibinfo {author} {\bibfnamefont {C.}~\bibnamefont
  {Herrero}}, \bibinfo {author} {\bibfnamefont {C.}~\bibnamefont {Scalliet}},
  \bibinfo {author} {\bibfnamefont {M.~D.}\ \bibnamefont {Ediger}}, \ and\
  \bibinfo {author} {\bibfnamefont {L.}~\bibnamefont {Berthier}},\ }\href
  {\doibase 10.1073/pnas.2220824120} {\bibfield  {journal} {\bibinfo  {journal}
  {Proc. Natl. Acad. Sci. U.S.A.}\ }\textbf {\bibinfo {volume} {120}},\
  \bibinfo {pages} {e2220824120} (\bibinfo {year} {2023})}\BibitemShut
  {NoStop}%
\bibitem [{\citenamefont {Keys}\ \emph {et~al.}(2011)\citenamefont {Keys},
  \citenamefont {Hedges}, \citenamefont {Garrahan}, \citenamefont {Glotzer},\
  and\ \citenamefont {Chandler}}]{keysExcitationsAreLocalized2011}%
  \BibitemOpen
  \bibfield  {author} {\bibinfo {author} {\bibfnamefont {A.~S.}\ \bibnamefont
  {Keys}}, \bibinfo {author} {\bibfnamefont {L.~O.}\ \bibnamefont {Hedges}},
  \bibinfo {author} {\bibfnamefont {J.~P.}\ \bibnamefont {Garrahan}}, \bibinfo
  {author} {\bibfnamefont {S.~C.}\ \bibnamefont {Glotzer}}, \ and\ \bibinfo
  {author} {\bibfnamefont {D.}~\bibnamefont {Chandler}},\ }\href {\doibase
  10.1103/PhysRevX.1.021013} {\bibfield  {journal} {\bibinfo  {journal} {Phys.
  Rev. X}\ }\textbf {\bibinfo {volume} {1}},\ \bibinfo {pages} {021013}
  (\bibinfo {year} {2011})}\BibitemShut {NoStop}%
\bibitem [{\citenamefont
  {Ediger}(2000)}]{edigerSpatiallyHeterogeneousDynamics2000}%
  \BibitemOpen
  \bibfield  {author} {\bibinfo {author} {\bibfnamefont {M.~D.}\ \bibnamefont
  {Ediger}},\ }\href {\doibase 10.1146/annurev.physchem.51.1.99} {\bibfield
  {journal} {\bibinfo  {journal} {Annual Review of Physical Chemistry}\
  }\textbf {\bibinfo {volume} {51}},\ \bibinfo {pages} {99} (\bibinfo {year}
  {2000})}\BibitemShut {NoStop}%
\bibitem [{\citenamefont {Donth}\ \emph {et~al.}(2001)\citenamefont {Donth},
  \citenamefont {Huth},\ and\ \citenamefont
  {Beiner}}]{donthCharacteristicLengthGlass2001}%
  \BibitemOpen
  \bibfield  {author} {\bibinfo {author} {\bibfnamefont {E.}~\bibnamefont
  {Donth}}, \bibinfo {author} {\bibfnamefont {H.}~\bibnamefont {Huth}}, \ and\
  \bibinfo {author} {\bibfnamefont {M.}~\bibnamefont {Beiner}},\ }\href
  {\doibase 10.1088/0953-8984/13/22/102} {\bibfield  {journal} {\bibinfo
  {journal} {J. Phys.: Condens. Matter}\ }\textbf {\bibinfo {volume} {13}},\
  \bibinfo {pages} {L451} (\bibinfo {year} {2001})}\BibitemShut {NoStop}%
\bibitem [{\citenamefont {Edera}\ \emph
  {et~al.}(2024{\natexlab{a}})\citenamefont {Edera}, \citenamefont {Brizioli},
  \citenamefont {Madani}, \citenamefont {Ngouamba}, \citenamefont {Coussot},
  \citenamefont {Trappe}, \citenamefont {Petekidis}, \citenamefont {Giavazzi},\
  and\ \citenamefont {Cerbino}}]{ederaYieldingMicroscopeMultiscale2024}%
  \BibitemOpen
  \bibfield  {author} {\bibinfo {author} {\bibfnamefont {P.}~\bibnamefont
  {Edera}}, \bibinfo {author} {\bibfnamefont {M.}~\bibnamefont {Brizioli}},
  \bibinfo {author} {\bibfnamefont {M.}~\bibnamefont {Madani}}, \bibinfo
  {author} {\bibfnamefont {E.}~\bibnamefont {Ngouamba}}, \bibinfo {author}
  {\bibfnamefont {P.}~\bibnamefont {Coussot}}, \bibinfo {author} {\bibfnamefont
  {V.}~\bibnamefont {Trappe}}, \bibinfo {author} {\bibfnamefont
  {G.}~\bibnamefont {Petekidis}}, \bibinfo {author} {\bibfnamefont
  {F.}~\bibnamefont {Giavazzi}}, \ and\ \bibinfo {author} {\bibfnamefont
  {R.}~\bibnamefont {Cerbino}},\ }\href@noop {} {\enquote {\bibinfo {title}
  {Yielding under the microscope: A multi-scale perspective on brittle and
  ductile behaviors in oscillatory shear},}\ } (\bibinfo {year}
  {2024}{\natexlab{a}}),\ \Eprint {http://arxiv.org/abs/2402.00221}
  {arXiv:2402.00221 [cond-mat]} \BibitemShut {NoStop}%
\bibitem [{\citenamefont {Ozawa}\ \emph {et~al.}(2018)\citenamefont {Ozawa},
  \citenamefont {Berthier}, \citenamefont {Biroli}, \citenamefont {Rosso},\
  and\ \citenamefont {Tarjus}}]{ozawaRandomCriticalPoint2018}%
  \BibitemOpen
  \bibfield  {author} {\bibinfo {author} {\bibfnamefont {M.}~\bibnamefont
  {Ozawa}}, \bibinfo {author} {\bibfnamefont {L.}~\bibnamefont {Berthier}},
  \bibinfo {author} {\bibfnamefont {G.}~\bibnamefont {Biroli}}, \bibinfo
  {author} {\bibfnamefont {A.}~\bibnamefont {Rosso}}, \ and\ \bibinfo {author}
  {\bibfnamefont {G.}~\bibnamefont {Tarjus}},\ }\href {\doibase
  10.1073/pnas.1806156115} {\bibfield  {journal} {\bibinfo  {journal} {Proc.
  Natl. Acad. Sci. U.S.A.}\ }\textbf {\bibinfo {volume} {115}},\ \bibinfo
  {pages} {6656} (\bibinfo {year} {2018})}\BibitemShut {NoStop}%
\bibitem [{\citenamefont {Pollard}\ and\ \citenamefont
  {Fielding}(2022)}]{pollardYieldingShearBanding2022}%
  \BibitemOpen
  \bibfield  {author} {\bibinfo {author} {\bibfnamefont {J.}~\bibnamefont
  {Pollard}}\ and\ \bibinfo {author} {\bibfnamefont {S.~M.}\ \bibnamefont
  {Fielding}},\ }\href {\doibase 10.1103/PhysRevResearch.4.043037} {\bibfield
  {journal} {\bibinfo  {journal} {Phys. Rev. Research}\ }\textbf {\bibinfo
  {volume} {4}},\ \bibinfo {pages} {043037} (\bibinfo {year}
  {2022})}\BibitemShut {NoStop}%
\bibitem [{\citenamefont {Pommella}\ \emph {et~al.}(2020)\citenamefont
  {Pommella}, \citenamefont {Cipelletti},\ and\ \citenamefont
  {Ramos}}]{pommellaRoleNormalStress2020}%
  \BibitemOpen
  \bibfield  {author} {\bibinfo {author} {\bibfnamefont {A.}~\bibnamefont
  {Pommella}}, \bibinfo {author} {\bibfnamefont {L.}~\bibnamefont
  {Cipelletti}}, \ and\ \bibinfo {author} {\bibfnamefont {L.}~\bibnamefont
  {Ramos}},\ }\href {\doibase 10.1103/PhysRevLett.125.268006} {\bibfield
  {journal} {\bibinfo  {journal} {Phys. Rev. Lett.}\ }\textbf {\bibinfo
  {volume} {125}},\ \bibinfo {pages} {268006} (\bibinfo {year}
  {2020})}\BibitemShut {NoStop}%
\bibitem [{\citenamefont {Le~Bouil}\ \emph {et~al.}(2014)\citenamefont
  {Le~Bouil}, \citenamefont {Amon}, \citenamefont {McNamara},\ and\
  \citenamefont {Crassous}}]{lebouilEmergenceCooperativityPlasticity2014}%
  \BibitemOpen
  \bibfield  {author} {\bibinfo {author} {\bibfnamefont {A.}~\bibnamefont
  {Le~Bouil}}, \bibinfo {author} {\bibfnamefont {A.}~\bibnamefont {Amon}},
  \bibinfo {author} {\bibfnamefont {S.}~\bibnamefont {McNamara}}, \ and\
  \bibinfo {author} {\bibfnamefont {J.}~\bibnamefont {Crassous}},\ }\href
  {\doibase 10.1103/PhysRevLett.112.246001} {\bibfield  {journal} {\bibinfo
  {journal} {Phys. Rev. Lett.}\ }\textbf {\bibinfo {volume} {112}},\ \bibinfo
  {pages} {246001} (\bibinfo {year} {2014})}\BibitemShut {NoStop}%
\bibitem [{\citenamefont {Houdoux}\ \emph {et~al.}(2018)\citenamefont
  {Houdoux}, \citenamefont {Nguyen}, \citenamefont {Amon},\ and\ \citenamefont
  {Crassous}}]{houdouxPlasticFlowLocalization2018}%
  \BibitemOpen
  \bibfield  {author} {\bibinfo {author} {\bibfnamefont {D.}~\bibnamefont
  {Houdoux}}, \bibinfo {author} {\bibfnamefont {T.~B.}\ \bibnamefont {Nguyen}},
  \bibinfo {author} {\bibfnamefont {A.}~\bibnamefont {Amon}}, \ and\ \bibinfo
  {author} {\bibfnamefont {J.}~\bibnamefont {Crassous}},\ }\href {\doibase
  10.1103/PhysRevE.98.022905} {\bibfield  {journal} {\bibinfo  {journal} {Phys.
  Rev. E}\ }\textbf {\bibinfo {volume} {98}},\ \bibinfo {pages} {022905}
  (\bibinfo {year} {2018})}\BibitemShut {NoStop}%
\bibitem [{\citenamefont {Argon}\ and\ \citenamefont
  {Kuo}(1979)}]{argonPlasticFlowDisordered1979}%
  \BibitemOpen
  \bibfield  {author} {\bibinfo {author} {\bibfnamefont {A.~S.}\ \bibnamefont
  {Argon}}\ and\ \bibinfo {author} {\bibfnamefont {H.~Y.}\ \bibnamefont
  {Kuo}},\ }\href {\doibase 10.1016/0025-5416(79)90174-5} {\bibfield  {journal}
  {\bibinfo  {journal} {Mater. Sci. Eng.}\ }\textbf {\bibinfo {volume} {39}},\
  \bibinfo {pages} {101} (\bibinfo {year} {1979})}\BibitemShut {NoStop}%
\bibitem [{\citenamefont {Maloney}\ and\ \citenamefont
  {Lema{\^i}tre}(2004)}]{maloneyUniversalBreakdownElasticity2004}%
  \BibitemOpen
  \bibfield  {author} {\bibinfo {author} {\bibfnamefont {C.}~\bibnamefont
  {Maloney}}\ and\ \bibinfo {author} {\bibfnamefont {A.}~\bibnamefont
  {Lema{\^i}tre}},\ }\href {\doibase 10.1103/PhysRevLett.93.195501} {\bibfield
  {journal} {\bibinfo  {journal} {Phys. Rev. Lett.}\ }\textbf {\bibinfo
  {volume} {93}},\ \bibinfo {pages} {195501} (\bibinfo {year}
  {2004})}\BibitemShut {NoStop}%
\bibitem [{\citenamefont {Dasgupta}\ \emph {et~al.}(2012)\citenamefont
  {Dasgupta}, \citenamefont {Hentschel},\ and\ \citenamefont
  {Procaccia}}]{dasguptaMicroscopicMechanismShear2012}%
  \BibitemOpen
  \bibfield  {author} {\bibinfo {author} {\bibfnamefont {R.}~\bibnamefont
  {Dasgupta}}, \bibinfo {author} {\bibfnamefont {H.~G.~E.}\ \bibnamefont
  {Hentschel}}, \ and\ \bibinfo {author} {\bibfnamefont {I.}~\bibnamefont
  {Procaccia}},\ }\href {\doibase 10.1103/PhysRevLett.109.255502} {\bibfield
  {journal} {\bibinfo  {journal} {Phys. Rev. Lett.}\ }\textbf {\bibinfo
  {volume} {109}},\ \bibinfo {pages} {255502} (\bibinfo {year}
  {2012})}\BibitemShut {NoStop}%
\bibitem [{\citenamefont {Antonaglia}\ \emph {et~al.}(2014)\citenamefont
  {Antonaglia}, \citenamefont {Wright}, \citenamefont {Gu}, \citenamefont
  {Byer}, \citenamefont {Hufnagel}, \citenamefont {LeBlanc}, \citenamefont
  {Uhl},\ and\ \citenamefont {Dahmen}}]{antonaglia_bulk_2014}%
  \BibitemOpen
  \bibfield  {author} {\bibinfo {author} {\bibfnamefont {J.}~\bibnamefont
  {Antonaglia}}, \bibinfo {author} {\bibfnamefont {W.~J.}\ \bibnamefont
  {Wright}}, \bibinfo {author} {\bibfnamefont {X.}~\bibnamefont {Gu}}, \bibinfo
  {author} {\bibfnamefont {R.~R.}\ \bibnamefont {Byer}}, \bibinfo {author}
  {\bibfnamefont {T.~C.}\ \bibnamefont {Hufnagel}}, \bibinfo {author}
  {\bibfnamefont {M.}~\bibnamefont {LeBlanc}}, \bibinfo {author} {\bibfnamefont
  {J.~T.}\ \bibnamefont {Uhl}}, \ and\ \bibinfo {author} {\bibfnamefont
  {K.~A.}\ \bibnamefont {Dahmen}},\ }\href
  {http://link.aps.org/doi/10.1103/PhysRevLett.112.155501} {\bibfield
  {journal} {\bibinfo  {journal} {Phys. Rev. Lett.}\ }\textbf {\bibinfo
  {volume} {112}},\ \bibinfo {pages} {155501} (\bibinfo {year}
  {2014})}\BibitemShut {NoStop}%
\bibitem [{\citenamefont {Alegre}\ \emph {et~al.}(2014)\citenamefont {Alegre},
  \citenamefont {Cuesta},\ and\ \citenamefont
  {Lorenzo}}]{alegreExtensionMonkmanGrantModel2014}%
  \BibitemOpen
  \bibfield  {author} {\bibinfo {author} {\bibfnamefont {J.~M.}\ \bibnamefont
  {Alegre}}, \bibinfo {author} {\bibfnamefont {I.~I.}\ \bibnamefont {Cuesta}},
  \ and\ \bibinfo {author} {\bibfnamefont {M.}~\bibnamefont {Lorenzo}},\ }\href
  {\doibase 10.1007/s11340-014-9927-6} {\bibfield  {journal} {\bibinfo
  {journal} {Exp Mech}\ }\textbf {\bibinfo {volume} {54}},\ \bibinfo {pages}
  {1441} (\bibinfo {year} {2014})}\BibitemShut {NoStop}%
\bibitem [{\citenamefont {Koivisto}\ \emph {et~al.}(2016)\citenamefont
  {Koivisto}, \citenamefont {Ovaska}, \citenamefont {Miksic}, \citenamefont
  {Laurson},\ and\ \citenamefont
  {Alava}}]{koivistoPredictingSampleLifetimes2016}%
  \BibitemOpen
  \bibfield  {author} {\bibinfo {author} {\bibfnamefont {J.}~\bibnamefont
  {Koivisto}}, \bibinfo {author} {\bibfnamefont {M.}~\bibnamefont {Ovaska}},
  \bibinfo {author} {\bibfnamefont {A.}~\bibnamefont {Miksic}}, \bibinfo
  {author} {\bibfnamefont {L.}~\bibnamefont {Laurson}}, \ and\ \bibinfo
  {author} {\bibfnamefont {M.~J.}\ \bibnamefont {Alava}},\ }\href {\doibase
  10.1103/PhysRevE.94.023002} {\bibfield  {journal} {\bibinfo  {journal}
  {Physical Review E}\ }\textbf {\bibinfo {volume} {94}},\ \bibinfo {pages}
  {23002} (\bibinfo {year} {2016})}\BibitemShut {NoStop}%
\bibitem [{\citenamefont {Garcimart{\'i}n}\ \emph {et~al.}(1997)\citenamefont
  {Garcimart{\'i}n}, \citenamefont {Guarino}, \citenamefont {Bellon},\ and\
  \citenamefont {Ciliberto}}]{garcimartinStatisticalPropertiesFracture1997}%
  \BibitemOpen
  \bibfield  {author} {\bibinfo {author} {\bibfnamefont {A.}~\bibnamefont
  {Garcimart{\'i}n}}, \bibinfo {author} {\bibfnamefont {A.}~\bibnamefont
  {Guarino}}, \bibinfo {author} {\bibfnamefont {L.}~\bibnamefont {Bellon}}, \
  and\ \bibinfo {author} {\bibfnamefont {S.}~\bibnamefont {Ciliberto}},\ }\href
  {\doibase 10.1103/PhysRevLett.79.3202} {\bibfield  {journal} {\bibinfo
  {journal} {Phys. Rev. Lett.}\ }\textbf {\bibinfo {volume} {79}},\ \bibinfo
  {pages} {3202} (\bibinfo {year} {1997})}\BibitemShut {NoStop}%
\bibitem [{\citenamefont {Kandula}\ \emph {et~al.}(2019)\citenamefont
  {Kandula}, \citenamefont {Cordonnier}, \citenamefont {Boller}, \citenamefont
  {Weiss}, \citenamefont {Dysthe},\ and\ \citenamefont
  {Renard}}]{kandulaDynamicsMicroscalePrecursors2019}%
  \BibitemOpen
  \bibfield  {author} {\bibinfo {author} {\bibfnamefont {N.}~\bibnamefont
  {Kandula}}, \bibinfo {author} {\bibfnamefont {B.}~\bibnamefont {Cordonnier}},
  \bibinfo {author} {\bibfnamefont {E.}~\bibnamefont {Boller}}, \bibinfo
  {author} {\bibfnamefont {J.}~\bibnamefont {Weiss}}, \bibinfo {author}
  {\bibfnamefont {D.~K.}\ \bibnamefont {Dysthe}}, \ and\ \bibinfo {author}
  {\bibfnamefont {F.}~\bibnamefont {Renard}},\ }\href {\doibase
  10.1029/2019JB017381} {\bibfield  {journal} {\bibinfo  {journal} {J. Geophys.
  Res. Solid Earth}\ }\textbf {\bibinfo {volume} {124}},\ \bibinfo {pages}
  {6121} (\bibinfo {year} {2019})}\BibitemShut {NoStop}%
\bibitem [{\citenamefont {Renard}\ \emph {et~al.}(2017)\citenamefont {Renard},
  \citenamefont {Cordonnier}, \citenamefont {Kobchenko}, \citenamefont
  {Kandula}, \citenamefont {Weiss},\ and\ \citenamefont
  {Zhu}}]{renardMicroscaleCharacterizationRupture2017}%
  \BibitemOpen
  \bibfield  {author} {\bibinfo {author} {\bibfnamefont {F.}~\bibnamefont
  {Renard}}, \bibinfo {author} {\bibfnamefont {B.}~\bibnamefont {Cordonnier}},
  \bibinfo {author} {\bibfnamefont {M.}~\bibnamefont {Kobchenko}}, \bibinfo
  {author} {\bibfnamefont {N.}~\bibnamefont {Kandula}}, \bibinfo {author}
  {\bibfnamefont {J.}~\bibnamefont {Weiss}}, \ and\ \bibinfo {author}
  {\bibfnamefont {W.}~\bibnamefont {Zhu}},\ }\href {\doibase
  10.1016/j.epsl.2017.08.002} {\bibfield  {journal} {\bibinfo  {journal} {Earth
  Planet. Sci. Lett.}\ }\textbf {\bibinfo {volume} {476}},\ \bibinfo {pages}
  {69} (\bibinfo {year} {2017})}\BibitemShut {NoStop}%
\bibitem [{\citenamefont {Lockner}\ \emph {et~al.}(1991)\citenamefont
  {Lockner}, \citenamefont {Byerlee}, \citenamefont {Kuksenko}, \citenamefont
  {Ponomarev},\ and\ \citenamefont
  {Sidorin}}]{locknerQuasistaticFaultGrowth1991}%
  \BibitemOpen
  \bibfield  {author} {\bibinfo {author} {\bibfnamefont {D.~A.}\ \bibnamefont
  {Lockner}}, \bibinfo {author} {\bibfnamefont {J.~D.}\ \bibnamefont
  {Byerlee}}, \bibinfo {author} {\bibfnamefont {V.}~\bibnamefont {Kuksenko}},
  \bibinfo {author} {\bibfnamefont {A.}~\bibnamefont {Ponomarev}}, \ and\
  \bibinfo {author} {\bibfnamefont {A.}~\bibnamefont {Sidorin}},\ }\href
  {\doibase 10.1038/350039a0} {\bibfield  {journal} {\bibinfo  {journal}
  {Nature}\ }\textbf {\bibinfo {volume} {350}},\ \bibinfo {pages} {39}
  (\bibinfo {year} {1991})}\BibitemShut {NoStop}%
\bibitem [{\citenamefont {Lockner}(1993)}]{locknerRoleAcousticEmission1993}%
  \BibitemOpen
  \bibfield  {author} {\bibinfo {author} {\bibfnamefont {D.}~\bibnamefont
  {Lockner}},\ }\href {\doibase 10.1016/0148-9062(93)90041-B} {\bibfield
  {journal} {\bibinfo  {journal} {Int. J. Rock Mech. Min. Sci.}\ }\textbf
  {\bibinfo {volume} {30}},\ \bibinfo {pages} {883} (\bibinfo {year}
  {1993})}\BibitemShut {NoStop}%
\bibitem [{\citenamefont {Schuh}\ and\ \citenamefont
  {Lund}(2003)}]{schuhAtomisticBasisPlastic2003}%
  \BibitemOpen
  \bibfield  {author} {\bibinfo {author} {\bibfnamefont {C.~A.}\ \bibnamefont
  {Schuh}}\ and\ \bibinfo {author} {\bibfnamefont {A.~C.}\ \bibnamefont
  {Lund}},\ }\href {\doibase 10.1038/nmat918} {\bibfield  {journal} {\bibinfo
  {journal} {Nature Mater}\ }\textbf {\bibinfo {volume} {2}},\ \bibinfo {pages}
  {449} (\bibinfo {year} {2003})}\BibitemShut {NoStop}%
\bibitem [{\citenamefont {Z{\"o}ller}\ \emph {et~al.}(2001)\citenamefont
  {Z{\"o}ller}, \citenamefont {Hainzl},\ and\ \citenamefont
  {Kurths}}]{zollerObservationGrowingCorrelation2001}%
  \BibitemOpen
  \bibfield  {author} {\bibinfo {author} {\bibfnamefont {G.}~\bibnamefont
  {Z{\"o}ller}}, \bibinfo {author} {\bibfnamefont {S.}~\bibnamefont {Hainzl}},
  \ and\ \bibinfo {author} {\bibfnamefont {J.}~\bibnamefont {Kurths}},\ }\href
  {\doibase 10.1029/2000JB900379} {\bibfield  {journal} {\bibinfo  {journal}
  {J. Geophys. Res.}\ }\textbf {\bibinfo {volume} {106}},\ \bibinfo {pages}
  {2167} (\bibinfo {year} {2001})}\BibitemShut {NoStop}%
\bibitem [{\citenamefont {Potirakis}\ \emph {et~al.}(2013)\citenamefont
  {Potirakis}, \citenamefont {Karadimitrakis},\ and\ \citenamefont
  {Eftaxias}}]{potirakisNaturalTimeAnalysis2013}%
  \BibitemOpen
  \bibfield  {author} {\bibinfo {author} {\bibfnamefont {S.~M.}\ \bibnamefont
  {Potirakis}}, \bibinfo {author} {\bibfnamefont {A.}~\bibnamefont
  {Karadimitrakis}}, \ and\ \bibinfo {author} {\bibfnamefont {K.}~\bibnamefont
  {Eftaxias}},\ }\href {\doibase 10.1063/1.4807908} {\bibfield  {journal}
  {\bibinfo  {journal} {Chaos}\ }\textbf {\bibinfo {volume} {23}},\ \bibinfo
  {pages} {023117} (\bibinfo {year} {2013})}\BibitemShut {NoStop}%
\bibitem [{\citenamefont {Girard}\ \emph {et~al.}(2012)\citenamefont {Girard},
  \citenamefont {Weiss},\ and\ \citenamefont
  {Amitrano}}]{girardDamageClusterDistributionsSize2012}%
  \BibitemOpen
  \bibfield  {author} {\bibinfo {author} {\bibfnamefont {L.}~\bibnamefont
  {Girard}}, \bibinfo {author} {\bibfnamefont {J.}~\bibnamefont {Weiss}}, \
  and\ \bibinfo {author} {\bibfnamefont {D.}~\bibnamefont {Amitrano}},\ }\href
  {\doibase 10.1103/PhysRevLett.108.225502} {\bibfield  {journal} {\bibinfo
  {journal} {Phys. Rev. Lett.}\ }\textbf {\bibinfo {volume} {108}},\ \bibinfo
  {pages} {225502} (\bibinfo {year} {2012})}\BibitemShut {NoStop}%
\bibitem [{\citenamefont
  {Sornette}(2002)}]{sornettePredictabilityCatastrophicEvents2002}%
  \BibitemOpen
  \bibfield  {author} {\bibinfo {author} {\bibfnamefont {D.}~\bibnamefont
  {Sornette}},\ }\href {\doibase 10.1073/pnas.022581999} {\bibfield  {journal}
  {\bibinfo  {journal} {Proc. Natl. Acad. Sci. U.S.A.}\ }\textbf {\bibinfo
  {volume} {99}},\ \bibinfo {pages} {2522} (\bibinfo {year}
  {2002})}\BibitemShut {NoStop}%
\bibitem [{\citenamefont {Sornette}\ and\ \citenamefont
  {Andersen}(1998)}]{sornetteScalingRespectDisorder1998a}%
  \BibitemOpen
  \bibfield  {author} {\bibinfo {author} {\bibfnamefont {D.}~\bibnamefont
  {Sornette}}\ and\ \bibinfo {author} {\bibfnamefont {J.}~\bibnamefont
  {Andersen}},\ }\href {\doibase 10.1007/s100510050194} {\bibfield  {journal}
  {\bibinfo  {journal} {Eur. Phys. J. B}\ }\textbf {\bibinfo {volume} {1}},\
  \bibinfo {pages} {353} (\bibinfo {year} {1998})}\BibitemShut {NoStop}%
\bibitem [{\citenamefont {Zapperi}\ \emph {et~al.}(1997)\citenamefont
  {Zapperi}, \citenamefont {Vespignani},\ and\ \citenamefont
  {Stanley}}]{zapperiPlasticityAvalancheBehaviour1997}%
  \BibitemOpen
  \bibfield  {author} {\bibinfo {author} {\bibfnamefont {S.}~\bibnamefont
  {Zapperi}}, \bibinfo {author} {\bibfnamefont {A.}~\bibnamefont {Vespignani}},
  \ and\ \bibinfo {author} {\bibfnamefont {H.~E.}\ \bibnamefont {Stanley}},\
  }\href {\doibase 10.1038/41737} {\bibfield  {journal} {\bibinfo  {journal}
  {Nature}\ }\textbf {\bibinfo {volume} {388}},\ \bibinfo {pages} {658}
  (\bibinfo {year} {1997})}\BibitemShut {NoStop}%
\bibitem [{\citenamefont {Johnson}\ \emph {et~al.}(2013)\citenamefont
  {Johnson}, \citenamefont {Ferdowsi}, \citenamefont {Kaproth}, \citenamefont
  {Scuderi}, \citenamefont {Griffa}, \citenamefont {Carmeliet}, \citenamefont
  {Guyer}, \citenamefont {Le~Bas}, \citenamefont {Trugman},\ and\ \citenamefont
  {Marone}}]{johnsonAcousticEmissionMicroslip2013}%
  \BibitemOpen
  \bibfield  {author} {\bibinfo {author} {\bibfnamefont {P.~A.}\ \bibnamefont
  {Johnson}}, \bibinfo {author} {\bibfnamefont {B.}~\bibnamefont {Ferdowsi}},
  \bibinfo {author} {\bibfnamefont {B.~M.}\ \bibnamefont {Kaproth}}, \bibinfo
  {author} {\bibfnamefont {M.}~\bibnamefont {Scuderi}}, \bibinfo {author}
  {\bibfnamefont {M.}~\bibnamefont {Griffa}}, \bibinfo {author} {\bibfnamefont
  {J.}~\bibnamefont {Carmeliet}}, \bibinfo {author} {\bibfnamefont {R.~A.}\
  \bibnamefont {Guyer}}, \bibinfo {author} {\bibfnamefont {P.-Y.}\ \bibnamefont
  {Le~Bas}}, \bibinfo {author} {\bibfnamefont {D.~T.}\ \bibnamefont {Trugman}},
  \ and\ \bibinfo {author} {\bibfnamefont {C.}~\bibnamefont {Marone}},\ }\href
  {\doibase 10.1002/2013GL057848} {\bibfield  {journal} {\bibinfo  {journal}
  {Geophys. Res. Lett.}\ }\textbf {\bibinfo {volume} {40}},\ \bibinfo {pages}
  {5627} (\bibinfo {year} {2013})}\BibitemShut {NoStop}%
\bibitem [{\citenamefont {Houdoux}\ \emph {et~al.}(2021)\citenamefont
  {Houdoux}, \citenamefont {Amon}, \citenamefont {Marsan}, \citenamefont
  {Weiss},\ and\ \citenamefont
  {Crassous}}]{houdouxMicroslipsExperimentalGranular2021}%
  \BibitemOpen
  \bibfield  {author} {\bibinfo {author} {\bibfnamefont {D.}~\bibnamefont
  {Houdoux}}, \bibinfo {author} {\bibfnamefont {A.}~\bibnamefont {Amon}},
  \bibinfo {author} {\bibfnamefont {D.}~\bibnamefont {Marsan}}, \bibinfo
  {author} {\bibfnamefont {J.}~\bibnamefont {Weiss}}, \ and\ \bibinfo {author}
  {\bibfnamefont {J.}~\bibnamefont {Crassous}},\ }\href {\doibase
  10.1038/s43247-021-00147-1} {\bibfield  {journal} {\bibinfo  {journal}
  {Commun Earth Environ}\ }\textbf {\bibinfo {volume} {2}},\ \bibinfo {pages}
  {1} (\bibinfo {year} {2021})}\BibitemShut {NoStop}%
\bibitem [{\citenamefont {Ostapchuk}\ and\ \citenamefont
  {Morozova}(2020)}]{ostapchukMechanismLaboratoryEarthquake2020}%
  \BibitemOpen
  \bibfield  {author} {\bibinfo {author} {\bibfnamefont {A.~A.}\ \bibnamefont
  {Ostapchuk}}\ and\ \bibinfo {author} {\bibfnamefont {K.~G.}\ \bibnamefont
  {Morozova}},\ }\href {\doibase 10.1038/s41598-020-64272-1} {\bibfield
  {journal} {\bibinfo  {journal} {Sci Rep}\ }\textbf {\bibinfo {volume} {10}},\
  \bibinfo {pages} {7245} (\bibinfo {year} {2020})}\BibitemShut {NoStop}%
\bibitem [{\citenamefont {Gvirtzman}\ and\ \citenamefont
  {Fineberg}(2021)}]{gvirtzmanNucleationFrontsIgnite2021}%
  \BibitemOpen
  \bibfield  {author} {\bibinfo {author} {\bibfnamefont {S.}~\bibnamefont
  {Gvirtzman}}\ and\ \bibinfo {author} {\bibfnamefont {J.}~\bibnamefont
  {Fineberg}},\ }\href {\doibase 10.1038/s41567-021-01299-9} {\bibfield
  {journal} {\bibinfo  {journal} {Nat. Phys.}\ }\textbf {\bibinfo {volume}
  {17}},\ \bibinfo {pages} {1037} (\bibinfo {year} {2021})}\BibitemShut
  {NoStop}%
\bibitem [{\citenamefont {Kromer}\ \emph {et~al.}(2015)\citenamefont {Kromer},
  \citenamefont {Hutchinson}, \citenamefont {Lato}, \citenamefont {Gauthier},\
  and\ \citenamefont {Edwards}}]{kromerIdentifyingRockSlope2015}%
  \BibitemOpen
  \bibfield  {author} {\bibinfo {author} {\bibfnamefont {R.~A.}\ \bibnamefont
  {Kromer}}, \bibinfo {author} {\bibfnamefont {D.~J.}\ \bibnamefont
  {Hutchinson}}, \bibinfo {author} {\bibfnamefont {M.~J.}\ \bibnamefont
  {Lato}}, \bibinfo {author} {\bibfnamefont {D.}~\bibnamefont {Gauthier}}, \
  and\ \bibinfo {author} {\bibfnamefont {T.}~\bibnamefont {Edwards}},\ }\href
  {\doibase 10.1016/j.enggeo.2015.05.012} {\bibfield  {journal} {\bibinfo
  {journal} {Eng. Geol.}\ }\textbf {\bibinfo {volume} {195}},\ \bibinfo {pages}
  {93} (\bibinfo {year} {2015})}\BibitemShut {NoStop}%
\bibitem [{\citenamefont {Scuderi}\ \emph {et~al.}(2016)\citenamefont
  {Scuderi}, \citenamefont {Marone}, \citenamefont {Tinti}, \citenamefont
  {Di~Stefano},\ and\ \citenamefont
  {Collettini}}]{scuderiPrecursoryChangesSeismic2016}%
  \BibitemOpen
  \bibfield  {author} {\bibinfo {author} {\bibfnamefont {M.~M.}\ \bibnamefont
  {Scuderi}}, \bibinfo {author} {\bibfnamefont {C.}~\bibnamefont {Marone}},
  \bibinfo {author} {\bibfnamefont {E.}~\bibnamefont {Tinti}}, \bibinfo
  {author} {\bibfnamefont {G.}~\bibnamefont {Di~Stefano}}, \ and\ \bibinfo
  {author} {\bibfnamefont {C.}~\bibnamefont {Collettini}},\ }\href {\doibase
  10.1038/ngeo2775} {\bibfield  {journal} {\bibinfo  {journal} {Nature Geosci}\
  }\textbf {\bibinfo {volume} {9}},\ \bibinfo {pages} {695} (\bibinfo {year}
  {2016})}\BibitemShut {NoStop}%
\bibitem [{\citenamefont {Dixon}\ \emph {et~al.}(2014)\citenamefont {Dixon},
  \citenamefont {Jiang}, \citenamefont {Malservisi}, \citenamefont {McCaffrey},
  \citenamefont {Voss}, \citenamefont {Protti},\ and\ \citenamefont
  {Gonzalez}}]{dixonEarthquakeTsunamiForecasts2014}%
  \BibitemOpen
  \bibfield  {author} {\bibinfo {author} {\bibfnamefont {T.~H.}\ \bibnamefont
  {Dixon}}, \bibinfo {author} {\bibfnamefont {Y.}~\bibnamefont {Jiang}},
  \bibinfo {author} {\bibfnamefont {R.}~\bibnamefont {Malservisi}}, \bibinfo
  {author} {\bibfnamefont {R.}~\bibnamefont {McCaffrey}}, \bibinfo {author}
  {\bibfnamefont {N.}~\bibnamefont {Voss}}, \bibinfo {author} {\bibfnamefont
  {M.}~\bibnamefont {Protti}}, \ and\ \bibinfo {author} {\bibfnamefont
  {V.}~\bibnamefont {Gonzalez}},\ }\href {\doibase 10.1073/pnas.1412299111}
  {\bibfield  {journal} {\bibinfo  {journal} {Proceedings of the National
  Academy of Sciences}\ }\textbf {\bibinfo {volume} {111}},\ \bibinfo {pages}
  {17039} (\bibinfo {year} {2014})}\BibitemShut {NoStop}%
\bibitem [{\citenamefont {Vasseur}\ \emph {et~al.}(2015)\citenamefont
  {Vasseur}, \citenamefont {Wadsworth}, \citenamefont {Lavall{\'e}e},
  \citenamefont {Bell}, \citenamefont {Main},\ and\ \citenamefont
  {Dingwell}}]{vasseurHeterogeneityKeyFailure2015}%
  \BibitemOpen
  \bibfield  {author} {\bibinfo {author} {\bibfnamefont {J.}~\bibnamefont
  {Vasseur}}, \bibinfo {author} {\bibfnamefont {F.~B.}\ \bibnamefont
  {Wadsworth}}, \bibinfo {author} {\bibfnamefont {Y.}~\bibnamefont
  {Lavall{\'e}e}}, \bibinfo {author} {\bibfnamefont {A.~F.}\ \bibnamefont
  {Bell}}, \bibinfo {author} {\bibfnamefont {I.~G.}\ \bibnamefont {Main}}, \
  and\ \bibinfo {author} {\bibfnamefont {D.~B.}\ \bibnamefont {Dingwell}},\
  }\href {\doibase 10.1038/srep13259} {\bibfield  {journal} {\bibinfo
  {journal} {Sci Rep}\ }\textbf {\bibinfo {volume} {5}},\ \bibinfo {pages}
  {13259} (\bibinfo {year} {2015})}\BibitemShut {NoStop}%
\bibitem [{\citenamefont {K{\'a}d{\'a}r}\ \emph {et~al.}(2020)\citenamefont
  {K{\'a}d{\'a}r}, \citenamefont {P{\'a}l},\ and\ \citenamefont
  {Kun}}]{kadarRecordStatisticsBursts2020}%
  \BibitemOpen
  \bibfield  {author} {\bibinfo {author} {\bibfnamefont {V.}~\bibnamefont
  {K{\'a}d{\'a}r}}, \bibinfo {author} {\bibfnamefont {G.}~\bibnamefont
  {P{\'a}l}}, \ and\ \bibinfo {author} {\bibfnamefont {F.}~\bibnamefont
  {Kun}},\ }\href {\doibase 10.1038/s41598-020-59333-4} {\bibfield  {journal}
  {\bibinfo  {journal} {Sci. Rep.}\ }\textbf {\bibinfo {volume} {10}},\
  \bibinfo {pages} {2508} (\bibinfo {year} {2020})}\BibitemShut {NoStop}%
\bibitem [{\citenamefont {Yu}\ \emph {et~al.}(2021)\citenamefont {Yu},
  \citenamefont {Datye}, \citenamefont {Chen}, \citenamefont {Zhou},
  \citenamefont {Dagdeviren}, \citenamefont {Schroers},\ and\ \citenamefont
  {Schwarz}}]{yuAtomicscaleHomogeneousPlastic2021}%
  \BibitemOpen
  \bibfield  {author} {\bibinfo {author} {\bibfnamefont {J.}~\bibnamefont
  {Yu}}, \bibinfo {author} {\bibfnamefont {A.}~\bibnamefont {Datye}}, \bibinfo
  {author} {\bibfnamefont {Z.}~\bibnamefont {Chen}}, \bibinfo {author}
  {\bibfnamefont {C.}~\bibnamefont {Zhou}}, \bibinfo {author} {\bibfnamefont
  {O.~E.}\ \bibnamefont {Dagdeviren}}, \bibinfo {author} {\bibfnamefont
  {J.}~\bibnamefont {Schroers}}, \ and\ \bibinfo {author} {\bibfnamefont
  {U.~D.}\ \bibnamefont {Schwarz}},\ }\href {\doibase
  10.1038/s43246-021-00124-3} {\bibfield  {journal} {\bibinfo  {journal}
  {Commun Mater}\ }\textbf {\bibinfo {volume} {2}},\ \bibinfo {pages} {1}
  (\bibinfo {year} {2021})}\BibitemShut {NoStop}%
\bibitem [{\citenamefont {Sun}\ \emph {et~al.}(2010)\citenamefont {Sun},
  \citenamefont {Yu}, \citenamefont {Jiao}, \citenamefont {Bai}, \citenamefont
  {Zhao},\ and\ \citenamefont {Wang}}]{sunPlasticityDuctileMetallic2010}%
  \BibitemOpen
  \bibfield  {author} {\bibinfo {author} {\bibfnamefont {B.~A.}\ \bibnamefont
  {Sun}}, \bibinfo {author} {\bibfnamefont {H.~B.}\ \bibnamefont {Yu}},
  \bibinfo {author} {\bibfnamefont {W.}~\bibnamefont {Jiao}}, \bibinfo {author}
  {\bibfnamefont {H.~Y.}\ \bibnamefont {Bai}}, \bibinfo {author} {\bibfnamefont
  {D.~Q.}\ \bibnamefont {Zhao}}, \ and\ \bibinfo {author} {\bibfnamefont
  {W.~H.}\ \bibnamefont {Wang}},\ }\href {\doibase
  10.1103/PhysRevLett.105.035501} {\bibfield  {journal} {\bibinfo  {journal}
  {Phys. Rev. Lett.}\ }\textbf {\bibinfo {volume} {105}},\ \bibinfo {pages}
  {035501} (\bibinfo {year} {2010})}\BibitemShut {NoStop}%
\bibitem [{\citenamefont {Idrissi}\ \emph {et~al.}(2019)\citenamefont
  {Idrissi}, \citenamefont {Ghidelli}, \citenamefont {B{\'e}ch{\'e}},
  \citenamefont {Turner}, \citenamefont {Gravier}, \citenamefont {Blandin},
  \citenamefont {Raskin}, \citenamefont {Schryvers},\ and\ \citenamefont
  {Pardoen}}]{idrissiAtomicscaleViscoplasticityMechanisms2019}%
  \BibitemOpen
  \bibfield  {author} {\bibinfo {author} {\bibfnamefont {H.}~\bibnamefont
  {Idrissi}}, \bibinfo {author} {\bibfnamefont {M.}~\bibnamefont {Ghidelli}},
  \bibinfo {author} {\bibfnamefont {A.}~\bibnamefont {B{\'e}ch{\'e}}}, \bibinfo
  {author} {\bibfnamefont {S.}~\bibnamefont {Turner}}, \bibinfo {author}
  {\bibfnamefont {S.}~\bibnamefont {Gravier}}, \bibinfo {author} {\bibfnamefont
  {J.-J.}\ \bibnamefont {Blandin}}, \bibinfo {author} {\bibfnamefont {J.-P.}\
  \bibnamefont {Raskin}}, \bibinfo {author} {\bibfnamefont {D.}~\bibnamefont
  {Schryvers}}, \ and\ \bibinfo {author} {\bibfnamefont {T.}~\bibnamefont
  {Pardoen}},\ }\href {\doibase 10.1038/s41598-019-49910-7} {\bibfield
  {journal} {\bibinfo  {journal} {Sci Rep}\ }\textbf {\bibinfo {volume} {9}},\
  \bibinfo {pages} {13426} (\bibinfo {year} {2019})}\BibitemShut {NoStop}%
\bibitem [{\citenamefont {Richard}\ \emph {et~al.}(2020)\citenamefont
  {Richard}, \citenamefont {Ozawa}, \citenamefont {Patinet}, \citenamefont
  {Stanifer}, \citenamefont {Shang}, \citenamefont {Ridout}, \citenamefont
  {Xu}, \citenamefont {Zhang}, \citenamefont {Morse}, \citenamefont {Barrat},
  \citenamefont {Berthier}, \citenamefont {Falk}, \citenamefont {Guan},
  \citenamefont {Liu}, \citenamefont {Martens}, \citenamefont {Sastry},
  \citenamefont {Vandembroucq}, \citenamefont {Lerner},\ and\ \citenamefont
  {Manning}}]{richardPredictingPlasticityDisordered2020}%
  \BibitemOpen
  \bibfield  {author} {\bibinfo {author} {\bibfnamefont {D.}~\bibnamefont
  {Richard}}, \bibinfo {author} {\bibfnamefont {M.}~\bibnamefont {Ozawa}},
  \bibinfo {author} {\bibfnamefont {S.}~\bibnamefont {Patinet}}, \bibinfo
  {author} {\bibfnamefont {E.}~\bibnamefont {Stanifer}}, \bibinfo {author}
  {\bibfnamefont {B.}~\bibnamefont {Shang}}, \bibinfo {author} {\bibfnamefont
  {S.~A.}\ \bibnamefont {Ridout}}, \bibinfo {author} {\bibfnamefont
  {B.}~\bibnamefont {Xu}}, \bibinfo {author} {\bibfnamefont {G.}~\bibnamefont
  {Zhang}}, \bibinfo {author} {\bibfnamefont {P.~K.}\ \bibnamefont {Morse}},
  \bibinfo {author} {\bibfnamefont {J.-L.}\ \bibnamefont {Barrat}}, \bibinfo
  {author} {\bibfnamefont {L.}~\bibnamefont {Berthier}}, \bibinfo {author}
  {\bibfnamefont {M.~L.}\ \bibnamefont {Falk}}, \bibinfo {author}
  {\bibfnamefont {P.}~\bibnamefont {Guan}}, \bibinfo {author} {\bibfnamefont
  {A.~J.}\ \bibnamefont {Liu}}, \bibinfo {author} {\bibfnamefont
  {K.}~\bibnamefont {Martens}}, \bibinfo {author} {\bibfnamefont
  {S.}~\bibnamefont {Sastry}}, \bibinfo {author} {\bibfnamefont
  {D.}~\bibnamefont {Vandembroucq}}, \bibinfo {author} {\bibfnamefont
  {E.}~\bibnamefont {Lerner}}, \ and\ \bibinfo {author} {\bibfnamefont {M.~L.}\
  \bibnamefont {Manning}},\ }\href {\doibase 10.1103/PhysRevMaterials.4.113609}
  {\bibfield  {journal} {\bibinfo  {journal} {Phys. Rev. Materials}\ }\textbf
  {\bibinfo {volume} {4}},\ \bibinfo {pages} {113609} (\bibinfo {year}
  {2020})}\BibitemShut {NoStop}%
\bibitem [{\citenamefont {Rossi}\ and\ \citenamefont
  {Tarjus}(2022)}]{rossiEmergenceRandomField2022}%
  \BibitemOpen
  \bibfield  {author} {\bibinfo {author} {\bibfnamefont {S.}~\bibnamefont
  {Rossi}}\ and\ \bibinfo {author} {\bibfnamefont {G.}~\bibnamefont {Tarjus}},\
  }\href {\doibase 10.1088/1742-5468/ac8741} {\bibfield  {journal} {\bibinfo
  {journal} {J. Stat. Mech.}\ }\textbf {\bibinfo {volume} {2022}},\ \bibinfo
  {pages} {093301} (\bibinfo {year} {2022})}\BibitemShut {NoStop}%
\bibitem [{\citenamefont {Rodney}\ \emph {et~al.}(2011)\citenamefont {Rodney},
  \citenamefont {Tanguy},\ and\ \citenamefont
  {Vandembroucq}}]{rodneyModelingMechanicsAmorphous2011}%
  \BibitemOpen
  \bibfield  {author} {\bibinfo {author} {\bibfnamefont {D.}~\bibnamefont
  {Rodney}}, \bibinfo {author} {\bibfnamefont {A.}~\bibnamefont {Tanguy}}, \
  and\ \bibinfo {author} {\bibfnamefont {D.}~\bibnamefont {Vandembroucq}},\
  }\href {\doibase 10.1088/0965-0393/19/8/083001} {\bibfield  {journal}
  {\bibinfo  {journal} {Modelling Simul. Mater. Sci. Eng.}\ }\textbf {\bibinfo
  {volume} {19}},\ \bibinfo {pages} {083001} (\bibinfo {year}
  {2011})}\BibitemShut {NoStop}%
\bibitem [{\citenamefont {Di~Dio}\ \emph {et~al.}(2022)\citenamefont {Di~Dio},
  \citenamefont {Khabaz}, \citenamefont {Bonnecaze},\ and\ \citenamefont
  {Cloitre}}]{didioTransientDynamicsSoft2022}%
  \BibitemOpen
  \bibfield  {author} {\bibinfo {author} {\bibfnamefont {B.~F.}\ \bibnamefont
  {Di~Dio}}, \bibinfo {author} {\bibfnamefont {F.}~\bibnamefont {Khabaz}},
  \bibinfo {author} {\bibfnamefont {R.~T.}\ \bibnamefont {Bonnecaze}}, \ and\
  \bibinfo {author} {\bibfnamefont {M.}~\bibnamefont {Cloitre}},\ }\href
  {\doibase 10.1122/8.0000448} {\bibfield  {journal} {\bibinfo  {journal} {J.
  Rheol.}\ }\textbf {\bibinfo {volume} {66}},\ \bibinfo {pages} {717} (\bibinfo
  {year} {2022})}\BibitemShut {NoStop}%
\bibitem [{\citenamefont {Edera}\ \emph
  {et~al.}(2024{\natexlab{b}})\citenamefont {Edera}, \citenamefont {Bantawa},
  \citenamefont {Aime}, \citenamefont {Bonnecaze},\ and\ \citenamefont
  {Cloitre}}]{ederaTuningResidualStress2024}%
  \BibitemOpen
  \bibfield  {author} {\bibinfo {author} {\bibfnamefont {P.}~\bibnamefont
  {Edera}}, \bibinfo {author} {\bibfnamefont {M.}~\bibnamefont {Bantawa}},
  \bibinfo {author} {\bibfnamefont {S.}~\bibnamefont {Aime}}, \bibinfo {author}
  {\bibfnamefont {R.~T.}\ \bibnamefont {Bonnecaze}}, \ and\ \bibinfo {author}
  {\bibfnamefont {M.}~\bibnamefont {Cloitre}},\ }\href {\doibase
  10.48550/arXiv.2402.08293} {\enquote {\bibinfo {title} {Tuning residual
  stress, directional memory and aging in soft glassy materials},}\ } (\bibinfo
  {year} {2024}{\natexlab{b}}),\ \Eprint {http://arxiv.org/abs/2402.08293}
  {arXiv:2402.08293 [cond-mat]} \BibitemShut {NoStop}%
\bibitem [{\citenamefont {Vasoya}\ \emph {et~al.}(2016)\citenamefont {Vasoya},
  \citenamefont {Rycroft},\ and\ \citenamefont
  {Bouchbinder}}]{vasoyaNotchFractureToughness2016}%
  \BibitemOpen
  \bibfield  {author} {\bibinfo {author} {\bibfnamefont {M.}~\bibnamefont
  {Vasoya}}, \bibinfo {author} {\bibfnamefont {C.~H.}\ \bibnamefont {Rycroft}},
  \ and\ \bibinfo {author} {\bibfnamefont {E.}~\bibnamefont {Bouchbinder}},\
  }\href {\doibase 10.1103/PhysRevApplied.6.024008} {\bibfield  {journal}
  {\bibinfo  {journal} {Phys. Rev. Appl.}\ }\textbf {\bibinfo {volume} {6}},\
  \bibinfo {pages} {024008} (\bibinfo {year} {2016})}\BibitemShut {NoStop}%
\bibitem [{\citenamefont {Fan}\ \emph {et~al.}(2017)\citenamefont {Fan},
  \citenamefont {Wang}, \citenamefont {Zhang}, \citenamefont {Liu},
  \citenamefont {Schroers}, \citenamefont {Shattuck},\ and\ \citenamefont
  {O'Hern}}]{fanEffectsCoolingRate2017}%
  \BibitemOpen
  \bibfield  {author} {\bibinfo {author} {\bibfnamefont {M.}~\bibnamefont
  {Fan}}, \bibinfo {author} {\bibfnamefont {M.}~\bibnamefont {Wang}}, \bibinfo
  {author} {\bibfnamefont {K.}~\bibnamefont {Zhang}}, \bibinfo {author}
  {\bibfnamefont {Y.}~\bibnamefont {Liu}}, \bibinfo {author} {\bibfnamefont
  {J.}~\bibnamefont {Schroers}}, \bibinfo {author} {\bibfnamefont {M.~D.}\
  \bibnamefont {Shattuck}}, \ and\ \bibinfo {author} {\bibfnamefont {C.~S.}\
  \bibnamefont {O'Hern}},\ }\href {\doibase 10.1103/PhysRevE.95.022611}
  {\bibfield  {journal} {\bibinfo  {journal} {Phys. Rev. E}\ }\textbf {\bibinfo
  {volume} {95}},\ \bibinfo {pages} {022611} (\bibinfo {year}
  {2017})}\BibitemShut {NoStop}%
\bibitem [{\citenamefont {Kumar}\ \emph {et~al.}(2013)\citenamefont {Kumar},
  \citenamefont {Neibecker}, \citenamefont {Liu},\ and\ \citenamefont
  {Schroers}}]{kumarCriticalFictiveTemperature2013}%
  \BibitemOpen
  \bibfield  {author} {\bibinfo {author} {\bibfnamefont {G.}~\bibnamefont
  {Kumar}}, \bibinfo {author} {\bibfnamefont {P.}~\bibnamefont {Neibecker}},
  \bibinfo {author} {\bibfnamefont {Y.~H.}\ \bibnamefont {Liu}}, \ and\
  \bibinfo {author} {\bibfnamefont {J.}~\bibnamefont {Schroers}},\ }\href
  {\doibase 10.1038/ncomms2546} {\bibfield  {journal} {\bibinfo  {journal}
  {Nat. Commun.}\ }\textbf {\bibinfo {volume} {4}},\ \bibinfo {pages} {1536}
  (\bibinfo {year} {2013})}\BibitemShut {NoStop}%
\bibitem [{\citenamefont {{\c S}opu}\ \emph {et~al.}(2023)\citenamefont {{\c
  S}opu}, \citenamefont {Spieckermann}, \citenamefont {Bian}, \citenamefont
  {Fellner}, \citenamefont {Wright}, \citenamefont {Cordill}, \citenamefont
  {Gammer}, \citenamefont {Wang}, \citenamefont {Stoica},\ and\ \citenamefont
  {Eckert}}]{sopuRejuvenationEngineeringMetallic2023}%
  \BibitemOpen
  \bibfield  {author} {\bibinfo {author} {\bibfnamefont {D.}~\bibnamefont {{\c
  S}opu}}, \bibinfo {author} {\bibfnamefont {F.}~\bibnamefont {Spieckermann}},
  \bibinfo {author} {\bibfnamefont {X.}~\bibnamefont {Bian}}, \bibinfo {author}
  {\bibfnamefont {S.}~\bibnamefont {Fellner}}, \bibinfo {author} {\bibfnamefont
  {J.}~\bibnamefont {Wright}}, \bibinfo {author} {\bibfnamefont
  {M.}~\bibnamefont {Cordill}}, \bibinfo {author} {\bibfnamefont
  {C.}~\bibnamefont {Gammer}}, \bibinfo {author} {\bibfnamefont
  {G.}~\bibnamefont {Wang}}, \bibinfo {author} {\bibfnamefont {M.}~\bibnamefont
  {Stoica}}, \ and\ \bibinfo {author} {\bibfnamefont {J.}~\bibnamefont
  {Eckert}},\ }\href {\doibase 10.1038/s41427-023-00509-5} {\bibfield
  {journal} {\bibinfo  {journal} {NPG Asia Mater}\ }\textbf {\bibinfo {volume}
  {15}},\ \bibinfo {pages} {1} (\bibinfo {year} {2023})}\BibitemShut {NoStop}%
\bibitem [{\citenamefont {Bian}\ \emph {et~al.}(2020)\citenamefont {Bian},
  \citenamefont {{\c S}opu}, \citenamefont {Wang}, \citenamefont {Sun},
  \citenamefont {Bednar{\v c}ik}, \citenamefont {Gammer}, \citenamefont
  {Zhai},\ and\ \citenamefont {Eckert}}]{bianSignatureLocalStress2020}%
  \BibitemOpen
  \bibfield  {author} {\bibinfo {author} {\bibfnamefont {X.}~\bibnamefont
  {Bian}}, \bibinfo {author} {\bibfnamefont {D.}~\bibnamefont {{\c S}opu}},
  \bibinfo {author} {\bibfnamefont {G.}~\bibnamefont {Wang}}, \bibinfo {author}
  {\bibfnamefont {B.}~\bibnamefont {Sun}}, \bibinfo {author} {\bibfnamefont
  {J.}~\bibnamefont {Bednar{\v c}ik}}, \bibinfo {author} {\bibfnamefont
  {C.}~\bibnamefont {Gammer}}, \bibinfo {author} {\bibfnamefont
  {Q.}~\bibnamefont {Zhai}}, \ and\ \bibinfo {author} {\bibfnamefont
  {J.}~\bibnamefont {Eckert}},\ }\href {\doibase 10.1038/s41427-020-00241-4}
  {\bibfield  {journal} {\bibinfo  {journal} {NPG Asia Mater}\ }\textbf
  {\bibinfo {volume} {12}},\ \bibinfo {pages} {1} (\bibinfo {year}
  {2020})}\BibitemShut {NoStop}%
\bibitem [{\citenamefont {Patinet}\ \emph {et~al.}(2016)\citenamefont
  {Patinet}, \citenamefont {Vandembroucq},\ and\ \citenamefont
  {Falk}}]{patinetConnectingLocalYield2016}%
  \BibitemOpen
  \bibfield  {author} {\bibinfo {author} {\bibfnamefont {S.}~\bibnamefont
  {Patinet}}, \bibinfo {author} {\bibfnamefont {D.}~\bibnamefont
  {Vandembroucq}}, \ and\ \bibinfo {author} {\bibfnamefont {M.~L.}\
  \bibnamefont {Falk}},\ }\href {\doibase 10.1103/PhysRevLett.117.045501}
  {\bibfield  {journal} {\bibinfo  {journal} {Phys. Rev. Lett.}\ }\textbf
  {\bibinfo {volume} {117}},\ \bibinfo {pages} {045501} (\bibinfo {year}
  {2016})}\BibitemShut {NoStop}%
\bibitem [{\citenamefont {Rossi}\ \emph {et~al.}(2023)\citenamefont {Rossi},
  \citenamefont {Biroli}, \citenamefont {Ozawa},\ and\ \citenamefont
  {Tarjus}}]{rossiFarfromequilibriumCriticalityRandomfield2023}%
  \BibitemOpen
  \bibfield  {author} {\bibinfo {author} {\bibfnamefont {S.}~\bibnamefont
  {Rossi}}, \bibinfo {author} {\bibfnamefont {G.}~\bibnamefont {Biroli}},
  \bibinfo {author} {\bibfnamefont {M.}~\bibnamefont {Ozawa}}, \ and\ \bibinfo
  {author} {\bibfnamefont {G.}~\bibnamefont {Tarjus}},\ }\href {\doibase
  10.1103/PhysRevB.108.L220202} {\bibfield  {journal} {\bibinfo  {journal}
  {Phys. Rev. B}\ }\textbf {\bibinfo {volume} {108}},\ \bibinfo {pages}
  {L220202} (\bibinfo {year} {2023})}\BibitemShut {NoStop}%
\bibitem [{\citenamefont {Sethna}\ \emph {et~al.}(2001)\citenamefont {Sethna},
  \citenamefont {Dahmen},\ and\ \citenamefont
  {Myers}}]{sethnaCracklingNoise2001}%
  \BibitemOpen
  \bibfield  {author} {\bibinfo {author} {\bibfnamefont {J.~P.}\ \bibnamefont
  {Sethna}}, \bibinfo {author} {\bibfnamefont {K.~A.}\ \bibnamefont {Dahmen}},
  \ and\ \bibinfo {author} {\bibfnamefont {C.~R.}\ \bibnamefont {Myers}},\
  }\href {\doibase 10.1038/35065675} {\bibfield  {journal} {\bibinfo  {journal}
  {Nature}\ }\textbf {\bibinfo {volume} {410}},\ \bibinfo {pages} {242}
  (\bibinfo {year} {2001})}\BibitemShut {NoStop}%
\bibitem [{\citenamefont {Rossi}\ \emph {et~al.}(2022)\citenamefont {Rossi},
  \citenamefont {Biroli}, \citenamefont {Ozawa}, \citenamefont {Tarjus},\ and\
  \citenamefont {Zamponi}}]{rossiFiniteDisorderCriticalPoint2022a}%
  \BibitemOpen
  \bibfield  {author} {\bibinfo {author} {\bibfnamefont {S.}~\bibnamefont
  {Rossi}}, \bibinfo {author} {\bibfnamefont {G.}~\bibnamefont {Biroli}},
  \bibinfo {author} {\bibfnamefont {M.}~\bibnamefont {Ozawa}}, \bibinfo
  {author} {\bibfnamefont {G.}~\bibnamefont {Tarjus}}, \ and\ \bibinfo {author}
  {\bibfnamefont {F.}~\bibnamefont {Zamponi}},\ }\href {\doibase
  10.1103/PhysRevLett.129.228002} {\bibfield  {journal} {\bibinfo  {journal}
  {Phys. Rev. Lett.}\ }\textbf {\bibinfo {volume} {129}},\ \bibinfo {pages}
  {228002} (\bibinfo {year} {2022})}\BibitemShut {NoStop}%
\bibitem [{\citenamefont {Parmar}\ \emph {et~al.}(2019)\citenamefont {Parmar},
  \citenamefont {Kumar},\ and\ \citenamefont
  {Sastry}}]{parmarStrainLocalizationYielding2019}%
  \BibitemOpen
  \bibfield  {author} {\bibinfo {author} {\bibfnamefont {A.~D.~S.}\
  \bibnamefont {Parmar}}, \bibinfo {author} {\bibfnamefont {S.}~\bibnamefont
  {Kumar}}, \ and\ \bibinfo {author} {\bibfnamefont {S.}~\bibnamefont
  {Sastry}},\ }\href {\doibase 10.1103/PhysRevX.9.021018} {\bibfield  {journal}
  {\bibinfo  {journal} {Phys. Rev. X}\ }\textbf {\bibinfo {volume} {9}},\
  \bibinfo {pages} {021018} (\bibinfo {year} {2019})}\BibitemShut {NoStop}%
\bibitem [{\citenamefont {Ozawa}\ \emph {et~al.}(2022)\citenamefont {Ozawa},
  \citenamefont {Berthier}, \citenamefont {Biroli},\ and\ \citenamefont
  {Tarjus}}]{ozawaRareEventsDisorder2022}%
  \BibitemOpen
  \bibfield  {author} {\bibinfo {author} {\bibfnamefont {M.}~\bibnamefont
  {Ozawa}}, \bibinfo {author} {\bibfnamefont {L.}~\bibnamefont {Berthier}},
  \bibinfo {author} {\bibfnamefont {G.}~\bibnamefont {Biroli}}, \ and\ \bibinfo
  {author} {\bibfnamefont {G.}~\bibnamefont {Tarjus}},\ }\href {\doibase
  10.1103/PhysRevResearch.4.023227} {\bibfield  {journal} {\bibinfo  {journal}
  {Phys. Rev. Res.}\ }\textbf {\bibinfo {volume} {4}},\ \bibinfo {pages}
  {023227} (\bibinfo {year} {2022})}\BibitemShut {NoStop}%
\bibitem [{\citenamefont {Berne}\ and\ \citenamefont
  {Pecora}(1990)}]{berneDynamicLightScattering1990}%
  \BibitemOpen
  \bibfield  {author} {\bibinfo {author} {\bibfnamefont {B.~J.}\ \bibnamefont
  {Berne}}\ and\ \bibinfo {author} {\bibfnamefont {R.}~\bibnamefont {Pecora}},\
  }\href@noop {} {\emph {\bibinfo {title} {Dynamic {{Light}} Scattering with
  Application to Chemistry, Biology, and Physics.}}}\ (\bibinfo  {publisher}
  {{John Wiley and Sons, INC}},\ \bibinfo {year} {1990})\BibitemShut {NoStop}%
\bibitem [{\citenamefont {Philippe}\ \emph {et~al.}(2018)\citenamefont
  {Philippe}, \citenamefont {Truzzolillo}, \citenamefont {{Galvan-Myoshi}},
  \citenamefont {{Dieudonn{\'e}-George}}, \citenamefont {Trappe}, \citenamefont
  {Berthier},\ and\ \citenamefont
  {Cipelletti}}]{philippeGlassTransitionSoft2018}%
  \BibitemOpen
  \bibfield  {author} {\bibinfo {author} {\bibfnamefont {A.-M.}\ \bibnamefont
  {Philippe}}, \bibinfo {author} {\bibfnamefont {D.}~\bibnamefont
  {Truzzolillo}}, \bibinfo {author} {\bibfnamefont {J.}~\bibnamefont
  {{Galvan-Myoshi}}}, \bibinfo {author} {\bibfnamefont {P.}~\bibnamefont
  {{Dieudonn{\'e}-George}}}, \bibinfo {author} {\bibfnamefont {V.}~\bibnamefont
  {Trappe}}, \bibinfo {author} {\bibfnamefont {L.}~\bibnamefont {Berthier}}, \
  and\ \bibinfo {author} {\bibfnamefont {L.}~\bibnamefont {Cipelletti}},\
  }\href {\doibase 10.1103/PhysRevE.97.040601} {\bibfield  {journal} {\bibinfo
  {journal} {Phys. Rev. E}\ }\textbf {\bibinfo {volume} {97}},\ \bibinfo
  {pages} {040601} (\bibinfo {year} {2018})}\BibitemShut {NoStop}%
\bibitem [{\citenamefont {Wyss}\ \emph {et~al.}(2007)\citenamefont {Wyss},
  \citenamefont {Miyazaki}, \citenamefont {Mattsson}, \citenamefont {Hu},
  \citenamefont {Reichman},\ and\ \citenamefont
  {Weitz}}]{wyssStrainRateFrequencySuperposition2007}%
  \BibitemOpen
  \bibfield  {author} {\bibinfo {author} {\bibfnamefont {H.~M.}\ \bibnamefont
  {Wyss}}, \bibinfo {author} {\bibfnamefont {K.}~\bibnamefont {Miyazaki}},
  \bibinfo {author} {\bibfnamefont {J.}~\bibnamefont {Mattsson}}, \bibinfo
  {author} {\bibfnamefont {Z.}~\bibnamefont {Hu}}, \bibinfo {author}
  {\bibfnamefont {D.~R.}\ \bibnamefont {Reichman}}, \ and\ \bibinfo {author}
  {\bibfnamefont {D.~A.}\ \bibnamefont {Weitz}},\ }\href {\doibase
  10.1103/PhysRevLett.98.238303} {\bibfield  {journal} {\bibinfo  {journal}
  {Phys. Rev. Lett.}\ }\textbf {\bibinfo {volume} {98}},\ \bibinfo {pages}
  {238303} (\bibinfo {year} {2007})}\BibitemShut {NoStop}%
\bibitem [{\citenamefont {Hess}\ and\ \citenamefont
  {Aksel}(2011)}]{hessYieldingStructuralRelaxation2011a}%
  \BibitemOpen
  \bibfield  {author} {\bibinfo {author} {\bibfnamefont {A.}~\bibnamefont
  {Hess}}\ and\ \bibinfo {author} {\bibfnamefont {N.}~\bibnamefont {Aksel}},\
  }\href {\doibase 10.1103/PhysRevE.84.051502} {\bibfield  {journal} {\bibinfo
  {journal} {Phys. Rev. E}\ }\textbf {\bibinfo {volume} {84}},\ \bibinfo
  {pages} {051502} (\bibinfo {year} {2011})}\BibitemShut {NoStop}%
\bibitem [{\citenamefont {Hallett}\ \emph {et~al.}(2018)\citenamefont
  {Hallett}, \citenamefont {Turci},\ and\ \citenamefont
  {Royall}}]{hallettLocalStructureDeeply2018}%
  \BibitemOpen
  \bibfield  {author} {\bibinfo {author} {\bibfnamefont {J.~E.}\ \bibnamefont
  {Hallett}}, \bibinfo {author} {\bibfnamefont {F.}~\bibnamefont {Turci}}, \
  and\ \bibinfo {author} {\bibfnamefont {C.~P.}\ \bibnamefont {Royall}},\
  }\href {\doibase 10.1038/s41467-018-05371-6} {\bibfield  {journal} {\bibinfo
  {journal} {Nat Commun}\ }\textbf {\bibinfo {volume} {9}},\ \bibinfo {pages}
  {3272} (\bibinfo {year} {2018})}\BibitemShut {NoStop}%
\bibitem [{\citenamefont {Gokhale}\ \emph {et~al.}(2014)\citenamefont
  {Gokhale}, \citenamefont {Hima~Nagamanasa}, \citenamefont {Ganapathy},\ and\
  \citenamefont {Sood}}]{gokhaleGrowingDynamicalFacilitation2014}%
  \BibitemOpen
  \bibfield  {author} {\bibinfo {author} {\bibfnamefont {S.}~\bibnamefont
  {Gokhale}}, \bibinfo {author} {\bibfnamefont {K.}~\bibnamefont
  {Hima~Nagamanasa}}, \bibinfo {author} {\bibfnamefont {R.}~\bibnamefont
  {Ganapathy}}, \ and\ \bibinfo {author} {\bibfnamefont {A.~K.}\ \bibnamefont
  {Sood}},\ }\href {\doibase 10.1038/ncomms5685} {\bibfield  {journal}
  {\bibinfo  {journal} {Nat Commun}\ }\textbf {\bibinfo {volume} {5}},\
  \bibinfo {pages} {4685} (\bibinfo {year} {2014})}\BibitemShut {NoStop}%
\bibitem [{\citenamefont {Mishra}\ \emph {et~al.}(2014)\citenamefont {Mishra},
  \citenamefont {Hima~Nagamanasa}, \citenamefont {Ganapathy}, \citenamefont
  {Sood},\ and\ \citenamefont
  {Gokhale}}]{mishraDynamicalFacilitationGoverns2014}%
  \BibitemOpen
  \bibfield  {author} {\bibinfo {author} {\bibfnamefont {C.~K.}\ \bibnamefont
  {Mishra}}, \bibinfo {author} {\bibfnamefont {K.}~\bibnamefont
  {Hima~Nagamanasa}}, \bibinfo {author} {\bibfnamefont {R.}~\bibnamefont
  {Ganapathy}}, \bibinfo {author} {\bibfnamefont {A.~K.}\ \bibnamefont {Sood}},
  \ and\ \bibinfo {author} {\bibfnamefont {S.}~\bibnamefont {Gokhale}},\ }\href
  {\doibase 10.1073/pnas.1413384111} {\bibfield  {journal} {\bibinfo  {journal}
  {Proc. Natl. Acad. Sci. U.S.A.}\ }\textbf {\bibinfo {volume} {111}},\
  \bibinfo {pages} {15362} (\bibinfo {year} {2014})}\BibitemShut {NoStop}%
\bibitem [{\citenamefont {Chandler}\ and\ \citenamefont
  {Garrahan}(2010)}]{chandlerDynamicsWayForming2010}%
  \BibitemOpen
  \bibfield  {author} {\bibinfo {author} {\bibfnamefont {D.}~\bibnamefont
  {Chandler}}\ and\ \bibinfo {author} {\bibfnamefont {J.~P.}\ \bibnamefont
  {Garrahan}},\ }\href {\doibase 10.1146/annurev.physchem.040808.090405}
  {\bibfield  {journal} {\bibinfo  {journal} {Annual Review of Physical
  Chemistry}\ }\textbf {\bibinfo {volume} {61}},\ \bibinfo {pages} {191}
  (\bibinfo {year} {2010})}\BibitemShut {NoStop}%
\bibitem [{\citenamefont {Speck}(2019)}]{speckDynamicFacilitationTheory2019}%
  \BibitemOpen
  \bibfield  {author} {\bibinfo {author} {\bibfnamefont {T.}~\bibnamefont
  {Speck}},\ }\href {\doibase 10.1088/1742-5468/ab2ace} {\bibfield  {journal}
  {\bibinfo  {journal} {J. Stat. Mech.}\ }\textbf {\bibinfo {volume} {2019}},\
  \bibinfo {pages} {084015} (\bibinfo {year} {2019})}\BibitemShut {NoStop}%
\bibitem [{\citenamefont {Landau}\ \emph {et~al.}(2011)\citenamefont {Landau},
  \citenamefont {Lif{\v s}ic}, \citenamefont {Pitaevskij},\ and\ \citenamefont
  {Landau}}]{landauStatisticalPhysicsLifshitz2011}%
  \BibitemOpen
  \bibfield  {author} {\bibinfo {author} {\bibfnamefont {L.~D.}\ \bibnamefont
  {Landau}}, \bibinfo {author} {\bibfnamefont {E.~M.}\ \bibnamefont {Lif{\v
  s}ic}}, \bibinfo {author} {\bibfnamefont {L.~P.}\ \bibnamefont {Pitaevskij}},
  \ and\ \bibinfo {author} {\bibfnamefont {L.~D.}\ \bibnamefont {Landau}},\
  }\href@noop {} {\emph {\bibinfo {title} {Statistical Physics. 1: By {{E}}.
  {{M}}. {{Lifshitz}} and {{L}}. {{P}}. {{Pitaevskii}}}}},\ \bibinfo {edition}
  {3rd}\ ed.,\ \bibinfo {series} {Course of Theoretical Physics / {{L}}. {{D}}.
  {{Landau}} and {{E}}. {{M}}. {{Lifshitz}}}\ No.~\bibinfo {number} {5}\
  (\bibinfo  {publisher} {Elsevier Butterworth Heinemann},\ \bibinfo {address}
  {Amsterdam Heidelberg},\ \bibinfo {year} {2011})\BibitemShut {NoStop}%
\bibitem [{See()}]{SeeSupplementalMaterial}%
  \BibitemOpen
  \href@noop {} {\bibinfo  {journal} {See {{Supplemental Material}} at [{{URL}}
  Will Be Inserted by Publisher] for {{I}}) {{Model}} with arbitrary range of
  facilitated advection coupling; {{II}}) {{Role}} of disorder: General
  considerations using the central limit theorem; {{III}}) {{Convergence}}
  kinetics; {{IV}}) {{Statistics}} at fixed disorder ({{Figure}} 4 Main Text);
  {{V}}) {{Spatial}} correlations for discontinuous yielding ({$\xi$} {$\sim$}
  lattice spacing)}\ }\BibitemShut {NoStop}%
\bibitem [{\citenamefont {Tong}\ \emph {et~al.}(2020)\citenamefont {Tong},
  \citenamefont {Sengupta},\ and\ \citenamefont
  {Tanaka}}]{tongEmergentSolidityAmorphous2020}%
  \BibitemOpen
\bibfield  {journal} {  }\bibfield  {author} {\bibinfo {author} {\bibfnamefont
  {H.}~\bibnamefont {Tong}}, \bibinfo {author} {\bibfnamefont {S.}~\bibnamefont
  {Sengupta}}, \ and\ \bibinfo {author} {\bibfnamefont {H.}~\bibnamefont
  {Tanaka}},\ }\href {\doibase 10.1038/s41467-020-18663-7} {\bibfield
  {journal} {\bibinfo  {journal} {Nat Commun}\ }\textbf {\bibinfo {volume}
  {11}},\ \bibinfo {pages} {4863} (\bibinfo {year} {2020})}\BibitemShut
  {NoStop}%
\bibitem [{\citenamefont {Colombo}\ \emph {et~al.}(2013)\citenamefont
  {Colombo}, \citenamefont {{Widmer-Cooper}},\ and\ \citenamefont
  {Del~Gado}}]{colomboMicroscopicPictureCooperative2013}%
  \BibitemOpen
  \bibfield  {author} {\bibinfo {author} {\bibfnamefont {J.}~\bibnamefont
  {Colombo}}, \bibinfo {author} {\bibfnamefont {A.}~\bibnamefont
  {{Widmer-Cooper}}}, \ and\ \bibinfo {author} {\bibfnamefont {E.}~\bibnamefont
  {Del~Gado}},\ }\href {\doibase 10.1103/PhysRevLett.110.198301} {\bibfield
  {journal} {\bibinfo  {journal} {Phys. Rev. Lett.}\ }\textbf {\bibinfo
  {volume} {110}},\ \bibinfo {pages} {198301} (\bibinfo {year}
  {2013})}\BibitemShut {NoStop}%
\bibitem [{\citenamefont {Ortlieb}\ \emph {et~al.}(2023)\citenamefont
  {Ortlieb}, \citenamefont {Ingebrigtsen}, \citenamefont {Hallett},
  \citenamefont {Turci},\ and\ \citenamefont
  {Royall}}]{ortliebProbingExcitationsCooperatively2023}%
  \BibitemOpen
  \bibfield  {author} {\bibinfo {author} {\bibfnamefont {L.}~\bibnamefont
  {Ortlieb}}, \bibinfo {author} {\bibfnamefont {T.~S.}\ \bibnamefont
  {Ingebrigtsen}}, \bibinfo {author} {\bibfnamefont {J.~E.}\ \bibnamefont
  {Hallett}}, \bibinfo {author} {\bibfnamefont {F.}~\bibnamefont {Turci}}, \
  and\ \bibinfo {author} {\bibfnamefont {C.~P.}\ \bibnamefont {Royall}},\
  }\href {\doibase 10.1038/s41467-023-37793-2} {\bibfield  {journal} {\bibinfo
  {journal} {Nat Commun}\ }\textbf {\bibinfo {volume} {14}},\ \bibinfo {pages}
  {2621} (\bibinfo {year} {2023})}\BibitemShut {NoStop}%
\bibitem [{\citenamefont {Cubuk}\ \emph {et~al.}(2017)\citenamefont {Cubuk},
  \citenamefont {Ivancic}, \citenamefont {Schoenholz}, \citenamefont
  {Strickland}, \citenamefont {Basu}, \citenamefont {Davidson}, \citenamefont
  {Fontaine}, \citenamefont {Hor}, \citenamefont {Huang}, \citenamefont
  {Jiang}, \citenamefont {Keim}, \citenamefont {Koshigan}, \citenamefont
  {Lefever}, \citenamefont {Liu}, \citenamefont {Ma}, \citenamefont
  {Magagnosc}, \citenamefont {Morrow}, \citenamefont {Ortiz}, \citenamefont
  {Rieser}, \citenamefont {Shavit}, \citenamefont {Still}, \citenamefont {Xu},
  \citenamefont {Zhang}, \citenamefont {Nordstrom}, \citenamefont {Arratia},
  \citenamefont {Carpick}, \citenamefont {Durian}, \citenamefont {Fakhraai},
  \citenamefont {Jerolmack}, \citenamefont {Lee}, \citenamefont {Li},
  \citenamefont {Riggleman}, \citenamefont {Turner}, \citenamefont {Yodh},
  \citenamefont {Gianola},\ and\ \citenamefont
  {Liu}}]{cubukStructurepropertyRelationshipsUniversal2017}%
  \BibitemOpen
  \bibfield  {author} {\bibinfo {author} {\bibfnamefont {E.~D.}\ \bibnamefont
  {Cubuk}}, \bibinfo {author} {\bibfnamefont {R.~J.~S.}\ \bibnamefont
  {Ivancic}}, \bibinfo {author} {\bibfnamefont {S.~S.}\ \bibnamefont
  {Schoenholz}}, \bibinfo {author} {\bibfnamefont {D.~J.}\ \bibnamefont
  {Strickland}}, \bibinfo {author} {\bibfnamefont {A.}~\bibnamefont {Basu}},
  \bibinfo {author} {\bibfnamefont {Z.~S.}\ \bibnamefont {Davidson}}, \bibinfo
  {author} {\bibfnamefont {J.}~\bibnamefont {Fontaine}}, \bibinfo {author}
  {\bibfnamefont {J.~L.}\ \bibnamefont {Hor}}, \bibinfo {author} {\bibfnamefont
  {Y.-R.}\ \bibnamefont {Huang}}, \bibinfo {author} {\bibfnamefont
  {Y.}~\bibnamefont {Jiang}}, \bibinfo {author} {\bibfnamefont {N.~C.}\
  \bibnamefont {Keim}}, \bibinfo {author} {\bibfnamefont {K.~D.}\ \bibnamefont
  {Koshigan}}, \bibinfo {author} {\bibfnamefont {J.~A.}\ \bibnamefont
  {Lefever}}, \bibinfo {author} {\bibfnamefont {T.}~\bibnamefont {Liu}},
  \bibinfo {author} {\bibfnamefont {X.-G.}\ \bibnamefont {Ma}}, \bibinfo
  {author} {\bibfnamefont {D.~J.}\ \bibnamefont {Magagnosc}}, \bibinfo {author}
  {\bibfnamefont {E.}~\bibnamefont {Morrow}}, \bibinfo {author} {\bibfnamefont
  {C.~P.}\ \bibnamefont {Ortiz}}, \bibinfo {author} {\bibfnamefont {J.~M.}\
  \bibnamefont {Rieser}}, \bibinfo {author} {\bibfnamefont {A.}~\bibnamefont
  {Shavit}}, \bibinfo {author} {\bibfnamefont {T.}~\bibnamefont {Still}},
  \bibinfo {author} {\bibfnamefont {Y.}~\bibnamefont {Xu}}, \bibinfo {author}
  {\bibfnamefont {Y.}~\bibnamefont {Zhang}}, \bibinfo {author} {\bibfnamefont
  {K.~N.}\ \bibnamefont {Nordstrom}}, \bibinfo {author} {\bibfnamefont {P.~E.}\
  \bibnamefont {Arratia}}, \bibinfo {author} {\bibfnamefont {R.~W.}\
  \bibnamefont {Carpick}}, \bibinfo {author} {\bibfnamefont {D.~J.}\
  \bibnamefont {Durian}}, \bibinfo {author} {\bibfnamefont {Z.}~\bibnamefont
  {Fakhraai}}, \bibinfo {author} {\bibfnamefont {D.~J.}\ \bibnamefont
  {Jerolmack}}, \bibinfo {author} {\bibfnamefont {D.}~\bibnamefont {Lee}},
  \bibinfo {author} {\bibfnamefont {J.}~\bibnamefont {Li}}, \bibinfo {author}
  {\bibfnamefont {R.}~\bibnamefont {Riggleman}}, \bibinfo {author}
  {\bibfnamefont {K.~T.}\ \bibnamefont {Turner}}, \bibinfo {author}
  {\bibfnamefont {A.~G.}\ \bibnamefont {Yodh}}, \bibinfo {author}
  {\bibfnamefont {D.~S.}\ \bibnamefont {Gianola}}, \ and\ \bibinfo {author}
  {\bibfnamefont {A.~J.}\ \bibnamefont {Liu}},\ }\href {\doibase
  10.1126/science.aai8830} {\bibfield  {journal} {\bibinfo  {journal}
  {Science}\ }\textbf {\bibinfo {volume} {358}},\ \bibinfo {pages} {1033}
  (\bibinfo {year} {2017})}\BibitemShut {NoStop}%
\bibitem [{\citenamefont {Tong}\ and\ \citenamefont
  {Tanaka}(2019)}]{tongStructuralOrderGenuine2019}%
  \BibitemOpen
  \bibfield  {author} {\bibinfo {author} {\bibfnamefont {H.}~\bibnamefont
  {Tong}}\ and\ \bibinfo {author} {\bibfnamefont {H.}~\bibnamefont {Tanaka}},\
  }\href {\doibase 10.1038/s41467-019-13606-3} {\bibfield  {journal} {\bibinfo
  {journal} {Nat Commun}\ }\textbf {\bibinfo {volume} {10}},\ \bibinfo {pages}
  {5596} (\bibinfo {year} {2019})}\BibitemShut {NoStop}%
\bibitem [{\citenamefont {Fleermann}\ and\ \citenamefont
  {Kirsch}(2022)}]{fleermannCENTRALLIMITTHEOREM}%
  \BibitemOpen
  \bibfield  {author} {\bibinfo {author} {\bibfnamefont {M.}~\bibnamefont
  {Fleermann}}\ and\ \bibinfo {author} {\bibfnamefont {W.}~\bibnamefont
  {Kirsch}},\ }\href@noop {} {\bibfield  {journal} {\bibinfo  {journal}
  {arXiv.2202.04717}\ ,\ \bibinfo {pages} {17}} (\bibinfo {year}
  {2022})}\BibitemShut {NoStop}%
\bibitem [{\citenamefont {Petrov}\ and\ \citenamefont
  {Fur{\'o}}(2011)}]{petrovStudyFreezingMelting2011}%
  \BibitemOpen
  \bibfield  {author} {\bibinfo {author} {\bibfnamefont {O.}~\bibnamefont
  {Petrov}}\ and\ \bibinfo {author} {\bibfnamefont {I.}~\bibnamefont
  {Fur{\'o}}},\ }\href {\doibase 10.1039/c1cp21902b} {\bibfield  {journal}
  {\bibinfo  {journal} {Phys. Chem. Chem. Phys.}\ }\textbf {\bibinfo {volume}
  {13}},\ \bibinfo {pages} {16358} (\bibinfo {year} {2011})}\BibitemShut
  {NoStop}%
\bibitem [{\citenamefont {Chikazumi}\ \emph {et~al.}(1997)\citenamefont
  {Chikazumi}, \citenamefont {Graham},\ and\ \citenamefont
  {Chikazumi}}]{chikazumiPhysicsFerromagnetism1997}%
  \BibitemOpen
  \bibfield  {author} {\bibinfo {author} {\bibfnamefont {S.}~\bibnamefont
  {Chikazumi}}, \bibinfo {author} {\bibfnamefont {C.~D.}\ \bibnamefont
  {Graham}}, \ and\ \bibinfo {author} {\bibfnamefont {S.}~\bibnamefont
  {Chikazumi}},\ }\href@noop {} {\emph {\bibinfo {title} {Physics of
  Ferromagnetism}}},\ \bibinfo {edition} {2nd}\ ed.,\ \bibinfo {series} {The
  International Series of Monographs on Physics}\ No.~\bibinfo {number} {94}\
  (\bibinfo  {publisher} {Clarendon Press ; Oxford University Press},\ \bibinfo
  {address} {Oxford : New York},\ \bibinfo {year} {1997})\BibitemShut {NoStop}%
\bibitem [{\citenamefont {Dang}\ \emph {et~al.}(2016)\citenamefont {Dang},
  \citenamefont {Denisov}, \citenamefont {Struth}, \citenamefont {Zaccone},\
  and\ \citenamefont {Schall}}]{dangReversibilityHysteresisSharp2016}%
  \BibitemOpen
  \bibfield  {author} {\bibinfo {author} {\bibfnamefont {M.~T.}\ \bibnamefont
  {Dang}}, \bibinfo {author} {\bibfnamefont {D.}~\bibnamefont {Denisov}},
  \bibinfo {author} {\bibfnamefont {B.}~\bibnamefont {Struth}}, \bibinfo
  {author} {\bibfnamefont {A.}~\bibnamefont {Zaccone}}, \ and\ \bibinfo
  {author} {\bibfnamefont {P.}~\bibnamefont {Schall}},\ }\href {\doibase
  10.1140/epje/i2016-16044-3} {\bibfield  {journal} {\bibinfo  {journal} {Eur.
  Phys. J. E}\ }\textbf {\bibinfo {volume} {39}},\ \bibinfo {pages} {44}
  (\bibinfo {year} {2016})}\BibitemShut {NoStop}%
\bibitem [{\citenamefont {{W.
  Weibull}}(1939)}]{weibullStatisticalTheoryStrength1939}%
  \BibitemOpen
  \bibfield  {author} {\bibinfo {author} {\bibnamefont {{W. Weibull}}},\
  }\href@noop {} {\bibfield  {journal} {\bibinfo  {journal} {IVB-Handl.}\
  }\textbf {\bibinfo {volume} {151}} (\bibinfo {year} {1939})}\BibitemShut
  {NoStop}%
\bibitem [{\citenamefont
  {Coleman}(1958)}]{colemanStatisticsTimeDependence1958}%
  \BibitemOpen
  \bibfield  {author} {\bibinfo {author} {\bibfnamefont {B.~D.}\ \bibnamefont
  {Coleman}},\ }\href {\doibase 10.1063/1.1723343} {\bibfield  {journal}
  {\bibinfo  {journal} {J. Appl. Phys.}\ }\textbf {\bibinfo {volume} {29}},\
  \bibinfo {pages} {968} (\bibinfo {year} {1958})}\BibitemShut {NoStop}%
\bibitem [{\citenamefont {Aime}\ \emph {et~al.}(2017)\citenamefont {Aime},
  \citenamefont {Eisenmenger},\ and\ \citenamefont
  {Engels}}]{aimeModelFailureThermoplastic2017}%
  \BibitemOpen
  \bibfield  {author} {\bibinfo {author} {\bibfnamefont {S.}~\bibnamefont
  {Aime}}, \bibinfo {author} {\bibfnamefont {N.~D.}\ \bibnamefont
  {Eisenmenger}}, \ and\ \bibinfo {author} {\bibfnamefont {T.~A.~P.}\
  \bibnamefont {Engels}},\ }\href {\doibase 10.1122/1.5000808} {\bibfield
  {journal} {\bibinfo  {journal} {J. Rheol.}\ }\textbf {\bibinfo {volume}
  {61}},\ \bibinfo {pages} {1329} (\bibinfo {year} {2017})}\BibitemShut
  {NoStop}%
\bibitem [{\citenamefont {Okni{\'n}ski}(1992)}]{okninskiCatastropheTheory1992}%
  \BibitemOpen
  \bibfield  {author} {\bibinfo {author} {\bibfnamefont {A.}~\bibnamefont
  {Okni{\'n}ski}},\ }\href@noop {} {\emph {\bibinfo {title} {Catastrophe
  Theory}}},\ \bibinfo {series} {Comprehensive Chemical Kinetics}\ No.\
  \bibinfo {number} {v. 33}\ (\bibinfo  {publisher} {Elsevier ; PWN--Polish
  Scientific Publishers},\ \bibinfo {address} {Amsterdam ; New York :
  Warszawa},\ \bibinfo {year} {1992})\BibitemShut {NoStop}%
\bibitem [{\citenamefont
  {Demazure}(2000)}]{demazureBifurcationsCatastrophes2000}%
  \BibitemOpen
  \bibfield  {author} {\bibinfo {author} {\bibfnamefont {M.}~\bibnamefont
  {Demazure}},\ }\href@noop {} {\emph {\bibinfo {title} {Bifurcations and
  {{Catastrophes}}}}}\ (\bibinfo  {publisher} {Springer},\ \bibinfo {address}
  {Berlin},\ \bibinfo {year} {2000})\BibitemShut {NoStop}%
\bibitem [{\citenamefont {{Widmer-Cooper}}\ \emph {et~al.}(2008)\citenamefont
  {{Widmer-Cooper}}, \citenamefont {Perry}, \citenamefont {Harrowell},\ and\
  \citenamefont
  {Reichman}}]{widmer-cooperIrreversibleReorganizationSupercooled2008}%
  \BibitemOpen
  \bibfield  {author} {\bibinfo {author} {\bibfnamefont {A.}~\bibnamefont
  {{Widmer-Cooper}}}, \bibinfo {author} {\bibfnamefont {H.}~\bibnamefont
  {Perry}}, \bibinfo {author} {\bibfnamefont {P.}~\bibnamefont {Harrowell}}, \
  and\ \bibinfo {author} {\bibfnamefont {D.~R.}\ \bibnamefont {Reichman}},\
  }\href {\doibase 10.1038/nphys1025} {\bibfield  {journal} {\bibinfo
  {journal} {Nature Phys}\ }\textbf {\bibinfo {volume} {4}},\ \bibinfo {pages}
  {711} (\bibinfo {year} {2008})}\BibitemShut {NoStop}%
\bibitem [{\citenamefont {Bellafard}\ \emph {et~al.}(2015)\citenamefont
  {Bellafard}, \citenamefont {Chakravarty}, \citenamefont {Troyer},\ and\
  \citenamefont {Katzgraber}}]{bellafardEffectQuenchedBond2015a}%
  \BibitemOpen
  \bibfield  {author} {\bibinfo {author} {\bibfnamefont {A.}~\bibnamefont
  {Bellafard}}, \bibinfo {author} {\bibfnamefont {S.}~\bibnamefont
  {Chakravarty}}, \bibinfo {author} {\bibfnamefont {M.}~\bibnamefont {Troyer}},
  \ and\ \bibinfo {author} {\bibfnamefont {H.~G.}\ \bibnamefont {Katzgraber}},\
  }\href {\doibase 10.1016/j.aop.2015.03.026} {\bibfield  {journal} {\bibinfo
  {journal} {Annals of Physics}\ }\textbf {\bibinfo {volume} {357}},\ \bibinfo
  {pages} {66} (\bibinfo {year} {2015})}\BibitemShut {NoStop}%
\end{thebibliography}
%
\end{document}



\title[]{Supplemental Material: Emergent scales and spatial correlations at the yielding transition of glassy materials}
\author{Stefano Aime}
\email{stefano.aime@espci.fr}
 \affiliation{Laboratoire Matière Molle et Chimie, (UMR 7167, ESPCI-CNRS) ESPCI ParisTech, 10 rue Vauquelin, 75005 Paris, France}
\author{Domenico Truzzolillo}
\email{domenico.truzzolillo@umontpellier.fr}
\affiliation{Laboratoire Charles Coulomb (L2C), UMR 5221, CNRS-Universit\'{e} de Montpellier, France}

\date{\today}
\maketitle
\tableofcontents
\newpage
\section{Model with arbitrary range of facilitated advection coupling}
Equation 4 of the main text takes a more general form when the dynamics of each site is coupled via bonds comprised within the $\beta-1$ shells centered on the site $i$. More in detail: The rate of each site $i$ is coupled to the all combinations of rates characterizing the nodes included in the shortest path going from the site $i$ to the site $j_{\beta-1}$, the latter being distant $(\beta-1)$ lattice spacings from the former. Therefore the dynamics of each cell is coupled to $N_z(N_z-1)^{\beta-2}$ neighboring $j$-cells via positive coupling constants $\alpha_{ij_1...j_{\beta-1}}$, where $2\leq\beta\leq \infty$ dictates the range of the interaction.  The most general form of this model then reads:\\ 
\begin{equation}\label{latticerange}
    \Gamma_i-\Gamma_0=\left[\frac{\omega K}{\frac{1}{\gamma_0^n}+\omega^{\beta}\frac{1}{N_z(N_z-1)^{\beta-2}}\sum\limits_{j_1=1}^{N_z}\sum\limits_{j_2=1}^{N_z-1}... ...\sum\limits_{j_{\beta-1}=1}^{N_z-1}\frac{\alpha_{ij_1...j_{\beta-1}}}{\Gamma_i\prod\limits_{k=1}^{\beta-1}\Gamma_{j_k}}}
    \right]
\end{equation}
where $\Gamma_{j_k}=1/\tau_{j_k}$ is the relaxation rate characterizing the site $j_k$. The term containing the sums running over the ${j_k}$ indices represents the general facilitated advection term, modeling the average coupling between adjacent mesoscopic regions of soft glasses relaxing at different rates.

\begin{figure}[htbp]
  \includegraphics[width=14cm]{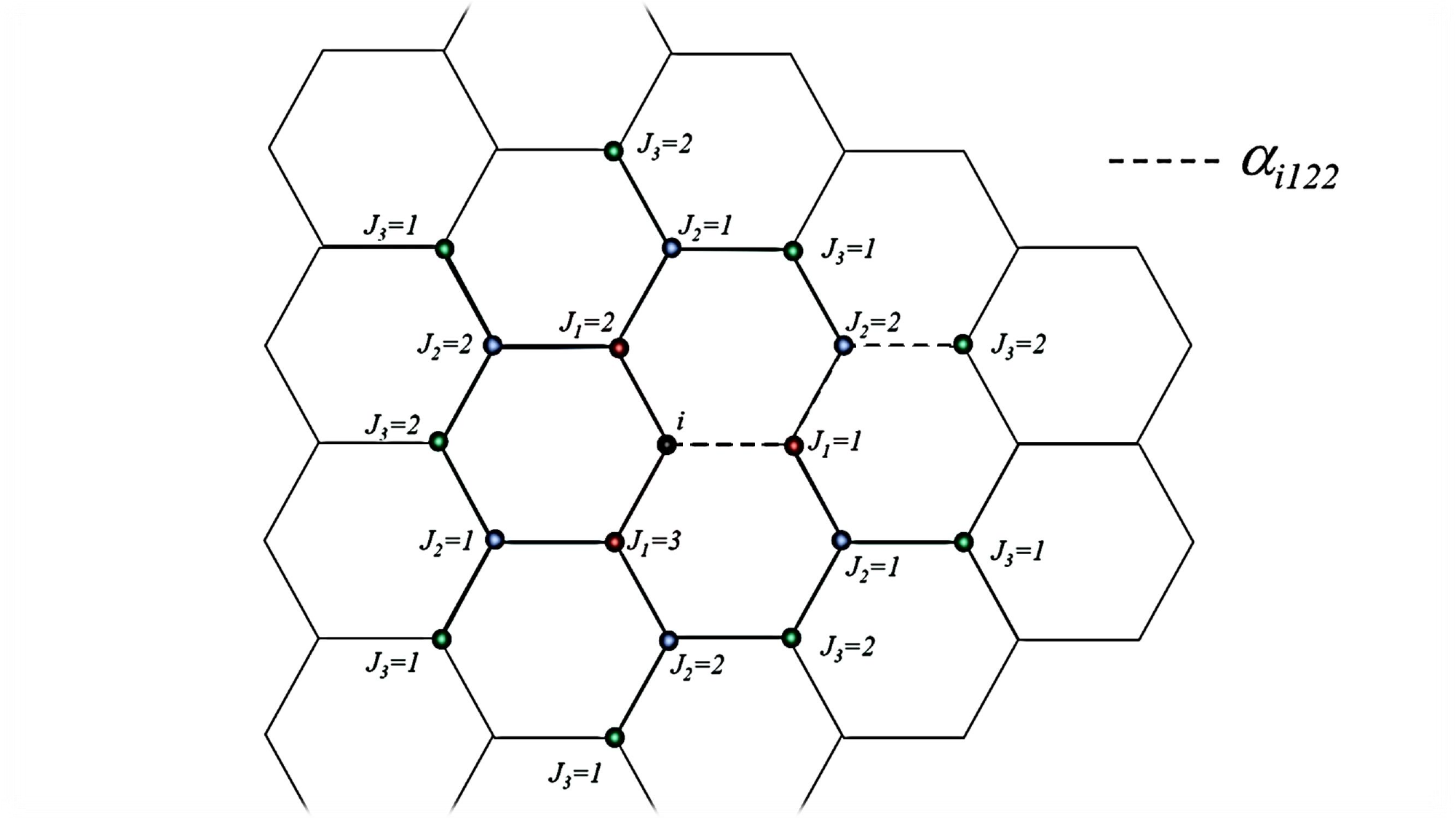}
\caption{Sketch of the interaction between sites on a hexagonal lattice for $\beta=4$. The site $i$ is coupled via the coupling constants $\alpha_{ij_1j_2j_3}$ with all sites comprised in the first three shells centered on the site $i$.}\label{lattice_sketch}
\end{figure}

The mean field approximation of equation \ref{latticerange} is Equation 7 of the main text and it is obtained by imposing $\Gamma_{i}=\Gamma_{j_k}=\Gamma$ and $\alpha_{ij_1...j_{\beta-1}}=\alpha$ for each site of index $i$.

\begin{figure}[htbp]
  \includegraphics[width=14cm]{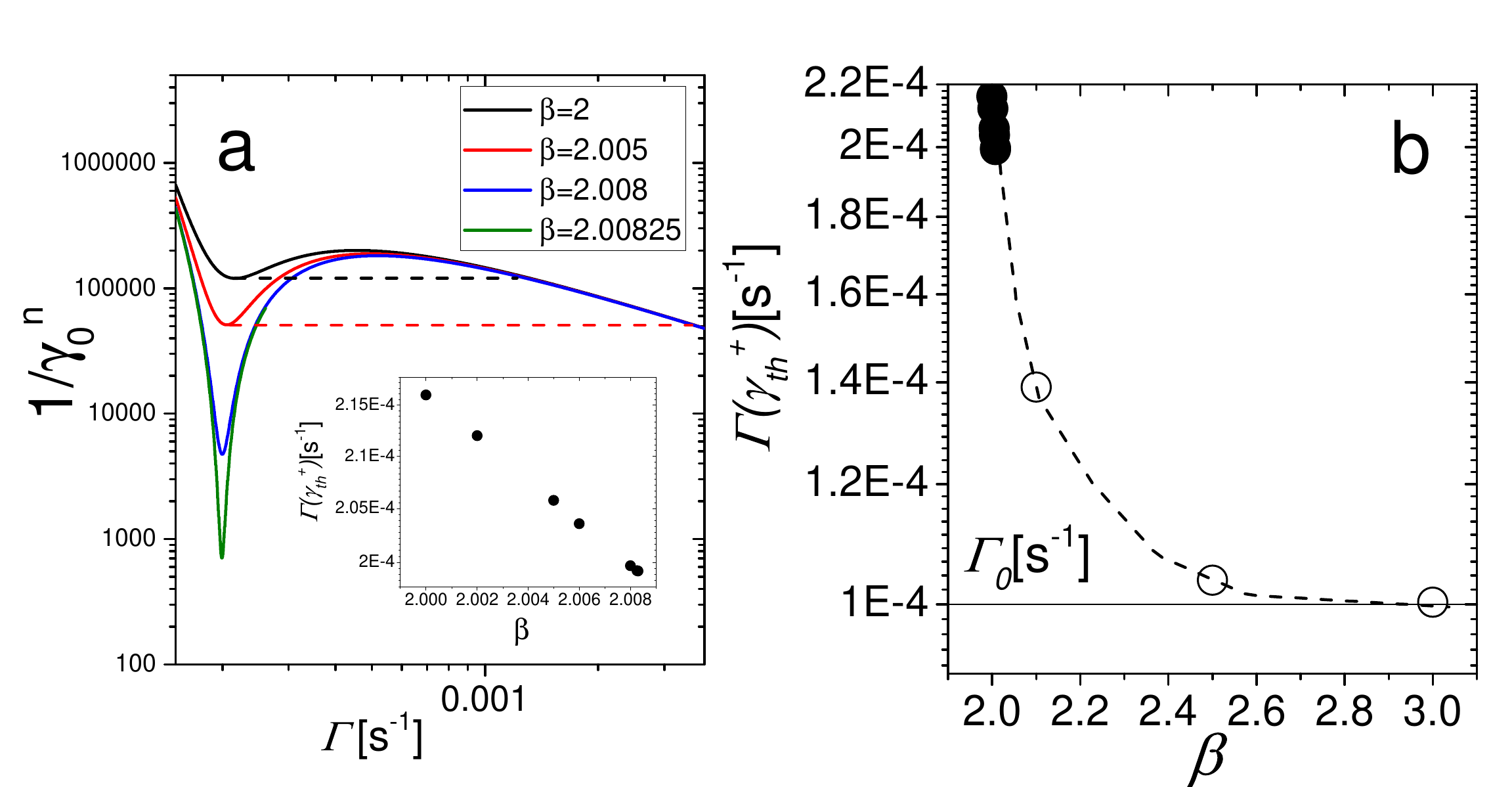}
\caption{Effect of the spatial range of facilitated advection (FA) on the mean field equation of state. a) Four selected equations of state with $\alpha=0.17$, $K=300$, $\omega=0.66$ Hz and different values of $\beta$ as indicated in the figure. The values of the relaxation rate at the local minimum of the EOS $\Gamma(\gamma_{th}^+)$ are shown in the inset. Here $\beta$ is such we obtain only positive $\gamma_{th}^+$. Horizontal dashed lines show two representative jump of relaxation rate at the yielding transition.  
b) $\Gamma(\gamma_{th}^+)$ in function of $\beta$. Full points are are the same shown in the inset of panel a) Empty points are obtained for larger values of $\beta$ for which $(\gamma_{th}^+)^{-n}$ is negative. The full horizontal line marks the limiting case $\Gamma(\gamma_th^+)=\Gamma_0$}\label{beta-effect}
\end{figure}

\section{Role of disorder: general considerations using the central limit theorem}
As for classical first order transition the question we ask is whether the macroscopic phases survive in the presence of disorder or the system forms domains (droplets) that follow the local value of the random-interaction. To this aim let's first make some general considerations on the effect of the disorder on the equation of state and consider a virtual single domain (of linear size L) of one phase (say fluid) embedded in the
other phase (say solid). Let's take equation \ref{latticerange} for $\beta=2$ and average over $N_D$ sites forming a fluid droplet:
\begin{equation}\label{droplet1}
   \frac{1}{N_D}\sum\limits_{i=1}^{N_D}(\Gamma_i-\Gamma_0)=\frac{1}{N_D}\sum\limits_{i=1}^{N_D}\left[\frac{\omega K}{\left(\frac{1}{\gamma_0^n}+\frac{\omega^{2}}{N_z}\sum\limits_{j=1}^{N_z}\frac{\alpha_{ij}}{\Gamma_{i}\Gamma_{j}}\right)}
    \right]=\frac{1}{N_D}\sum\limits_{i=1}^{N_D}Y_i
\end{equation}
If $\alpha_{ij}$ is extracted from a distribution defined on a bounded support [0,$\alpha_M$], then $Y=\{Y_i\}$ has also a pdf $P(Y)$ defined on a bounded support $\frac{\omega K}{\left(\gamma_0^{-n}+\left(\frac{\omega}{\Gamma_0}\right)^{2}N_z\alpha_M\right)}$ $\leq Y_i\leq$  $K\omega\gamma_0^n$.
Hence, the variance of $P(Y)$ ($\sigma_Y^2$) exists and is finite if $P(Y)$ calculated at the two bounds is finite. For $N_D\gg 1$ we can apply the central limit theorem for weakly dependent random variables \cite{fleermannCENTRALLIMITTHEOREM} to the expression (\ref{droplet1}) and write
\begin{equation}\label{droplet2}
   \Gamma_D-\Gamma_0=\frac{1}{N_D}\left[N_DE[Y]\pm\sigma_Y\sqrt{N_D}\right]=\\E[Y]\pm\frac{\sigma_Y}{\sqrt{N_D}}\sim E[Y]\pm\frac{\sigma_Y}{L^{d/2}}
\end{equation}
where $\Gamma_D$ is the average relaxation rate in the droplet,
$E[Y]$ is the expectation value of $Y$ and $d$ is the dimension of the space. The expectation value $E[Y]$ is equal to the arithmetic mean in the limit $N_D\rightarrow\infty$.\\
Equation \ref{droplet2} can be viewed as a local MF equation for the droplet where now the effective spontaneous dynamics is characterized by a relaxation rate $\left(\Gamma_0\pm \frac{\sigma_Y}{L^{d/2}}\right)$ depending on the disorder.
It is important to point out that the expectation value $E[Y]$ is not given by the mean field value obtained by fixing all coupling constants to one value $\alpha_{ij}=\alpha$, since it depends on the specific form of the probability density $P(Y)$ which in turn depends on the chosen pdf of the coupling constants $p(\alpha_{ij})$. Equation \ref{droplet2} shows that that for $\sigma_Y=0$ the value of the average relaxation rate of the droplet does not depend on its size: the dynamics coincides with that predicted by the MF approximation. On the contrary, when $\sigma_{Y}>0$ the system dynamics may split into fluid and solid droplets depending on whether locally $Y$ undergoes positive or negative fluctuations.\\

Equation \ref{droplet2}, however, is not enough to inspect the effect of disorder and the occurrence of fluid nucleation and heterogeneous dynamics. For this purpose we re-write equation \ref{droplet1} with the random coupling constants $\alpha_{ij}=\langle\alpha\rangle+\varepsilon_{ij}$, where $\varepsilon_{ij}$ are random deviations with zero mean that here for simplicity we assume gaussian: $\varepsilon_{ij}=\mathcal{N}(0,\sigma_{\alpha})$.
\begin{equation}\label{droplet3}
\Gamma_D-\Gamma_0=\frac{1}{N_D}\sum\limits_{i=1}^{N_D}\left[\frac{\omega K(\varphi)}{\left(\frac{1}{\gamma_0^n}+\frac{\omega^{2}}{N_z}\sum\limits_{j=1}^{N_z}\frac{\langle\alpha\rangle+\varepsilon_{ij}}{\Gamma_{i}\Gamma_{j}}\right)}\right]
\end{equation}
Since the system will split into fluid-like and solid-like droplets we further simplify the problem and we seek for a solution of equation \ref{droplet3} by adopting the approximation of homogeneous droplets, namely $\langle\Gamma_i\Gamma_j\rangle=\langle\Gamma\rangle^2=\Gamma_D^2$. By doing so and expanding for small $\epsilon_{ij}$, we obtain
\begin{equation}\label{droplet4}
\Gamma_D-\Gamma_0\simeq\frac{K\omega}{\gamma_0^{-n}+\frac{\omega^2\langle\alpha\rangle}{\Gamma_D^2}}\left[1-\frac{1}{N_D}\sum\limits_{i=1}^{N_D}\frac{\frac{\omega^2}{\Gamma_D^2N_z}\sum\limits_{j=1}^{N_z}\varepsilon_{ij}}{\gamma_0^{-n}+\frac{\omega^2\langle\alpha\rangle}{\Gamma_D^2}}\right]\end{equation}.
We can now apply the central limit theorem, knowing that the mean of the deviations over the nearest neighbor sites is still normally distributed with zero mean and standard deviation $\sigma_{\alpha}/\sqrt{N_z}$. Equation \ref{droplet4} becomes
\begin{equation}\label{droplet5}
\Gamma_D-\Gamma_0\simeq\frac{K\omega}{\gamma_0^{-n}+\frac{\omega^2\langle\alpha\rangle}{\Gamma_D^2}}\left[1\pm\frac{\frac{\omega^2\sigma_{\alpha}}{\sqrt{N_z N_D}}}{\Gamma_D^2\gamma_0^{-n}+\omega^2\langle\alpha\rangle}\right]\end{equation}.
From Equation \ref{droplet5} therefore we reach the following conclusions.\\ 
i) By increasing the number of nearest neighbors $N_z$ or the lattice sites $N_D$ belonging to a droplet the effect of disorder weakens. The mean field equation becomes thus exact in presence of disorder in infinite dimensions.\\ 
ii) In the presence of disorder ($\sigma_{\alpha}>0$) the equation of state differs from the meanfield one and the system splits in solid and fluid regions, corresponding respectively to positive and negative local fluctuations of the coupling constants $\alpha_{ij}$ over droplets of size $N_D$. 
Fluid nucleation occurs for strain amplitudes $\gamma_0$ lower than $\gamma_{th}^{+}$, correspondent to the local minimum of the equation of state obtained for $\sigma_{\alpha}=0$. Similarly, for $\gamma_0>\gamma_{th}^{+}$ we will still have solid regions, since part of the system will follow equation \ref{droplet5} with the negative sign in its r.h.s. This simplifying approach, although it is nothing but a two-state model for nucleation as it considers only the average fluctuations via the central limit theorem, it allows to draw some important conclusions more quantitatively.

\begin{figure}[htbp]
  \includegraphics[width=14cm]{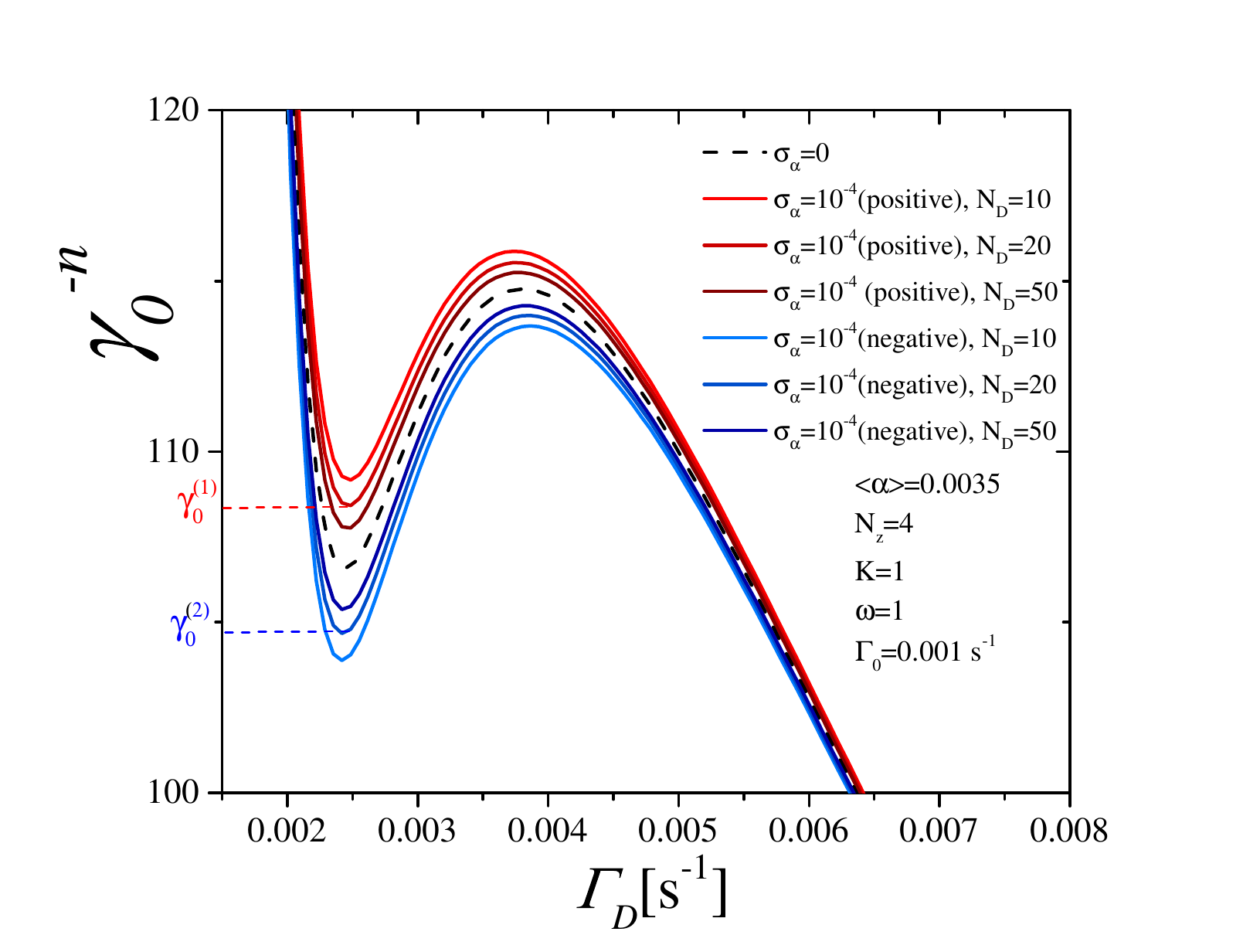}
  \caption{Equations of state drawn from equation \ref{droplet5} in the presence of finite disorder with positive and negative average fluctuations of coupling constants in droplets of size $N_D$ as indicated in the figure. All model parameters are indicated in the figure as well.}\label{disorderEOS}
\end{figure}

Figure \ref{disorderEOS} shows different equations of state for zero disorder and both positive and negative fluctuations of the coupling constants in the droplets. 
Let's suppose to progressively increase the strain so to move downwards along the solid branch of the curves and reach the point $\gamma_0=\gamma_0^{(1)}$. In absence of disorder, this point would correspond to a homogeneous solid-like state, since the system did not reach the local minimum. In the presence of quenched disorder however small droplets can nucleate and reach, on average, a typical size equal to $N_D=20$ for the case shown in figure \ref{disorderEOS}, while other regions will stay solid with an average dynamics slower than that predicted by the MF equation. On the other hand some regions of the system will stay solid even beyond the yield strain predicted for $\sigma_{\alpha}=0$, since they have not yet reached the critical set, namely the minimum of the (blue) mean field curves in figure \ref{disorderEOS}. This critical strain, we repeat for clarity, depends on the droplet size. Such a condition, for example, occurs at $\gamma_0=\gamma_0^{(2)}$, where only solid regions smaller than $N_D=20$ survive. This two-state approximation therefore confirms that i) a yielding transition starts at lower strain with respect to the MF prediction when disorder is taken into account, and ii) that solid-like domains persist well beyond the yielding threshold that the system would have in absence of disorder. This scenario is confirmed by simulations discussed in section (V) and (VII) of the main manuscript.

\section{Convergence kinetics}

\begin{figure}[htbp]
  \includegraphics[width=0.9\columnwidth]{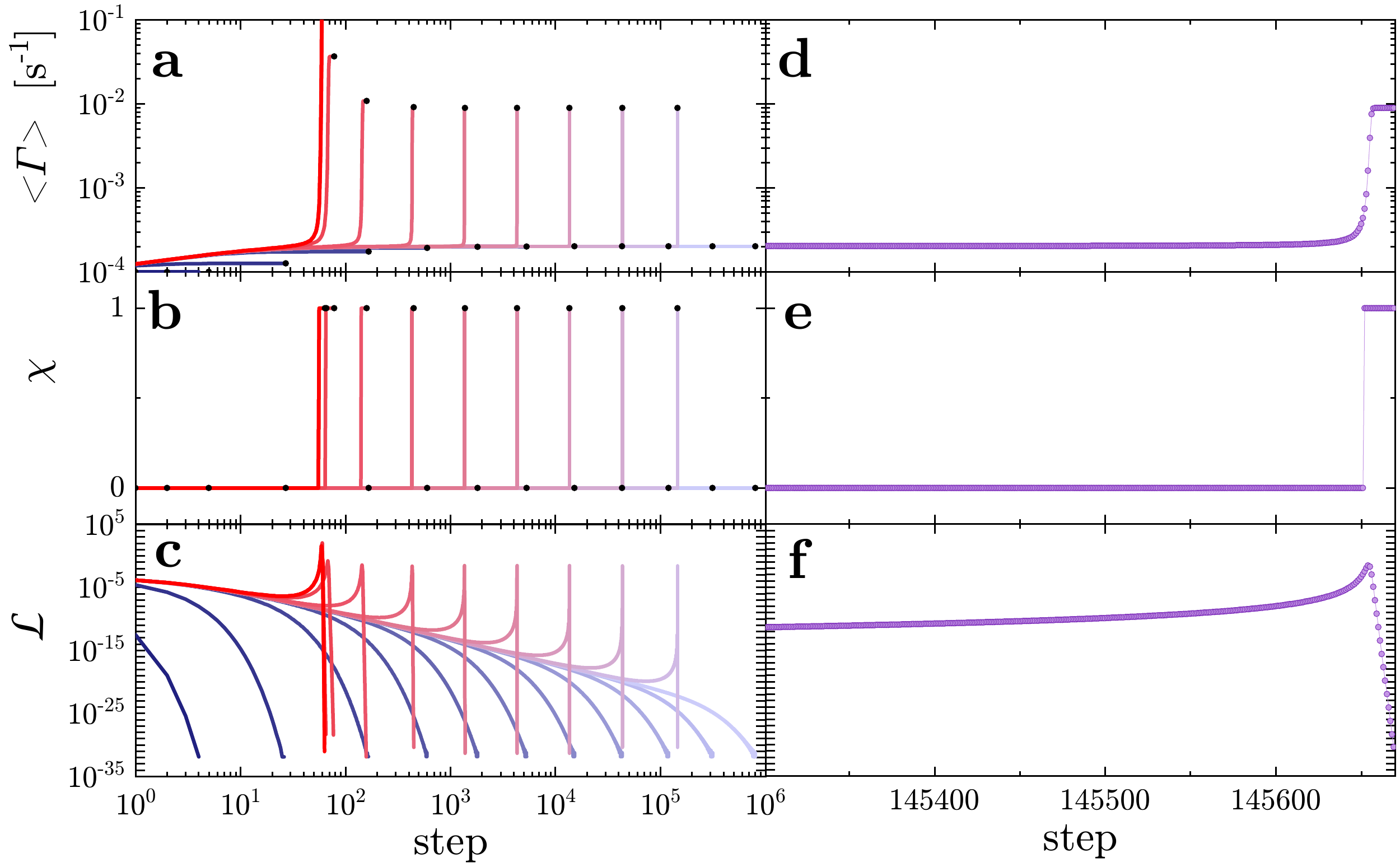}
\caption{Left column (a-c): evolution of the average relaxation rate $\langle \Gamma \rangle$ (a), the fraction of fast-relaxing sites $\chi$ (b) and the value of the loss function $\mathscr{L}$ (c), as a function of the number of iterations, for different values of the strain amplitude, increasing from blue to red shades. Black circles denote the convergence to the numerical solution. Right column (d-f): magnified view of the last iterations for the first strain amplitude above yielding.}\label{convergence_1}
\end{figure}

\begin{figure}[htbp]
  \includegraphics[width=0.8\columnwidth]{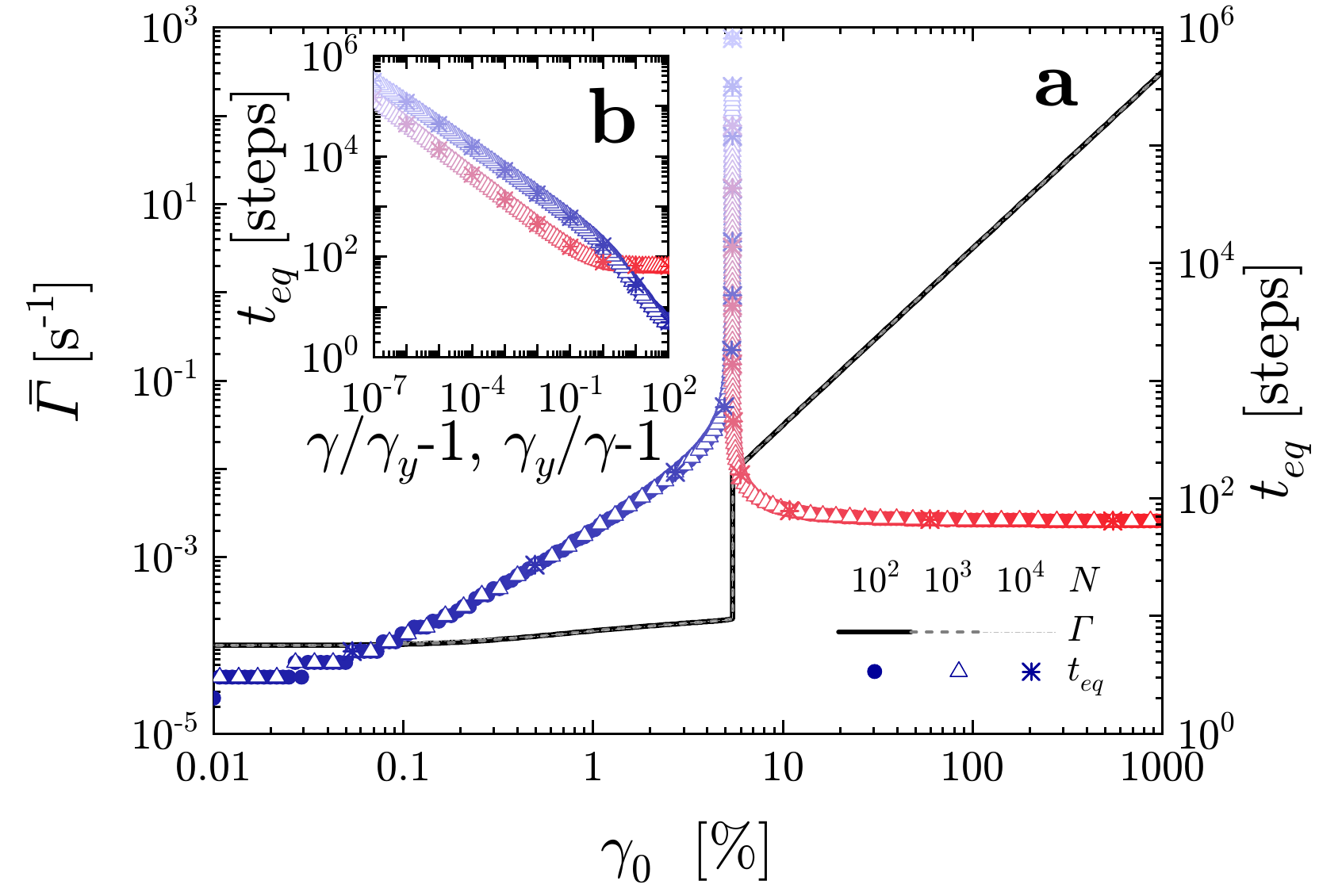}
\caption{(a) Black lines, left axis: average relaxation rate as a function of $\gamma_0$. Solid, dashed and dotted lines correspond to three system sizes, $N=10^2, 10^3, 10^4$, as specified in the legend. Blue to red symbols, right axis: iterations required to reach convergence. Symbol color changes with $\gamma_0$, and correspond to line colors in Figure \ref{convergence_1} (b) To highlight the power-law divergence at yielding, the same data from panel a are plotted in logarithmic scale, as a function of the distance from yielding: $\gamma_y/\gamma-1$ and $\gamma/\gamma_y-1$ for data below and above yielding, respectively.}\label{convergence_2}
\end{figure}

Figure \ref{convergence_1} shows the evolution of the average relaxation rate $\langle \Gamma \rangle$, the fraction of fast-relaxing sites $\chi$ and the value of the loss function $H$, as a function of the number of iterations. Figure \ref{convergence_2} shows the number of iterations needed to reach convergence as a function of strain amplitude, $\gamma_0$. 

\section{Statistics at fixed disorder (Figure 4 main text)}

\begin{figure}[htbp]
  \includegraphics[width=0.8\columnwidth]{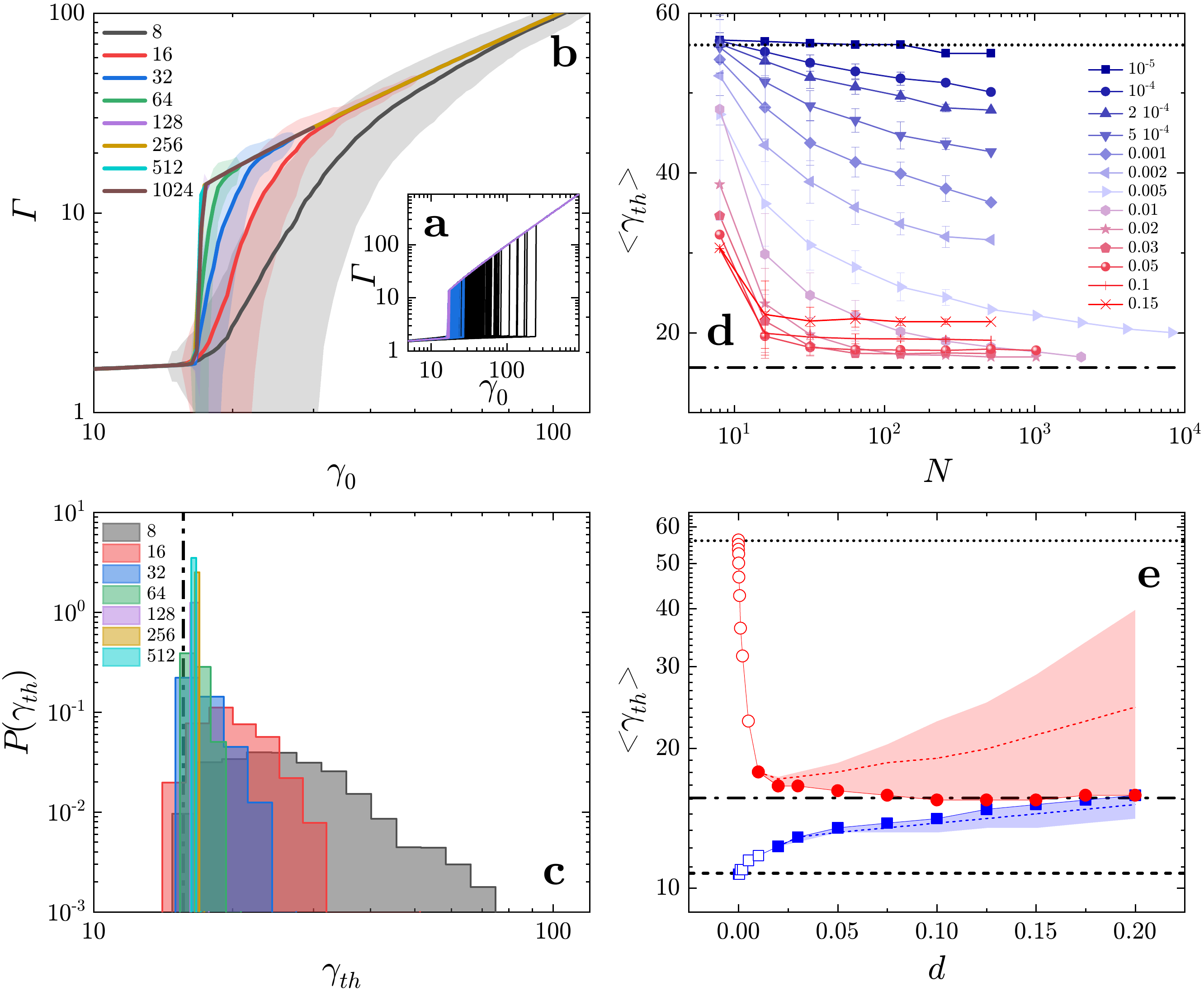}
\caption{(a) Average rate, $\Gamma$, as a function of $\gamma_0$, for a selection of independent realizations of $N\times N$ simulation lattices, with $N=8,32,128$ for black, blue and purple lines, respectively. (b) Solid lines: average on many independent simulations for different $N$, as specified in the legend. Shaded areas represent the dispersion of individual datapoints (c) Histograms of threshold strains, $\gamma_{th}$, defined at $\chi=0.5$, for many independent $N\times N$ simulations, with $N$ increasing from 8 to 512 as specified in the label. Dash-dotted line: $\gamma_y^{(M)}$. (d) average threshold strain, $\langle \gamma_{th} \rangle$, as a function of simulation lattice size, $N$, for various $d$ increasing from $10^{-5}$ to 0.15, as specified in the label. Error bars represent the standard variation of the threshold strain distribution. Dotted black line: mean-field threshold strain (delay convention). Dash-dotted line: $\gamma_y^{(M)}$. (e) Red solid line: $\langle \gamma_{th} \rangle$ measured with $N=512$ as a function of $d$. Red datapoints: threshold strain measured at $\chi=0.1$. Open symbols: simulations with abrupt yielding. Red shaded area: span between threshold strain defined at $\chi=0.1$ and $\chi=0.9$. Dotted black line: $\gamma_{th}^+$, dash-dotted black line: $\gamma_y^{(M)}$. Blue line, symbols and shaded area: threshold strain measured in a descending strain amplitude ramp. Dashed black line: mean-field prediction for the threshold strain in a descending ramp}\label{statistics_size}
\end{figure}

Figure \ref{statistics_size} reports the analysis of multiple independent $N\times N$ simulations at a fixed disorder, to quantify the variability and assess finite size effects.

